\documentclass[prd,nofootinbib]{revtex4}

\usepackage{amsmath,amssymb}
\usepackage{mathrsfs}
\usepackage{mathtools}
\usepackage{rotating}
\usepackage{mathrsfs}
\usepackage{pstricks}
\usepackage{pst-func}
\usepackage{graphicx}
\usepackage{cases}

\usepackage{soul}
\usepackage{xcolor}

\newcommand{\hll}[1]{\colorbox{yellow}{$\displaystyle #1$}}

\usepackage{enumitem}

\newtheorem{innercustomthm}{Theorem}
\newenvironment{customthm}[1]
  {\renewcommand\theinnercustomthm{#1}\innercustomthm}
  {\endinnercustomthm}

\def\BibTeX{{\rm B\kern-.05em{\sc i\kern-.025em b}\kern-.08em
    T\kern-.1667em\lower.7ex\hbox{E}\kern-.125emX}}

\begin{document}

\title{Response to `Comment on ``Quantum correlations are weaved by the spinors of the~Euclidean primitives'''}

\author{Joy Christian}

\email{jjc@bu.edu}

\affiliation{Einstein Centre for Local-Realistic Physics, Oxford OX2 6LB, United Kingdom}

\begin{abstract}
In this paper I respond to a critique of one of my papers previously published in the {\it Royal Society Open Science} entitled ``Quantum correlations are weaved by the spinors of the Euclidean primitives.'' Without engaging with the geometrical framework presented in my paper, the critique incorrectly claims that there are mathematical errors in it. I demonstrate that the critique is based on a series of misunderstandings, and refute each of its claims of error. I also bring out a number of logical, mathematical, and conceptual errors from the critique and the critiques it relies on.
\end{abstract}

\maketitle

\section{Introduction}\label{Intro}

The geometric framework proposed in my paper \cite{RSOS} is based on a Clifford-algebraic interplay between a quaternionic 3-sphere, or $S^3$, which I have taken to model the geometry of three-dimensional physical space, and an octonion-like 7-sphere, or $S^7$, which is an algebraic representation space of this quaternionic 3-sphere. This framework overcomes Bell's theorem by reproducing quantum correlations local-realistically as geometrical effects, without resorting to backward causation, superdeterminism, or any conspiracy loophole. It is summed up in the following theorem proven in \cite{RSOS}:
\begin{customthm}{3.1}\label{Th1}
{\it Every quantum mechanical correlation can be understood as a classical, local, realistic, and deterministic correlation among a set of points of an octonion-like 7-sphere.}
\end{customthm}
It is important to note that, contrary to the impression given in the critique \cite{Gill-RSOS}, it has not engaged with the actual contents of the framework presented in \cite{RSOS}. The readers are urged to read \cite{RSOS} for themselves to appreciate this fact. For a comprehensive review and detailed mathematical foundations of the 7-sphere framework for reproducing quantum correlations as geometrical effects, see also \cite{Reply-to-Lasenby}.

It is also important to note that I have already refuted many incorrect claims made by the same author of the critique \cite{Gill-RSOS} in several of my previous publications \cite{Aaronson,Disproof,IEEE-3,IEEE-4}. But since these claims have been repeated in \cite{Gill-RSOS}, some overlap in my point by point response to them below is not avoidable.

\section{Point by point response to the critique \cite{Gill-RSOS}} \label{Sec4}

\subsection{Incorrect claims in the first paragraph of the critique}

Let me begin with a number of incorrect claims and mistakes in the first paragraph of the critique. 
\begin{enumerate}[label=(\arabic*)]
\item It is claimed in the first paragraph of \cite{Gill-RSOS} that in my papers published between 2007 and 2021 I have proposed ``a {\it local hidden variable model} (but not always the same one)...'' However, in all my publications on the subject \cite{Disproof,IEEE-3,IEEE-4,IJTP,IEEE-1,IEEE-2}, what I have proposed is one and the same quaternionic 3-sphere model mentioned above. The critique's claim is thus an early indication that it has not quite understood what I have proposed in my papers \cite{Disproof,IEEE-3,IEEE-4,IJTP,IEEE-1,IEEE-2}.
\item This fact is confirmed as we read the paragraph further, which claims: ``Christian argues that Bell's proof of his theorem is mathematically wrong.'' But nowhere have I made such a claim. The mathematical inequalities on which Bell's argument depends are trivially correct (cf. Eq.~(4.9) of \cite{RSOS}). In fact, they were discovered and proven by Boole some one hundred and eleven years before the work of Bell \cite{Boole-1,Boole-2}. I have never questioned Boole's inequalities in any of my publications. Thus the claim made in the critique is not correct.
\item The next mistake in the Introduction is more serious. It is claimed that in my paper \cite{RSOS} I connect the 3-sphere, or $S^3$, ``to special relativity, specifically to the solution of Einstein's field equations known as Friedmann-Robertson-Walker spacetime with a constant spatial curvature." The same claim appears in several preprint versions of the critique \cite{Gill-RSOS} that have been posted on arXiv as well as in online discussions. The quoted sentence from \cite{Gill-RSOS} thus exhibits a lack of understanding of the difference between the special and general theories of relativity and how my proposed quaternionic 3-sphere model fits into the Friedmann-Robertson-Walker solution of Einstein's field equations of {\it general} relativity. In fact, much of the confusion in \cite{Gill-RSOS} stems from its failure to understand the difference between strong correlations within flat spacetime ${\mathrm{I\!R}}\times{\mathrm{I\!R}}^3$ and curved spacetime ${\mathrm{I\!R}}\times S^3$.
\item There is a further mathematical error in the first paragraph of \cite{Gill-RSOS}. Its last sentence reads: ``He furthermore connects it to the 7-sphere $S^7$, thought of as a quaternionic 3-sphere rather than a real 3-sphere." The two parts of the quoted sentence are even dimensionally incorrect. A 7-sphere cannot be thought of as 3-sphere in any sense. Their dimensions are not the same. Such misreadings of elementary facts indicate that critique's reading of \cite{RSOS} is mistaken and what it claims to be errors in \cite{RSOS} are, in fact, errors in the reading of \cite{RSOS}. 
\end{enumerate}
\subsection{Less than correct narrative and claims in the rest of its Introduction}
In the rest of its Introduction, the critique \cite{Gill-RSOS} makes several claims of equally questionable merit:
\begin{enumerate}[label=(\arabic*)]
\item It claims that in \cite{Gill-Entropy} a critique of my work was published. However, what is published in \cite{Gill-Entropy} is not a critique of my work at all. In it, a different model is presented based on a flat space, and using matrices and vector algebra, instead of the quaternionic 3-sphere model using Geometric Algebra I have proposed in my work \cite{Disproof,IEEE-3,IEEE-4,IJTP,IEEE-1,IEEE-2}.  This alternative model is then criticised, claiming the criticism to be that of my model. The critique in \cite{Gill-Entropy} is thus oblivious to the difference between its own flat space model and my 3-sphere model, and to the fact that such a comparison amounts to committing a logical fallacy. Moreover, in the abstract of \cite{Gill-Entropy} it is claimed that ``Christian's fundamental idea is simple and quite original: he gives a probabilistic interpretation of the fundamental GA equation $a\cdot b = (ab + ba)/2$.'' However, I have not proposed any such interpretation anywhere in my writings. Unfortunately, this is not the only thing that is mistaken in \cite{Gill-Entropy}. Several other equations attributed to my work are also misconstrued in \cite{Gill-Entropy}. For example, nowhere in my writings there appear anything like Eq.~(22) stipulated in \cite{Gill-Entropy}. It also does not follow mathematically from any other equations I have written down anywhere without violating the conservation of the initial zero spin angular momentum, as I have explained, for example, in Subsection~IV~E of \cite{IEEE-3}, Subsection~III~E of \cite{IEEE-4}, and Section~VIII of \cite{IEEE-1}. There are also other oversights in \cite{Gill-Entropy}, which I have brought out in \cite{Aaronson,Disproof,IEEE-3,IEEE-4}. In particular, \cite{Gill-Entropy} makes the same mathematical mistakes I bring out in detail below in the Subsection~\ref{Sec2.7}
\item In the Introduction of \cite{Gill-RSOS} it is claimed that other papers have also refuted my work. But to date, no one has refuted any part of my work, or undermined it in any way. To be sure, there have been attempts of refutation, but I have elucidated the errors in all such claims, for example in \cite{Aaronson,Disproof,IEEE-3,IEEE-4} and references cited therein. See, especially, Chapters 9 to 12 in \cite{Disproof}.
\item It is further claimed in the Introduction of \cite{Gill-RSOS} that Lasenby in \cite{Lasenby-AACA} has independently made the same claims about the algebraic core of the 7-sphere framework proposed in \cite{RSOS}. However, in \cite{Lasenby-AACA} it is acknowledged that ``... several of the points made [in \cite{Lasenby-AACA}] have been made independently by [the author of \cite{Gill-RSOS}] and others in the discussion thread attached to the Royal Society paper ...'' Thus it is not surprising that the claims in \cite{Lasenby-AACA} are similar to those in \cite{Gill-RSOS}. More importantly, in \cite{Reply-to-Lasenby} and \cite{Reply-Lasenby} I have refuted all of the claims made in \cite{Lasenby-AACA}.
\item The Introduction in \cite{Gill-RSOS} is summed up by quoting a few disjoint sentences out of context from Section~II of my reply in \cite{IEEE-3}. This allows \cite{Gill-RSOS} to misrepresent the argument I have presented in Section~II of \cite{IEEE-3} concerning Bell's so-called ``theorem", to which I now turn.
\end{enumerate}

\subsection{Bell's ``theorem" is not a theorem in the mathematical sense} \label{2.3}

One of the most serious misconceptions exhibited in the critique \cite{Gill-RSOS} is its presumption that Bell's so-called ``theorem'' is a proven theorem in the mathematical sense and therefore any critique of it must be wrong. But it is important to appreciate the difference between the mathematical inequalities used by Bell and his physical argument based on those inequalities. As noted above, the mathematical inequalities discovered by Boole \cite{Boole-1,Boole-2} on which Bell's theorem depends are trivially correct. Moreover, while even proven mathematical theorems may not be immune to refutations as so lucidly explained by Lakatos \cite{Lakatos}, Bell's theorem is not a theorem in the mathematical sense to begin with. It depends on a number of implicit and explicit {\it physical} assumptions, which can be and have been questioned before, not only by me \cite{Disproof,IJTP,RSOS,IEEE-1,IEEE-2,IEEE-3,IEEE-4}, but also by many others (cf. footnote~1 in \cite{IEEE-1}). If it were a theorem in mathematical sense, then it would not require physical experiments for its validity and any loophole (or ``gap") would render it invalid. 

It is important to note that  Bell's own writings do not exhibit any such misconception. Indeed, he actively sought strategies to overcome his ``theorem.'' In the concluding sentence of Chapter~17 of his book \cite{Bell-2004}, Bell reminds us that ``...what is proved by impossibility proofs is lack of imagination.'' More seriously, in Section 8 of Chapter 7 and Section 10 of Chapter 24 of his book, Bell points out that his theorem depends on the assumption of experimenters' ``free will'', which may turn out to be illusory. Thus Bell was well aware of the other physical and metaphysical assumptions that are necessary to support his theorem, in addition to those of locality and realism.

In the Introduction and Section~4.2 of \cite{RSOS}, and in Section~II of \cite{IEEE-3} and Section~III of \cite{IEEE-4}, I have highlighted several other physical assumptions that are necessary for Bell's theorem to hold. Among these, there are two assumptions that have been hitherto underappreciated, even though Bell himself has discussed them at least indirectly. Let me bring them out in some detail for clarity:

\subsection{Assumption of the additivity of expectation values}

The first among the two assumptions is the assumption of the additivity of expectation values:
\begin{align}
&\int_{\Lambda}{\mathscr A}({\bf a},\lambda){\mathscr B}({\bf b},\lambda)\,p(\lambda)\,d\lambda+\!\!\int_{\Lambda}{\mathscr A}({\bf a},\lambda){\mathscr B}({\bf b'},\lambda)\,p(\lambda)\,d\lambda+\!\!\!\int_{\Lambda}\!\!{\mathscr A}({\bf a'},\lambda){\mathscr B}({\bf b},\lambda)\,p(\lambda)\,d\lambda-\!\!\int_{\Lambda}\!\!{\mathscr A}({\bf a'},\lambda){\mathscr B}({\bf b'},\lambda)\,p(\lambda)\,d\lambda \notag \\
&\;\;\;\;=\!\!\int_{\Lambda}\!\Big\{\,{\mathscr A}({\bf a},\lambda)\,{\mathscr B}({\bf b},\lambda)+{\mathscr A}({\bf a},\lambda)\,{\mathscr B}({\bf b'},\lambda)+{\mathscr A}({\bf a'},\lambda)\,{\mathscr B}({\bf b},\lambda)-{\mathscr A}({\bf a'},\lambda)\,{\mathscr B}({\bf b'},\lambda)\,\Big\}\;p(\lambda)\,d\lambda, \label{id100}
\end{align}
where ${\Lambda}$ is the space of hidden variables ${\lambda}$ and ${p(\lambda)}$ is the probability distribution of ${\lambda}$, so that
\begin{equation}
{\cal E}({\mathbf a},{\mathbf b})=\int_{\Lambda}
{\mathscr A}({\mathbf a},\,\lambda)\,{\mathscr B}({\mathbf b},\,\lambda)\;p(\lambda)\,d\lambda,
\end{equation}
which is the usual expression of expectation value in the context of Bell's theorem \cite{Bell-2004}. Now it is universally accepted that without the validity of Eq.~(\ref{id100}), the Bell-CHSH inequalities cannot be derived. On the other hand, given the assumption of experimenters' ability to freely, randomly, or spontaneously choose the detector directions ${\mathbf a}$ and ${\mathbf b}$, which amounts to assuming $p(\lambda\,|\,{\mathbf a},{\mathbf b})=p(\lambda)$, mathematically Eq.~(\ref{id100}) follows at once. However, physically Eq.~(\ref{id100}) harbours a non-trivial assumption. This is obliquely recognized in the critique \cite{Gill-RSOS} in the last paragraph of its Section~2(b):
\begin{quote}
He also argues [in \cite{Oversight}] that Bell's proof contains a fundamental error in reasoning: the Bell-CHSH inequality involves correlations obtained from different sub-experiments involving measurements of non-commuting observables, and (he says) therefore cannot be combined. However, in quantum mechanics, even if two observables do not commute, a real linear combination of those observables is another observable. By the linearity encapsulated in the basic rules of quantum mechanics, expectation values of linear combinations of non-commuting observables are the same linear combination of the expectation values of each observable separately. If a local hidden variables model reproduces the statistical predictions of quantum mechanics, then it must reproduce this linearity.
\end{quote}
But the last sentence of the quoted paragraph is manifestly incorrect. It reveals a profound lack of understanding of what is meant by a hidden variable theory since the pioneering work of von~Neumann \cite{vonNeumann}. I have discussed the problem with Eq.~(\ref{id100}) in Section~II of \cite{IEEE-3} and in \cite{Oversight}. While mathematically correct, Eq.~(\ref{id100}) is physically meaningless within any hidden variable theory. It is an assumption over and above those of locality and realism. In fact, it is the same physical mistake that von~Neumann's former theorem against general hidden variable theories harboured. For observables that are not simultaneously measurable, such as those involved in Bell-test experiments, the replacement of the sum of expectation values with the expectation value of the sum, although respected in quantum mechanics, does not hold for hidden variable theories. This was pointed out by Einstein and Grete Hermann in the 1930s within the context of von~Neumann's theorem, and thirty years later by Bell \cite{Bell-1966} and others, as I have explained in \cite{Oversight}.

The example Bell \cite{Bell-1966} gives to illustrate this problem is that of the spin components of a fermion. A measurement of $\sigma_x$ can be made with a suitably oriented Stern-Gerlach magnet. But the measurement of $\sigma_y$ would require a different orientation of the magnet. And the measurement of the sum $(\sigma_x+\sigma_y)$ would again require a third and quite a different orientation of the magnet from the previous two. Consequently, the result of the last measurement --- {\it i.e.}, an eigenvalue of $(\sigma_x+\sigma_y)$ --- will not be the sum of an eigenvalue of $\sigma_x$ plus that of $\sigma_y$. The additivity of the expectation values, namely, $\langle\,\psi\,|\,\sigma_x\,|\,\psi\,\rangle+\langle\,\psi\,|\,\sigma_y\,|\,\psi\,\rangle=\langle\,\psi\,|\,\sigma_x+\,\sigma_y\,|\,\psi\,\rangle$, is a peculiar property of the quantum states $|\,\psi\,\rangle$. It would not hold for individual eigenvalues of non-commuting observables in a dispersion-free state of a hidden variable theory. In a dispersion-free state, every observable would have a unique value equal to one of its eigenvalues. And since there can be no linear relationship between the eigenvalues of non-commuting observables, the additivity relation that holds for quantum mechanical states would not hold for dispersion-free states \cite{Oversight}. 

In summary, the problem with Eq.~(\ref{id100}) is that, while the sum of expectation values appearing on its left-hand side is mathematically equal to the expectation value of the sum appearing on its right-hand side, and while this equality holds in quantum mechanics, it does not hold for hidden variable theories based on dispersion-free states \cite{Bell-1966}. That is because the eigenvalue of a sum of operators is not the sum of eigenvalues when the constituent operators are non-commuting, as in Bell-test experiments. In other words, while the joint results ${\mathscr A}(\mathbf{a},\lambda){\mathscr B}(\mathbf{b},\lambda)$, {\it etc.}, on the left-hand side of Eq.~(\ref{id100}) are possible eigenvalues of the spin operators ${\boldsymbol\sigma}_1\cdot{\bf a}\,\otimes\,{\boldsymbol\sigma}_2\cdot{\bf b}$, {\it etc.}, their summation
\begin{align}
{\mathscr A}({\bf a},\,\lambda)\,{\mathscr B}({\bf b},\,\lambda)+{\mathscr A}({\bf a},\,\lambda)\,{\mathscr B}({\bf b'},\,\lambda)+{\mathscr A}({\bf a'},\,\lambda)\,{\mathscr B}({\bf b},\,\lambda)-{\mathscr A}({\bf a'},\,\lambda)\,{\mathscr B}({\bf b'},\,\lambda) \label{sumof}
\end{align}
appearing as the integrand on the right-hand side of Eq.~(\ref{id100}) is {\it not} an eigenvalue of the operator
\begin{equation}
{\boldsymbol\sigma}_1\cdot{\bf a}\,\otimes\,{\boldsymbol\sigma}_2\cdot{\bf b}+{\boldsymbol\sigma}_1\cdot{\bf a}\,\otimes\,{\boldsymbol\sigma}_2\cdot{\bf b'}+{\boldsymbol\sigma}_1\cdot{\bf a'}\,\otimes\,{\boldsymbol\sigma}_2\cdot{\bf b}-{\boldsymbol\sigma}_1\cdot{\bf a'}\,\otimes\,{\boldsymbol\sigma}_2\cdot{\bf b'}, \label{spinop}
\end{equation}
because the joint operators ${\boldsymbol\sigma}_1\cdot{\bf a}\,\otimes\,{\boldsymbol\sigma}_2\cdot{\bf b}$, {\it etc.}, do not commute with each other. On the other hand, the very meaning of a hidden variable theory dictates simultaneous assignment of definite eigenvalues to {\it all} observables of the singlet system \cite{Oversight}, {\it including} the one in (\ref{spinop}), whether or not they are actually measured (albeit this assignment has to be contextual in the light of the Kochen-Specker theorem). But since the sum of results (\ref{sumof}) is not one of the eigenvalues of the summed operator in (\ref{spinop}), its appearance on the right-hand side of (\ref{id100}) is incorrect, making the replacement of the left-hand side of Eq.~(\ref{id100}) with its right-hand side at least physically invalid. But without this replacement the absolute upper bound of 2 on the left-hand side of Eq.~(\ref{id100}) cannot be derived.

Once this oversight is removed from Bell's ``theorem'' \cite{Bell-2004} and local realism is implemented correctly by using the correct eigenvalue of (\ref{spinop}) (which I have worked out explicitly in Appendix~A of \cite{Oversight}) instead of (\ref{sumof}) on the right-hand side of (\ref{id100}), the bounds on the left-hand side of (\ref{id100}) work out to be $\pm2\sqrt{2}$ instead of $\pm2$ (as I have demonstrated, for example, in Section~V of \cite{Oversight}), thereby mitigating the conclusions of Bell's theorem \cite{Oversight}. Consequently, what is ruled out by the Bell-test experiments is not local realism as widely believed, but the assumption of the additivity of expectation values, which does not hold in general for any hidden variable theories to begin with.

\subsection{Assumption of a flat and immutable spacetime} \label{Gravity}

The second assumption necessary to support Bell's theorem is that of immutable spacetime. The formulation of the theorem thus neglects the mutable spacetime geometries of Einstein's theory of gravity. Note that it is not the strength of gravity that is at stake here but the qualitative differences between immutable flat spacetime and mutable curved spacetimes \cite{Newton-Cartan}. This is hinted at by Bell himself. In Section 8 of Chapter 7 of his book \cite{Bell-2004}, while exploring possible strategies that may be used to overcome his theorem, he writes: ``The space time structure has been taken as given here. How then about gravitation?'' Thus Bell seems to have anticipated using a solution of Einstein's field equations of general relativity to overcome his theorem, as I have proposed in \cite{Disproof,IJTP,RSOS,IEEE-1,IEEE-2,IEEE-3,IEEE-4,Symmetric}.

By contrast, in the critique \cite{Gill-RSOS} Bell's theorem is stated using ``ordinary 3D Euclidean space'':
\begin{quote}
Suppose that $X_a$ and $Y_b$ are a family of random variables on a single probability space, taking values in the set $\{-1,+1\}$, and where $a$ and $b$ denote directions in ordinary 3D Euclidean space, represented by unit vectors $a$, $b$. Then it is not possible that ${\mathbb{E}(X_aY_b)=-a\cdot b}$ for all $a$ and $b$.
\end{quote}
Stated thus, the physical limitation of Bell's theorem is conspicuous. It is revealed in the following explicit assumption about the geometry of the three-dimensional physical space:
``... where $a$ and $b$ denote directions in ordinary 3D Euclidean space, represented by unit vectors $a$, $b$.'' But why must we allow such an outmoded view of physical space to make radical claims about the fundamental nature of locality and reality after more than a century of general-relativistic revolution in which spacetime geometry is rendered dynamical and malleable? And why must we use vector ``algebra'' after more than a century and a half of insights from Grassmann and Clifford on the correct algebraic representation of the three-dimensional physical space? As I have proposed in \cite{Disproof,IJTP,RSOS,IEEE-1,IEEE-2,IEEE-3,IEEE-4}, there are both theoretical and observational reasons that compel us to model physical space as a closed and compact quaternionic 3-sphere, or $S^3$, instead of a flat Euclidean space, or ${\mathrm{I\!R}^3}$, both being admissible spatial parts of one of the well-known cosmological solutions of Einstein's field equations of general relativity. Moreover, as explained in \cite{Disproof,IJTP,RSOS,IEEE-1,IEEE-2,IEEE-3,IEEE-4}, the correct language to model $S^3$ as physical space is Geometric Algebra, not vector ``algebra.'' But once the physical space is modelled as $S^3$ instead of $\mathrm{I\!R}^3$ using the powerful language of Geometric Algebra, the correlation between the local results ${\mathscr{A}_{\bf a}}$ and ${\mathscr{B}_{\bf b}}$ observed by Alice and Bob inevitably turns out to be ${\cal E}({\mathscr{A}_{\bf a}}{\mathscr{B}_{\bf b}})=-{\bf a}\cdot{\bf b}$ as I have proved in \cite{RSOS,Disproof,IJTP,IEEE-1,IEEE-2,IEEE-3,IEEE-4,Symmetric}, contrary to the claims made in \cite{Gill-RSOS}. For a comprehensive proof of ${\cal E}({\mathscr{A}_{\bf a}}{\mathscr{B}_{\bf b}})=-{\bf a}\cdot{\bf b}$ within the local-realistic geometry of $S^3$, I especially recommend the derivations of the singlet correlations in \cite{Reply-to-Lasenby} and \cite{Symmetric}.

Given the many implicit assumptions required to prove Bell's theorem in addition to locality and realism I have brought out in the Introduction and Section~4.2 of \cite{RSOS}, in Section~II of \cite{IEEE-3}, and in Section~III of \cite{IEEE-4}, together with the two assumptions discussed above, the infallibilist view of the theorem adhered to in the critique \cite{Gill-RSOS} (to borrow the term coined by Lakatos\cite{Lakatos}) is not justified. 

\subsection{Orientation $\lambda$ of ${\cal K}^{\lambda}$ acts as a hidden variable in the $S^7$ model}

Another issue raised in the critique \cite{Gill-RSOS} concerns the orientation of the vector space ${\cal K}^{\lambda}$ used in \cite{RSOS}:
\begin{quote}
A curious elementary mathematical error is that he defines two algebras, built from two 8-dimensional real vector spaces ${\cal K}^{+}$ and ${\cal K}^{-}$ by specifying a vector space basis for each algebra and multiplication tables for the 8 basis elements of each algebra. But they are the {\it same} algebra. The linear spans of those two bases are trivially the same. The multiplication operation is the same.
\end{quote}
However, there is no such error in \cite{RSOS}. I have not defined ``two algebras'' built from ``two'' vector spaces ${\cal K}^{+}$ and ${\cal K}^{-}$ and multiplication tables for ``each algebra.'' I have defined only {\it one} algebra, namely the even subalgebra ${\cal K}^{\lambda}$ of the Clifford algebra $\mathrm{Cl}_{4,0}$, with only {\it one} multiplication table for that algebra, namely the Table~1 on page~8 of \cite{RSOS}. The superscript ${\lambda=\pm}$ over ${\cal K}$ refers to the orientation (or handedness) of the corresponding vector space. Moreover, there is no claim in \cite{RSOS} that the linear spans of ${\cal K}^{+}$ and ${\cal K}^{-}$ are different. ${\cal K}^{\lambda}$ is one and the same algebra, or vector space, with two possible orientations. There is nothing unusual or unorthodox about this concept. It is clearly explained in Subsection 2.3 of \cite{RSOS} that ${\cal K}^+$ and ${\cal K}^-$ differ only in their orientations: $\lambda=+$ or $\lambda=-$. If one of them (say ${\cal K}^+$) is deemed right-handed, then the other one (${\cal K}^-$) is left-handed, and vice versa, with the definition of orientation from Milnor \cite{Milnor} reproduced on page 7 of \cite{RSOS}:
\begin{quote}
{\it Definition of Orientation}: An orientation of a finite dimensional vector space ${{\cal V}_n}$ is an equivalence class of ordered basis, say ${\left\{b_1,\,\dots,\,b_n\right\}}$, which determines the same orientation of ${\,{\cal V}_n}$ as the basis ${\left\{b'_1,\,\dots,\,b'_n\right\}}$ if ${b'_i =  \omega_{ij}\, b_j}$ holds with ${{\rm det}(\omega_{ij})>0}$, and the opposite orientation of ${{\cal V}_n}$ as the basis ${\left\{b'_1,\,\dots,\,b'_n\right\}}$ if ${b'_i = \omega_{ij}\, b_j}$ holds with ${{\rm det}(\omega_{ij}) < 0}$.
\end{quote}
On the page 8 of \cite{RSOS}, I have stated: ``It is easy to verify that the bases of ${{\cal K}^+}$ and ${{\cal K}^-}$ are indeed related by an ${8\times 8}$ diagonal matrix whose determinant is ${(-1)^7 < 0}$. Consequently, ${{\cal K}^+}$ and ${{\cal K}^-}$ indeed represent right-oriented and left-oriented vector spaces, respectively, in accordance with our definition of orientation. We can therefore leave the orientation unspecified and write ${{\cal K}^{\pm}}$ as ${\cal K}^{\lambda}=\,{\rm span}\left\{1,\,\lambda{\bf e}_x{\bf e}_y,\,\lambda{\bf e}_z{\bf e}_x,\,\lambda{\bf e}_y{\bf e}_z,\,\lambda{\bf e}_x{\bf e}_{\infty},\,\lambda{\bf e}_y{\bf e}_{\infty},\,\lambda{\bf e}_z{\bf e}_{\infty},\,\lambda I_3{\bf e}_{\infty}\right\},\;\lambda^2=1\Longleftrightarrow\lambda=\pm1.$''

In the later sections in \cite{RSOS}, ${\lambda}$ is taken as a Bell-type hidden variable, with physical consequences. Therefore ${{\cal K}^+}$ and ${{\cal K}^-}$ are physically not identical within the 7-sphere framework proposed in \cite{RSOS}.

\subsection{Algebra ${{\cal K}^{\lambda}}$ used in \cite{RSOS} is not incompatible with Hurwitz's theorem} \label{Sec2.7}

In its Section~2, the critique \cite{Gill-RSOS} claims that Hurwitz's theorem contradicts the mathematical claims in \cite{RSOS}. However, the critique seems to have missed Appendix~A of \cite{RSOS} where I have explicitly discussed Hurwitz's theorem in detail, as well as its significance for the 7-sphere framework presented in \cite{RSOS}. Overlooking this discussion, the critique claims a counterexample to Eq.~(2.40) of \cite{RSOS}, which says that for any multivectors $X$ and $Y$ in ${\cal K}^{\lambda}$, the following composition law holds:
\begin{equation}
||XY|| = ||X||\,||Y||, \label{not8}
\end{equation}
where the norms defined in \cite{RSOS} as $||X||:=\sqrt{XX^{\dagger}\,}$ using geometric products are positive definite:
\begin{equation}
||X||=0 \iff X=0\,. \label{pd}
\end{equation}
This definition is consistent with how norms are defined for complex numbers $c$, quaternions ${\bf q}$, and octonions ${\bf O}$. Namely, by $||{c}||=\sqrt{{c}\,{c}^{\dagger}}$, $||{\bf q}||=\sqrt{{\bf q}{\bf q}^{\dagger}}$, and $||{\bf O}||=\sqrt{{\bf O}{\bf O}^{\dagger}}$, respectively \cite{RSOS,Dray}. Moreover, a geometric product between $X$ and $X^{\dagger}$ in the definition $||X||:=\sqrt{XX^{\dagger}\,}$ is necessary for maintaining consistency between the two sides of Eq.~(\ref{not8}) because only a geometric product between $X$ and $Y$ can produce a new multivector $Z=XY$ in ${\cal K}^{\lambda}$ appearing on its left-hand side. On the other hand, because the algebra ${\cal K}^{\lambda}$ is a tensor product ${\mathbb H}\otimes\mathbb{C}'$ of quaternions with split complex numbers, its elements are of the form $X=\, {\bf q}_{r} + {\bf q}_{d}\,\varepsilon$ with $\varepsilon^2=1$, and consequently the quadratic form ${XX^{\dagger}}$ in general takes values in split complex numbers $\mathbb{C}'$ instead of real numbers. 

Challenging the norm relation (\ref{not8}) (which I have proved in Section~2.5 of \cite{RSOS}, Appendix~A.1 of \cite{Reply-to-Lasenby}, and Appendix~B of \cite{Eight}), the critique in \cite{Gill-RSOS} alleges the following counterexample without providing a definition of norm or specifying what it means by ``taking norms'' mathematically:
\begin{quote}
However, the author's algebra has an element called the `pseudo-scalar', I will denote it by $M$, such that $M^2 = 1$. It follows that $0 =M^2 - 1 = (M - 1)(M + 1)$. Taking norms, $0 = ||M - 1||.||M + 1||$. Hence $||M - 1|| = 0$ or
$||M + 1|| = 0$. Therefore $M - 1 = 0$ or $M + 1 = 0$, which implies that $M = 1$ or $M = -1$. That is a contradiction.
\end{quote}
The same claim appears in \cite{Gill-Entropy} and \cite{Gill-apology}, after the correction of a mistake I had pointed out in \cite{Aaronson}.

But it is easy to see that this alleged counterexample continues to harbour several mistakes. The first mistake in the quoted claim is immediately obvious. It starts out with the equation $M^2=1$ and ends with the equations $M=1$ or $M=-1$. And then it claims that ``That is a contradiction.'' However, the argument $M^2=1 \implies M=1$ or $M=-1$ by itself is not a contradiction. To allege a contradiction one must assume that $M$ is a pseudoscalar. But then it by no means follows from $||M - 1||\,||M + 1||=0$ that either $||M - 1|| = 0$ or $||M + 1|| = 0$ as alleged in the critique, unless two different product rules are employed on the two sides of equation (\ref{not8}) --- a geometric product, namely $(M - 1)(M + 1)=M^2-1$, to derive $0$ on the left-hand side of (\ref{not8}) and scalar products such as in $\sqrt{(M - 1)\cdot(M - 1)^{\dagger}}$ to evaluate the norms on its right-hand side. Needless to say, one can always derive a contradiction from {\it any} mathematical equation by employing two different product rules on the two sides of that equation\footnote{Some reviewers unjustifiably defended this counterexample during the review process of this paper. In addition to \cite{Reply-to-Lasenby}, my detailed rebuttal to their defence is available online in the Review History published along with this paper. All variants of the alleged counterexample depend on inconsistent application of product rules on the two sides of equation (\ref{not8}). For example, in the counterexample alleged in \cite{Lasenby-AACA}, a contradiction is achieved by employing geometric product between $X=(1+\varepsilon)/\sqrt{2}$ and $Y=(1-\varepsilon)/\sqrt{2}$ on the left-hand side of equation (\ref{not8}), whereas scalar products are used to evaluate its right-hand side.}. Barring that inconsistency, what the equation $||M - 1||\,||M + 1||=0$ says is that the {\it geometric} product between $||M - 1||$  and $||M + 1||$ must vanish, precisely because $M$, and therefore $||M - 1||$ and $||M + 1||$, are no longer scalar quantities, and a geometric product, namely $(M-1)(M+1)$, is used to derive 0 on the left-hand side of (\ref{not8}).

Indeed, the notation used in the critique for its equation ``$0 = ||M - 1||.||M + 1||$'' seems to tacitly acknowledge that $||M - 1||$ and $||M + 1||$ are not scalars. For otherwise there would be no need to introduce a ``dot'' between $||M - 1||$ and $||M + 1||$ indicating a scalar product between them when no such dot appears in the relation (\ref{not8}). But even with the scalar product introduced between $||M - 1||$ and $||M + 1||$ in an {\it ad hoc} manner, the critique's alleged conclusion that either $||M - 1|| = 0$ or $||M + 1|| = 0$ does not follow. Therefore the critique's claim of contradiction fails.

Now it is easy to prove that geometric product between $||M - 1||$  and $||M + 1||$ does vanish. In \cite{RSOS}, I have denoted the pseudoscalar in ${{\cal K}^{\lambda}}$ by ${\varepsilon:={\bf e}_1{\bf e}_2{\bf e}_3{\bf e}_{\infty}\not=\pm1}$. It satisfies the properties $\varepsilon^{\dagger}=\varepsilon$ and $\varepsilon^2=1$. Given this, what the critique has considered are the following elements in ${{\cal K}^{\lambda}}$:
\begin{equation}
X=\varepsilon-1\;\;\;\;\text{and}\;\;\;\;Y=\varepsilon+1. \label{not7}
\end{equation}
But, to begin with, no such two-dimensional multivectors play any role whatsoever in the 7-sphere framework presented in \cite{RSOS}. Therefore, even if such {\it ad hoc} two-dimensional objects lead to ``contradiction'' as alleged in the critique \cite{Gill-RSOS}, that would have no effect on or consequences for the 7-sphere framework. Nevertheless, it is instructive to play along with (\ref{not7}). The question then is: Do such multivectors in ${{\cal K}^{\lambda}}$ satisfy the norm relation (\ref{not8})? If they do not, then the claim made in the critique would be correct. But if they do, then the claim made in my paper would be correct. To investigate the question, we begin with evaluating the left-hand side of the norm relation (\ref{not8}):
\begin{equation}
||XY||=||(\varepsilon-1)(\varepsilon+1)||=\left|\left|\varepsilon^2-1\right|\right|=\left|\left|1-1\right|\right|=||\,0\,||=0, \label{12}
\end{equation}
where $\varepsilon^2=1$ is used. Next, using $\varepsilon^{\dagger}=\varepsilon$, we evaluate the right-hand side of the norm relation (\ref{not8}):
\begin{equation}
||X||=||(\varepsilon-1)||=\sqrt{(\varepsilon-1)(\varepsilon-1)^{\dagger}}=\sqrt{(\varepsilon-1)(\varepsilon-1)}=\sqrt{2\,(1-\varepsilon)}=1-\varepsilon \label{30X}
\end{equation}
and
\begin{equation}
||Y||=||(\varepsilon+1)||=\sqrt{(\varepsilon+1)(\varepsilon+1)^{\dagger}}=\sqrt{(\varepsilon+1)(\varepsilon+1)}=\sqrt{2\,(1+\varepsilon)}=1+\varepsilon, \label{30Y}
\end{equation}
and therefore
\begin{equation}
||X||\,||Y||=(1-\varepsilon)(1+\varepsilon)=1-\varepsilon^2=1-1=0. \label{20XY}
\end{equation}
Comparing (\ref{12}) and (\ref{20XY}) we see that the norm relation (\ref{not8}) is satisfied for the multivectors $X$ and $Y$ considered in (\ref{not7}). Thus, contrary to the claim in \cite{Gill-RSOS}, $X$ and $Y$ considered in (\ref{not7}) do not entail a counterexample to (\ref{not8}). The contradiction to (\ref{not8}) is achieved in \cite{Gill-RSOS} by computing norms incorrectly. Moreover, the non-scalar values of $||X||$ and $||Y||$ arrived at in (\ref{30X}) and (\ref{30Y}) reiterate the fact that $X$ and $Y$ considered in the critique \cite{Gill-RSOS} are not parts of the 7-sphere framework proposed in \cite{RSOS}.

\subsection{Proof of the norm relation (\ref{not8}) for the eight-dimensional algebra ${\cal K}^{\lambda}$}

Elsewhere \cite{Reply-to-Lasenby,Eight}, I have proved that the norm relation (\ref{not8}) holds, without exception, for arbitrary $X$ and $Y$ in ${\cal K}^{\lambda}$, and, as a special case, reduces to the one with scalar values for $||X||$, $||Y||$, and $||XY||$. The proofs of (\ref{not8}) and (\ref{pd}) are straightforward\footnote{During the review process of this paper, one of the reviewers independently verified this proof of the norm relations (\ref{not8})  and (\ref{R-11}) with detailed calculations and comments, which are available online in the Review History published with this paper.}. As noted above, the algebra ${\cal K}^{\lambda}$ is a tensor product $\mathbb{H}\otimes\mathbb{C}'$ of quaternions with split complex numbers \cite{Dray}, but with conjugation (or ``reverse") affecting only the quaternions \cite{RSOS}. In other words, any multivectors $X$ and $Y$ in ${\cal K}^{\lambda}$ are of the form $X=\, {\bf q}_{r1} + {\bf q}_{d1}\,\varepsilon$ and $Y=\, {\bf q}_{r2} + {\bf q}_{d2}\,\varepsilon$ with $\varepsilon^2=+1$ and $\varepsilon^{\dagger}=\varepsilon$, where ${\bf q}_{r1}$ and ${\bf q}_{d1}$ constituting $X$, for example, are two independent quaternions, which can be written as ${\bf q}_{r1}=g_1+I_3{\bf u}_1$ and ${\bf q}_{d1}=h_1+I_3{\bf v}_1$, where $g_1$ and $h_1$ are scalars, ${I_3={\bf e}_1{\bf e}_2{\bf e}_3}$ is the standard pseudoscalar in three dimensions, and ${{\bf u}_1=u_{1x}{\bf e}_x+u_{1y}{\bf e}_y+u_{1z}{\bf e}_z}$ and ${{\bf v}_1=v_{1x}{\bf e}_x+v_{1y}{\bf e}_y+v_{1z}{\bf e}_z}$ are Cartesian vectors. As a result, the geometric product $XX^{\dagger}$ between $X$ and $X^{\dagger}$ works out to be
\begin{align}
XX^{\dagger}\,&=\left({\bf q}_{r1} + {\bf q}_{d1}\,\varepsilon\right)\left({\bf q}_{r1} + {\bf q}_{d1}\,\varepsilon\right)^{\dagger}
\label{10a}\\
&=\left({\bf q}_{r1}\,{\bf q}^{\dagger}_{r1}\,+\,{\bf q}_{d1}\,{\bf q}^{\dagger}_{d1}\right)+\left({\bf q}_{r1}\,{\bf q}^{\dagger}_{d1}\,+\,{\bf q}_{d1}\,{\bf q}^{\dagger}_{r1}\right)\varepsilon \label{10b} \\
&=(g_1^2 + {\bf u}_1\cdot{\bf u}_1 + h_1^2 + {\bf v}_1\cdot{\bf v}_1) + ( 2\,g_1h_1 + 2\,{\bf u}_1\cdot{\bf v}_1)\,\varepsilon \label{10bbb} \\
&=\,\text{(a scalar)}+\text{(a scalar)}\;\varepsilon\,.
\end{align}
And likewise, using $Y=\, {\bf q}_{r2} + {\bf q}_{d2}\,\varepsilon$, together with ${\bf q}_{r2}=g_2+I_3{\bf u}_2$ and ${\bf q}_{d2}=h_2+I_3{\bf v}_2$, the geometric product $YY^{\dagger}$ between $Y$ and $Y^{\dagger}$ works out to be
\begin{align}
YY^{\dagger}\,&=\left({\bf q}_{r2} + {\bf q}_{d2}\,\varepsilon\right)\left({\bf q}_{r2} + {\bf q}_{d2}\,\varepsilon\right)^{\dagger}
\label{10ab}\\
&=\left({\bf q}_{r2}\,{\bf q}^{\dagger}_{r2}\,+\,{\bf q}_{d2}\,{\bf q}^{\dagger}_{d2}\right)+\left({\bf q}_{r2}\,{\bf q}^{\dagger}_{d2}\,+\,{\bf q}_{d2}\,{\bf q}^{\dagger}_{r2}\right)\varepsilon \label{10bc} \\
&=(g_2^2 + {\bf u}_2\cdot{\bf u}_2 + h_2^2 + {\bf v}_2\cdot{\bf v}_2) + ( 2\,g_2h_2 + 2\,{\bf u}_2\cdot{\bf v}_2)\,\varepsilon \label{10bcd} \\
&=\,\text{(a scalar)}+\text{(a scalar)}\;\varepsilon\,.
\end{align}

The positive definiteness (\ref{pd}) is now easy to prove. Since $X=\, {\bf q}_{r1} + {\bf q}_{d1}\,\varepsilon$, it can be zero only if both ${\bf q}_{r1}=0$ and ${\bf q}_{d1}=0$. But from (\ref{10b}) that implies $XX^{\dagger}=0$, and therefore $\sqrt{XX^{\dagger}}=\|X\|=0$. Thus $X=0\Longrightarrow\|X\|=0$ is proved. Conversely, suppose $\|X\|=0$. Then again from (\ref{10b}) we have $\|X\|=\sqrt{XX^{\dagger}}=\sqrt{\left({\bf q}_{r1}\,{\bf q}^{\dagger}_{r1}+\,{\bf q}_{d1}\,{\bf q}^{\dagger}_{d1}\right)+\left({\bf q}_{r1}\,{\bf q}^{\dagger}_{d1}+\,{\bf q}_{d1}\,{\bf q}^{\dagger}_{r1}\right)\varepsilon\,}=0$. But that is possible only if the quantities appearing in both parentheses under the square root are zero. However, both ${\bf q}_{r1}\,{\bf q}^{\dagger}_{r1}$ and ${\bf q}_{d1}\,{\bf q}^{\dagger}_{d1}$ in the first parenthesis $\left({\bf q}_{r1}\,{\bf q}^{\dagger}_{r1}+\,{\bf q}_{d1}\,{\bf q}^{\dagger}_{d1}\right)$ are positive definite scalars, and therefore they both must be zero for ${\bf q}_{r1}\,{\bf q}^{\dagger}_{r1}+\,{\bf q}_{d1}\,{\bf q}^{\dagger}_{d1}$ to be zero. But since ${\bf q}_{r1}\,{\bf q}^{\dagger}_{r1}$ and ${\bf q}_{d1}\,{\bf q}^{\dagger}_{d1}$ are positive definite scalars, they can be zero only if ${\bf q}_{r1}$ and ${\bf q}_{d1}$ are zero. In other words, both ${\bf q}_{r1}$ and ${\bf q}_{d1}$ must be zero for $\|X\|=0$ to hold. But ${\bf q}_{r1}=0$ and ${\bf q}_{d1}=0$ implies that $X=\, {\bf q}_{r1} + {\bf q}_{d1}\,\varepsilon=0$. Therefore, the converse $\|X\|=0\Longrightarrow X=0$ holds. The positive definiteness (\ref{pd}) is thus proved.

The proof of the norm relation (\ref{not8}) is also straightforward. Evidently, the geometric products $XX^{\dagger}$ and $YY^{\dagger}$ written above resemble split complex numbers because for any two quaternions (such as ${\bf q}_{r1}$ and ${\bf q}_{d1}$) the quantities appearing in the parentheses in (\ref{10bbb}) and (\ref{10bcd}) are scalar quantities. Thus the products are of the form $XX^{\dagger}=a+b\,\varepsilon$ and $YY^{\dagger}=c+d\,\varepsilon$, where $a={\bf q}_{r1}\,{\bf q}^{\dagger}_{r1}+{\bf q}_{d1}\,{\bf q}^{\dagger}_{d1}$, $b={\bf q}_{r1}\,{\bf q}^{\dagger}_{d1}+{\bf q}_{d1}\,{\bf q}^{\dagger}_{r1}$, $c={\bf q}_{r2}\,{\bf q}^{\dagger}_{r2}+{\bf q}_{d2}\,{\bf q}^{\dagger}_{d2}$, and $d={\bf q}_{r2}\,{\bf q}^{\dagger}_{d2}+{\bf q}_{d2}\,{\bf q}^{\dagger}_{r2}$ are scalar quantities, with $\varepsilon^{\dagger}=+\,\varepsilon$ instead of $-\,\varepsilon$. Consequently, for the left-hand side of (\ref{not8}) we have
\begin{align}
||XY||=\sqrt{(XY)(XY)^{\dagger}}=\sqrt{(XY)\left(Y^{\dagger}X^{\dagger}\right)}=\sqrt{X\left(YY^{\dagger}\right)X^{\dagger}}&=\sqrt{X(c+d\,\varepsilon)X^{\dagger}} \\
&=\sqrt{\left(XX^{\dagger}\right)(c+d\,\varepsilon)} \\
&=\sqrt{(a+b\,\varepsilon)(c+d\,\varepsilon)} \\
&=\sqrt{(ac+bd)+(ad+bc)\,\varepsilon\,}, \label{LH}
\end{align}
because, as noted, the pseudoscalar $\varepsilon$ satisfies $\varepsilon^2=1$ and commutes with every element of ${\cal K}^{\lambda}$, and consequently the identity $(XY)^{\dagger}=\left(Y^{\dagger}X^{\dagger}\right)$ for $X$ and $Y$ in ${\cal K}^{\lambda}$ is straightforward to verify. On the other hand, the right-hand side of the equation (\ref{not8}) also works out to give the same quantity:
\begin{equation}
||X||\,||Y||=\left(\sqrt{XX^{\dagger}}\;\right)\left(\sqrt{YY^{\dagger}}\;\right)=\sqrt{(a+b\,\varepsilon)(c+d\,\varepsilon)}=\sqrt{(ac+bd)+(ad+bc)\,\varepsilon\,}. \label{RH}
\end{equation}

Comparing (\ref{LH}) and (\ref{RH}) we see that when product rules are applied consistently on the two sides of equation (\ref{not8}) using geometric products, both of its sides work out to be identical to the square-root of the following quantity, which also resembles a split complex or hyperbolic number, 
\begin{align}
&\left\{\left(\varrho^2_{r1}+\varrho^2_{d1}\right)\left(\varrho^2_{r2}+\varrho^2_{d2}\right) \,+\,\left({\bf q}_{r1}\,{\bf q}^{\dagger}_{d1}+{\bf q}_{d1}\,{\bf q}^{\dagger}_{r1}\right)\left({\bf q}_{r2}\,{\bf q}^{\dagger}_{d2}+{\bf q}_{d2}\,{\bf q}^{\dagger}_{r2}\right)\right\} \notag \\
&\;\;\;\;\;\;\;\;\;\;\;\;\;\;\;\;\;\;\;\;+\left\{\left( \varrho^2_{r1}\,+\,\varrho^2_{d1}\right) \left({\bf q}_{r2}\,{\bf q}^{\dagger}_{d2}+{\bf q}_{d2}\,{\bf q}^{\dagger}_{r2}\right)
+ \left({\bf q}_{r1}\,{\bf q}^{\dagger}_{d1}+{\bf q}_{d1}\,{\bf q}^{\dagger}_{r1}\right)\left( \varrho^2_{r2}\,+\,\varrho^2_{d2}\right)\right\}\,\varepsilon\,, \label{not16}
\end{align}
thus proving the norm relation (\ref{not8}), where $\varrho_{r1}=\sqrt{{\bf q}_{r1}\,{\bf q}^{\dagger}_{r1}}\,$, {\it etc.}, and the quantities appearing in the two curly brackets are scalar quantities. Comparing (\ref{10b}), (\ref{10bc}), and (\ref{not16}), we thus see that the coefficient algebra underlying ${\cal K}^{\lambda}$ resembles that of split complex numbers\footnote{Hurwitz's theorem, as usually stated, assumes that the coefficient algebra underlying the algebras such as $\mathbb{R}$, $\mathbb{C}$, $\mathbb{H}$, and $\mathbb{O}$ is that of real numbers. It is known, however, that if the norms are generalized to take values in split complex numbers $\mathbb{C}'=\{a+b\,\varepsilon\,|\,\varepsilon^2=1\}$, then new composition algebras can be formed. See the discussion, for example, in Section~5.3 of \cite{Dray}.} instead of real numbers, and therefore what is just proved does not contradict Hurwitz's theorem \cite{RSOS,Dray}, contrary to the claims made in the critiques \cite{Gill-RSOS,Lasenby-AACA}. In fact, the result (\ref{not16}) holds for any multivectors $X$ and $Y$ in ${\cal K}^{\lambda}$, and therefore the norm relation (\ref{not8}) holds for any multivectors $X$ and $Y$ in the algebra ${\cal K}^{\lambda}$, because it is possible to work out square-root of a hyperbolic number such as (\ref{not16}), as shown in Appendix~A.1 of \cite{Reply-to-Lasenby}. The normalization or orthogonality condition specified in Eq.~(2.54) of \cite{RSOS} --- which amounts to setting the coefficient ${\bf q}_{r1}\,{\bf q}^{\dagger}_{d1}+{\bf q}_{d1}\,{\bf q}^{\dagger}_{r1}=0$ in (\ref{10b}) and ${\bf q}_{r2}\,{\bf q}^{\dagger}_{d2}+{\bf q}_{d2}\,{\bf q}^{\dagger}_{r2}=0$ in (\ref{10bc}) so that the norms $||X||$ and $||Y||$ reduce, respectively, to scalar values $\sqrt{\varrho_{r1}^2+\varrho_{d1}^2\,}$ and $\sqrt{\varrho_{r2}^2+\varrho_{d2}^2\,}$ --- then reduces the square-root of the hyperbolic number (\ref{not16}) to the scalar quantity
\begin{equation}
\sqrt{\left(\varrho^2_{r1}+\varrho^2_{d1}\right)\left(\varrho^2_{r2}+\varrho^2_{d2}\right)}\,.
\end{equation}
We have thus proved the norm relation (\ref{not8}) with scalar values, as originally proved in \cite{RSOS} and \cite{Eight}:
\begin{equation}
||XY|| = \sqrt{\left(\varrho^2_{r1}+\varrho^2_{d1}\right)\left(\varrho^2_{r2}+\varrho^2_{d2}\right)\,} = ||X||\,||Y||. \label{R-11}
\end{equation}
In particular, since the norm relation (\ref{not8}) holds for any multivectors $X$ and $Y$ in ${\cal K}^{\lambda}$ and therefore also for those with scalar values for their norms, if $X$ and $Y$ happen to be unit multivectors so that $||X||=1$ and $||Y||=1$, then (\ref{R-11}) necessitates that their product $Z = XY$ will also be a unit multivector, $||Z||=||XY||=||X||\,||Y||=1\times1=1$, contrary to the claims made in \cite{Gill-RSOS} and \cite{Lasenby-AACA}.

Eq.~(\ref{R-11}) also leads us to define a scalar-valued norm for any multivector ${\mathbb Q}_z=\, {\bf q}_{r} + {\bf q}_{d}\,\varepsilon$ in ${\cal K}^{\lambda}$,
\begin{equation}
||{\mathbb Q}_z||:=\sqrt{\left({\mathbb Q}_{z}{\mathbb Q}^{\dagger}_{z}\right)\!\Big|_{\,\varepsilon\cdot\left({\mathbb Q}_{z}{\mathbb Q}^{\dagger}_{z}\right)\,=\,0}\,}\;, \label{defnorm}
\end{equation}
which gives the same value for the norm as that obtained using the traditional {\it ad hoc} definition in Geometric Algebra (cf. Eq.~(2.8) in \cite{RSOS}). This is explained also with more detail in \cite{Eight}. As explained in Eqs.~(2.54) to (2.56) in \cite{RSOS}, the quadratic form in the definition (\ref{defnorm}) with split-complex image is
\begin{equation}
{\mathbb Q}_{z}{\mathbb Q}^{\dagger}_{z}=\left({\bf q}_{r}\,{\bf q}^{\dagger}_{r}+{\bf q}_{d}\,{\bf q}^{\dagger}_{d}\right)+\left({\bf q}_{r}\,{\bf q}^{\dagger}_{d}+{\bf q}_{d}\,{\bf q}^{\dagger}_{r}\right)\varepsilon=\left(\varrho_r^2+\varrho_d^2\right)+\left({\bf q}_{r}\,{\bf q}^{\dagger}_{d}+{\bf q}_{d}\,{\bf q}^{\dagger}_{r}\right)\varepsilon
\end{equation}
and $\varepsilon\cdot\left({\mathbb Q}_{z}{\mathbb Q}^{\dagger}_{z}\right)={\bf q}_{r}\,{\bf q}^{\dagger}_{d}+{\bf q}_{d}\,{\bf q}^{\dagger}_{r}$ is the scalar coefficient of $\varepsilon$. It is easy to see that definition (\ref{defnorm}) gives a scalar value for the norm as the Pythagorean distance from the origin in eight dimensions,
\begin{equation}
||{\mathbb Q}_z||=\sqrt{\varrho_r^2+\varrho_d^2\,}=\sqrt{g^2+u_x^2+u_y^2+u_z^2+h^2+v_x^2+v_y^2+v_z^2\,},  
\end{equation}
with the notation ${\bf q}_{r}=g+I_3{\bf u}$ and ${\bf q}_{d}=h+I_3{\bf v}$, and therefore the norms defined by (\ref{defnorm}) are also positive definite: $||{\mathbb Q}_{z}||=0\iff{\mathbb Q}_{z}=0$. More importantly, definition (\ref{defnorm}) immediately leads to the composition law (\ref{R-11}) with scalar values. Consequently, the algebra ${\cal K}^{\lambda}$ considered in \cite{RSOS} is a {\it normed division algebra} with respect to the norm defined in (\ref{defnorm}), as I have proved in \cite{Reply-to-Lasenby} and \cite{Eight}. 

Moreover, the definition (\ref{defnorm}) allows us to construct the following 7-sphere embedded in ${\cal K}^{\lambda}$:
\begin{equation}
{\cal K}^{\lambda}\hookleftarrow S^7=\,\left\{\,{\mathbb Q}_z=\,{\bf q}_r + {\bf q}_d\,\varepsilon\;\Big|\;||{\mathbb Q}_z||=\sqrt{\left({\mathbb Q}_{z}{\mathbb Q}^{\dagger}_{z}\right)\!\Big|_{\,{\bf q}_{r}\,{\bf q}^{\dagger}_{d}+{\bf q}_{d}\,{\bf q}^{\dagger}_{r}\,=\,0}\,}\,=\sqrt{\varrho_r^2+\varrho_d^2\,}\right\}. \label{sevsp}
\end{equation}
Theorem~\ref{Th1} stated in the Introduction and proved in \cite{RSOS} refers to this 7-sphere. It is also worth noting that definition (\ref{defnorm}) of the norm is valid also for the real and complex numbers, quaternions, and octonions, because for them the coefficient of $\varepsilon$ is naturally zero, where the complex numbers and quaternions are isomorphic, respectively, to the even sub-algebras of the Clifford algebras ${\mathrm{Cl}_{2,0}}$ and ${\mathrm{Cl}_{3,0}}$, whereas the algebra ${\cal K}^{\lambda}$ is isomorphic to the even subalgebra of the sixteen-dimensional algebra ${\mathrm{Cl}_{4,0}}$. Thus, the even sub-algebras of the Clifford algebras ${\mathrm{Cl}_{1,0}}$, ${\mathrm{Cl}_{2,0}}$, ${\mathrm{Cl}_{3,0}}$, and ${\mathrm{Cl}_{4,0}}$ form {\it associative} norm division algebras with respect to the norm defined in (\ref{defnorm}), isomorphic, respectively, to ${\mathbb R}$, ${\mathbb C}$, ${\mathbb H}$, and ${\cal K}^{\lambda}$, in the only possible dimensions 1, 2, 4, and 8 \cite{RSOS,Reply-to-Lasenby,Eight}. 

\subsection{The $S^7$ model concerns the workings of {\it Nature}, not computers}

The critique \cite{Gill-RSOS} puts forward two arguments that it claims are computer science versions of ``Bell's core result.'' But these arguments are as devoid of physical content as ``Bell's core result'', which is simply Boole's inequalities discussed in Subsection~\ref{2.3} above. They ignore the obvious fact that Alice and Bob are confined to perform their experiments within a three-dimensional physical space that is common to both of them. If this space is implicitly assumed to be a flat Euclidean space in the arguments put forward in \cite{Gill-RSOS}, then they are subject to the same criticism I have presented in Subsection~\ref{Gravity}. In particular, there is no indication that the geometry and topology of the quaternionic 3-sphere shared between Alice and Bob have been taken into account in either of the arguments. Therefore, they are not relevant for the physical experiments performed by Alice and Bob even within $\mathrm{I\!R}^3$. To put this more charitably, at best the arguments in \cite{Gill-RSOS} prove the poverty of computer programs that fail to capture the correct geometrical properties of $S^3$ and $S^7$, and therefore it is not surprising that they fail to reproduce the observed strong correlations.

By contrast, it is evident from the detailed, line-by-line annotations included in the codes of the two numerical simulations presented in \cite{RSOS} that they are faithful implementations of the physically motivated $S^7$ model in which $S^7$ is the algebraic representation space of a quaternionic 3-sphere. In particular, contrary to the claim made in the critique  \cite{Gill-RSOS}, the following line in the code,
\begin{equation}
\texttt{if(lambda==1) \{q=NA NB;\} else \{q=NB NA;\}},  \notag
\end{equation}
is a correct implementation of the orientation ${\lambda}$ in the 7-sphere model presented in \cite{RSOS}. It switches the order of the spin bivectors with respect to that of the detector bivectors, thereby shuffling the alternative orientations of the 7-sphere, precisely as necessitated in the model. In other words, the line in question is necessitated by the model itself. In fact, the line in question is {\it the very essence} of the 7-sphere model. Moreover, by now the codes have been independently translated by several programmers into different computer languages, such as {\it Python}, {\it Maple}, {\it R}, and {\it Mathematica} \cite{IEEE-3,IEEE-4}.

It is, however, important to note that the computer codes included in \cite{RSOS} are, by themselves, {\it not} the model, or even proofs of the model. Their purpose is to demonstrate how the model works. They are pedagogical tools that demonstrate the analytical derivations presented in \cite{RSOS}. Needless to say, the analytical derivations stand on their own and do not require numerical simulations for their validity. On the other hand, the numerical simulations in \cite{RSOS} provide additional support to the analytical derivations, because they are both pedagogically and statistically illuminating.

Ultimately, the merits of the geometrical framework presented in \cite{RSOS} can only be judged through Nature. Therefore, in \cite{IJTP,Symmetric} I have proposed an experiment, set in a macroscopic domain, that may be able to falsify the 3-sphere hypothesis underlying the framework presented in \cite{RSOS}. 

\section{Conclusion}\label{Sec5}

Without engaging with the actual local-realistic framework for reproducing quantum correlations I have proposed in \cite{RSOS}, the critique in \cite{Gill-RSOS} has claimed that there are errors in my papers published in \cite{RSOS,Disproof,IEEE-3,IEEE-4,IJTP,IEEE-1,IEEE-2}. In this response, I have demonstrated that there are no such errors in \cite{RSOS,Disproof,IEEE-3,IEEE-4,IJTP,IEEE-1,IEEE-2}. In fact, all of the claims made in \cite{Gill-RSOS} are either mistaken or irrelevant to the 7-sphere framework I have presented in \cite{RSOS}. Moreover, I have brought out a number of logical, mathematical, physical, and conceptual mistakes from the critique \cite{Gill-RSOS} and other arguments it has relied on (such as those in \cite{Lasenby-AACA}). As a result, the conclusions drawn in \cite{Gill-RSOS} are neither valid nor justified. I urge the readers to also consult my refutations in \cite{IEEE-3,IEEE-4} of the similar claims made by the author of \cite{Gill-RSOS}. Unfortunately, in the critiques \cite{Gill-RSOS} and \cite{Lasenby-AACA} the main message of \cite{RSOS} has been misunderstood and sidelined. Therefore, in \cite{Reply-to-Lasenby} and \cite{Symmetric} I have once again reviewed the mathematical and conceptual bases of the local-realistic framework presented in \cite{RSOS,Disproof,IEEE-3,IEEE-4,IJTP,IEEE-1,IEEE-2} to make its message more transparent. 

\vspace{-0.7cm}

\section*{Acknowledgements}

\vspace{-0.2cm}

The author thanks Prof. Tevian Dray for his insightful comments on Ref.~\cite{Eight}

\pagebreak

\parskip 0.25cm

\parindent 0.0cm

\baselineskip 0.476cm

\begin{center} 
\underbar{\bf\Large Peer Review History of the Paper:}
\end{center}

\begin{center}
\underbar{\bf\Large Response to Reviewers --- Round \# 1:}
\end{center}

\underbar{\textcolor{blue}{{\large\bf Reviewer \# 1}}}{\color{blue}{

Sections 1 and 3 are introduction and conclusion, comments will focus on section 2: 

Section 2.1: All points made here are valid and show that Gill has indeed misunderstood a lot.

{\color{black}{\underbar{\bf Author response}}: I thank the reviewer for this comment.

\underbar{\bf Author action}: No action.}

Section 2.2 (1): The response here regarding Gill's critique in reference 12 is largely valid except for the comments regarding Eq. (22) ``It also does not follow mathematically from any other equations I have written down anywhere." However Eq. (22) follows from equations (1) and (2) of your 2011 preprint for example ... 

{\color{black}{\underbar{\bf Author response}}: Eq. (22) of [12] does not follow from equations (1) and (2) of my 2011 preprint, which is a one-page summary I produced to meet frequent request by some that they do not have time to read my longer papers.

The 2011 preprint cites {\it six} other much longer and detailed preprints I had written prior to 2011, explaining all aspects of the quaternionic 3-sphere model. By 2020 when Gill's paper [12] was published, I had posted {\it ten} other papers on arXiv, again explaining all aspects of the 3-sphere model in more detail, including the {\it RSOS} paper [1] and two papers published in {\it IEEE Access}. What is more, my 2011 preprint itself was updated in 2015. That Gill has been aware of all of these developments is evident from the first line in the abstract of [12] and numerous online discussions. And yet, Gill's paper [12] cites the unpublished first version of my 2011 preprint, disregarding sixteen other papers I had written prior to 2020, to support its claim that Eq.~(22) of [12] is what the quaternionic 3-sphere model predicts.

\underbar{\bf Author action}: No action.}

... and if this is being misunderstood it indicates that you are not using mathematical notation clearly and correctly. I suggest that the discussion of Eq. (22) be removed from the paper.

{\color{black}{\underbar{\bf Author response}}: Equations (1) and (2) in my 2011 preprint are indeed misconstrued by Gill in [12]. Eq.~(22) of [12] does not follow from them since they are defined as products of two bivectors, representing the physical interaction during the detection processes between detector bivectors chosen by Alice and Bob and the spin bivectors originating at the source. This is evident from the definitions (3) and (4) of the bivectors and the notation $\lambda$, which, in the literature on Bell's theorem, represents the initial state of the spin system originating at the source. As a result, using (1) and (2), the product $AB$ of the measurement outcomes is a product of two non-pure quaternions in general, and therefore cannot be equal to $-1$ for all settings ${\bf a}$ and ${\bf b}$ as claimed in [12]. This is evident from the correlations (7), which are computed in the preprint using the {\it Pearson correlation coefficient}. All this is misunderstood in [12].  

Thus the notations in my 2011 preprint, or in any other paper, are just fine, and I have also used them correctly. The real reason for misunderstanding is elsewhere and it is conceptual. What is not understood is that the quaternionic 3-sphere model is {\it not} a model of after-the-event analysis followed by experimenters to compute correlations, but a {\it theoretical} model that predicts that the correlations will be strong if produced within a quaternionic 3-sphere (rather than a flat Euclidean space), taken as the 3D physical space. I have explained this conceptual difference many times before, especially in my replies [5] and [6] to Gill, and references therein. Once this conceptual difference is understood, it becomes evident that Eq.~(22) of [12] does not follow from the definitions (1) and (2) stated in my 2011 preprint.

\underbar{\bf Author action}: I have revised the last but one sentence of Section~2.2 (1) to: ``It also does not follow mathematically from any other equations I have written down anywhere, as I have explained in [6].''}

Section 2.2 (2) - (4) are subjective - readers can judge for themselves.

Section 2.3 makes valid points, ``Bell's Theorem" and indeed other similar ``no-go theorems" are indeed not mathematical theorems and are typically flim-flam based on erroneously and irrationally claiming that realist models must have a structure of a single joint probability space despite such a model neither being implied by nor implying realism. Indeed ``Bell's Theorem" and other ``no-go theorems" have been thoroughly and repeatedly debunked in the physics literature and are considered to have no relevance to modern approaches to QM such as the Consistent Histories formalism which explicitly rejects the idea that QM is non-realist or non-local while recognizing the need for different probability spaces for different frameworks.

{\color{black}{\underbar{\bf Author response}}: In my approach, I have not followed the Consistent Histories formalism or any other version or interpretation of quantum mechanics. Not a single concept from quantum mechanics is used in the 3-sphere model. As such, it is independent of any interpretation of quantum mechanics. The Consistent History formalism is not a universally accepted formalism among the experts in foundations of quantum mechanics.

\underbar{\bf Author action}: No action.}

Section 2.4 starts off making correct observations but becomes confused towards the end. The notion of Eq. (1) being mathematically valid but not physically valid makes no sense.

{\color{black}{\underbar{\bf Author response}}: There is no confusion of any kind in Section~2.4. The notion of Eq.~(1) being mathematically valid but physically invalid makes perfect sense. It is not something I have invented myself. It is an argument well known and near universally accepted in the foundations of quantum mechanics within the context of von~Neumann's theorem against hidden variable theories. The problem with Eq.~(1) was recognized by Einstein in the mid 1930s (see Ref.~[1] and [2] below). It was also recognized by Grete Hermann around the same time (see Ref.~[3] below). Thus the problem has been well known in foundations of quantum mechanics for nearly nine decades. One of the most lucid explanations of the problem can be found in Section~3 of Chapter~1 in Bell's book. The chapter has been reprinted from one of his papers (see Ref.~[4] below). The same problem with Eq.~(1) in the context of von~Neumann's theorem is also discussed by several other well known experts in foundations of quantum mechanics (see Refs.~[5] to [10] below).

Briefly, the problem is that, while the sum of expectation values is mathematically the same as the expectation value of the sum, as in the assumption that allows the replacement of the left-hand side of Eq.~(1) in the manuscript by its right-hand side, and while this assumption is valid in quantum mechanics, it is not valid for any hidden variable theory involving dispersion-free states, because the eigenvalue of a sum of operators is not the sum of individual eigenvalues when the constituent operators do not commute, as in the case of Bell-test experiments. This makes Eq.~(1) physically invalid. But without it the absolute upper bound of 2 on the CHSH correlator cannot be derived. Therefore the problem I have pointed out in Section~2.4 is very serious. It invalidates all derivations of the bound~of~2.

[1] J. von Neumann, {\it Mathematical Foundations of Quantum Mechanics} (Princeton University Press, Princeton, 1955).

[2] A. Shimony, {\it Search for a Naturalistic World View}, vol. II (Cambridge University Press, Cambridge, 1993), p. 89.

[3] G. Hermann, Die naturphilosophischen Grundlagen der Quantenmechanik, Abhandlungen der Fries’schen Schule {\bf 6}, 75 (1935).

[4] J. S. Bell, On the problem of hidden variables in quantum mechanics, Rev. Mod. Phys. {\bf 38}, 447 (1966).

[5] J. M. Jauch and C. Piron, Can hidden variables be excluded in quantum mechanics?, Helv. Phys. Acta {\bf 36}, 827 (1963).  

[6] A. Siegel, in {\it Differential Space, Quantum Systems, and Prediction}, eds: N. Wiener, A. Siegel, B. Rankin, and W. T. Martin (MIT Press, Cambridge, MA, 1966).

[7] S. Kochen and E. P. Specker, The problem of hidden variables in quantum mechanics, J. Math. Mech. {\bf 17}, 59 (1967).

[8] M. Jammer, {\it The Philosophy of Quantum Mechanics} (John Wiley and Sons, New York, NY, 1974).

[9] N. D. Mermin, Hidden variables and the two theorems of John Bell, Rev. Mod. Phys. {\bf 65}, 803 (1993).

[10] N. D. Mermin and R. Schack, Homer nodded: von Neumann's surprising oversight, Found. Phys. {\bf 48}, 1007 (2018).

\underbar{\bf Author action}: I have revised Section~2.4 to explain the problem and the above points more clearly.}

However it is indeed the case that Eq. (1) cannot be derived for sensible realist models of Bell experiments but this is because different probability spaces are required for incompatible measurement directions so that neither of the (mathematically equal) sides of Eq. (1) would be valid. And indeed that such different probability spaces are required is precisely what the standard QM formalism tells us and this is indeed related to the fact that spin is described by spin operators and their eigenvalues with non-commuting operators producing different probability spaces even for a fixed wavefunction - the eigenvalues are not simple random variables defined on a single joint probability space and cannot be treated as such. 

{\color{black}{\underbar{\bf Author response}}: The argument mentioned by the reviewer is a different argument. Elsewhere I have presented similar arguments, but without appealing to any quantum mechanical concepts (see, for example, references [1], [5], [6], [8], and [9] cited in the manuscript). Since Bell's theorem is not a theorem within quantum mechanics, it cannot be undermined by arguments that use quantum mechanical concepts. That would only reinforce the view that overcoming\break Bell's argument necessitates quantum concepts. In any event, discussing such an argument again is beyond the scope of the current manuscript, the purpose of which is to respond to the specific issues brought up in the critique by Gill. 

\underbar{\bf Author action}: No action.}

I would recommend that the discussion of mathematically valid vs physically valid be removed from this section and replaced with comments emphasizing the need for different distributions for different measurement directions. Rather than indicating that realism is wrong, this is a simple consequence of the fact that when dealing with complete states described by a function on a continuum instead of a finite sequence of numbers (as in classical mechanics) there is no single consistent way to measure the size of subsets of states corresponding to various measurement outcomes - a purely mathematical fact.

{\color{black}{\underbar{\bf Author response}}: That is not the argument I have put forward in the manuscript. Although, as noted above, elsewhere I have presented similar arguments, but without appealing to any quantum mechanical concepts (see, for example, references [1], [5], [6], [8], and [9] cited in the manuscript). Moreover, I have not claimed anywhere that realism is wrong. Physical invalidity of Eq.~(1) does not imply denial of realism. On the contrary, what I have argued is that Eq.~(1), and consequently Bell's argument, does not implement realism correctly by assigning definite values to {\it all} possible observables of the spin system, regardless of whether they are actually observed. In particular, it assigns incorrect definite value to the observable that is a sum of those observables appearing on the left-hand side of Eq.~(1). 

There is no reason to remove the argument I have presented in Section~2.4. It is one of the best and ironic arguments against Bell's theorem. The argument is ironic because Bell himself has made the problem with Eq.~(1) well known among the experts in the foundations of quantum mechanics within the context of von~Neumann's theorem (see the references I have listed above). What I have shown in Section~2.4 is that this problem also undermines Bell's theorem because Eq.~(1) --- encoding the assumption of the additivity of expectation values --- is essential for both theorems.   

\underbar{\bf Author action}: I have revised Section~2.4 to explain the problem and the above points more clearly.}

Section 2.5 makes a valid point, Gill ignores your space-time model and argues using a different model. 

{\color{black}{\underbar{\bf Author response}}: Indeed. That is the main defect of the critiques of my 3-sphere model by Gill and others.

\underbar{\bf Author action}: No action.}

However whether your proofs of the correlations using your model are mathematically sensible and rigorous is questionable, but this remains for the reader to decide themselves.

{\color{black}{\underbar{\bf Author response}}: Fair enough. While everything in science is questionable, in my view the 3-sphere model is not only mathematically sensible and rigorous but also adopted by Nature, which is the ultimate judge of its validity via experiment. 

\underbar{\bf Author action}: I have added reference [21] in the revised version of the manuscript. It discusses anew a macroscopic experiment I have proposed in [7] to test the quaternionic 3-sphere hypothesis. Thus it is a falsifiable model. It is my latest preprint on the arXiv that summarizes my approach to quantum correlations, and presents a self-contained and simpler derivation of the singlet correlations within $S^3$ that, I hope, will remove any remaining doubts in the validity of the 3-sphere model.}

Section 2.6 makes a valid point, both Lasenby and Gill have misunderstood that you are referring to different oriented vector spaces not to different vector spaces regardless of orientation.

{\color{black}{\underbar{\bf Author response}}: I thank the reviewer for this observation.

\underbar{\bf Author action}: No action.}

Section 2.7 starts with a valid point, your original paper shows that you are well aware of Hurwitz's theorem.

{\color{black}{\underbar{\bf Author response}}: I thank the reviewer for this observation. I have discuss Hurwitz's theorem in Appendix~A of [1].

\underbar{\bf Author action}: No action.}

However your reply to Gill's counterexample is not mathematically correct 

{\color{black}{\underbar{\bf Author response}}: My reply to Gill's counterexample is mathematically correct. It is Gill's counterexample that is based on a number of elementary mistakes, as I have demonstrated in both versions of the manuscript. 

I have addressed this issue with considerably more detail in my responses below to Review \# 2 and Review \# 6.

\underbar{\bf Author action}: In the revised manuscript I have improved my refutation of Gill's alleged counterexample.}

and shows that there is a problem in your definition of the algebra $K^{\lambda}$ - 

{\color{black}{\underbar{\bf Author response}}: There is no problem with the definition of the algebra ${\cal K}^{\lambda}$. It is simply the eight-dimensional even subalgebra of the algebra $\mathrm{Cl}_{4,0}$. Thus the algebra ${\cal K}^{\lambda}$ is a perfectly well defined, associative Clifford algebra. 

\underbar{\bf Author action}: No action.}

this is the issue that Lasenby raised in his critique and although he has misunderstood parts of your paper such as the fact that $K^+$ and $K^-$ are different oriented spaces, he is correct on his point about the properties of your pseudoscalar. 

{\color{black}{\underbar{\bf Author response}}: Lasenby's critique based on a variant of the alleged counterexample is also incorrect. In it, the contradiction is achieved by applying two different product rules on the two sides of the relation $||XY||= ||X||\,||Y||$.

I have addressed this issue with considerable more detail in my responses below to Review \# 2 and Review \# 6.  

\underbar{\bf Author action}: No action.}

The last expressions in (6) and (7) of your reply show that the epsilon must in fact be a scalar. This attempt to refute Gill's counterexample should be removed.

{\color{black}{\underbar{\bf Author response}}: The last expressions in (6) and (7) by no means show that $\varepsilon$ must be a scalar. They are results of elementary computation of norms and work out to be split complex numbers, with a scalar part and a pseudoscalar part. The fact that they are not scalars only indicates that they are not parts of the 7-sphere framework proposed in my paper [1]. The {\it ad hoc} two-dimensional objects considered by Gill and Lasenby play no role whatsoever in the 7-sphere framework proposed in [1]. For a fuller explanation see my responses below to Review \# 2 and Review \# 6.  

\underbar{\bf Author action}: I have revised Section~2.7 to make the above points more clear.}

Section 2.8 Gill's so called computer science no-go theorems are indeed not relevant as they apply to computer simulations that are obviously too limited to model the differing probability spaces for different measurement directions implied by QM. However the code routines you provide in the original paper are not actually simulations of Bell experiments they merely calculate the individual correlations. Here mention should be made to your RPubs code which is a simulation.

{\color{black}{\underbar{\bf Author response}}: That is correct. The codes presented in the original 2018 paper are not meant to be simulations of Bell-test experiments. They are included in the paper as validations of the theoretical calculations presented therein. 

\underbar{\bf Author action}: I have revised Section~2.8 considerably to make this point more clear.}

\bigskip

\bigskip

\hrule 

\bigskip 

\underbar{\textcolor{blue}{{\large\bf Reviewer \# 2}}}

Dear Editor,

I must recommend that this paper will not be accepted for publication.
The paper aims to provide a causal explanation for the violation of the Bell inequality observed in experiments. However, the arguments presented by the author are either unconvincing or plainly wrong.

For example, the arguments presented by the author in section 2.4 of the paper seem to rule out his own model. 

{\color{black}{\underbar{\bf Author response}}: On the contrary, what is argued in Section~2.4 of the manuscript is that the bounds of $\pm2$ on the CHSH correlator are not correct because they are based on the assumption of the additivity of expectation values, which is well known to be invalid for any hidden variable theory involving dispersion-free states. But since the bounds of $\pm2$ are not valid, Bell's theorem does not rule out models like the one I have presented in my original Royal Society paper published in 2018. The correct bounds on the CHSH correlator are  $\pm2\sqrt{2}$, which are respected by my model.

\underbar{\bf Author action}: I have revised Section~2.4 of the manuscript to make this point more clear.}

Actual experiments measure the correlation between the outcomes of two detectors for certain settings of the latter. The CHSH correlator is obtained after all these four needed individual correlations have been measured. The author argues that in his model the CHSH correlator is not an additive function of the four involved correlations, hence implying that his model does not reproduce the individual correlations measured in experiments.

{\color{black}{\underbar{\bf Author response}}: What is claimed by the reviewer above is not what is presented in Section~2.4 of the manuscript. Please refer to my previous answer for a brief summary of what is actually presented in Section~2.4 of the manuscript.

\underbar{\bf Author action}: No action.}

Neither the hand-waving arguments provided by the author in section 2.5 are more convincing. I agree with the author that any implicit or explicit assumption needed in the derivation of Bell's theorem must be carefully examined before radical claims are made about the nature of physical reality. But how is the geometry of space-time relevant for explaining Bell's experiments? How are the cosmological solutions of GR relevant to tabletop Bell's experiments? Why are high-energy particle physics experiments precisely reproduced by quantum field theories formulated in flat Minkowski space-time while general relativity would be needed in order to reproduce the Bell experiments?

{\color{black}{\underbar{\bf Author response}}: The arguments presented in Section~2.5 of the manuscript are hardly ``hand-waving.'' They are meant to address a specific statement made in the critique [2]. On the other hand, I do appreciate that more detailed answers to the questions raised above by the reviewer are desirable. Unfortunately, these questions cannot be addressed adequately in a Reply to Comment paper, the length of which is editorially restricted to the length of the Comment paper [2]. But I have indeed addressed all of the questions raised by the reviewer in extensive details for the past fifteen\break years, in the references [1], [3], [4], [5], [6], [7], [8], [9], [14], [17], [19], and [21] of the revised version of the manuscript. 

\underbar{\bf Author action}: I have added reference [21] in the paper. It addresses some of the questions raised by the reviewer.}

Finally, I wish to add that the author's argument in section 2.7 seems to be intentionally misleading.  
The author claims that his eq.(3) holds for any elements X, Y belonging to a certain algebra. Furthermore, the author claims that the said algebra contains an element epsilon, other than the identity I and its opposite -I, which satisfies the constraint epsilon$^2$=I. Nonetheless, in the comment to his paper by R.Gill, it has been noticed to the author that these two claims do contradict each other, as it can be readily shown as follows. First, epsilon$^2$ = I implies that (epsilon-I)(epsilon+I)=0, while eq.(3) then implies that $||(\text{epsilon-I})(\text{epsilon+I})||=||\text{epsilon-I}||\, ||\text{epsilon+I}||=0$, from which it follows that either epsilon = I or epsilon = -I, in contradiction with the starting assumptions.

In section 2.7 the author claims to refute this trivial proof. Instead, the author waves his hands and misleads the reader to think that the criticism is about if eq.(3) either holds or not in his algebra. The author then shows that eq.(3) holds in a particular case and he claims to have refuted the criticism. I find this author's strategy rather troublesome. 

{\color{black}{\underbar{\bf Author response}}: I am  afraid the above argument by the reviewer has missed the point I have made entirely. To begin with, if (as the reviewer claims) the argument in the critique [2] is {\it not} ``about if eq.(3) either holds or not in'' the algebra ${\cal K}^{\lambda}$ considered in my original 2018 paper, then the critique's point is irrelevant and there is no reason for me address it. But of course the critique is indeed about whether eq.(3), namely the norm relation $||XY||= ||X||\,||Y||$, holds or not in the algebra ${\cal K}^{\lambda}$ considered in my original 2018 paper.

Secondly, the innocuous looking starting equation
\begin{equation}
\varepsilon^2=1 \label{0-res}
\end{equation}
in the critique [2] is not as innocent as the reviewer seems to think. It displays a scalar value, $1$, of a {\it geometric} product of the pseudoscalar $\varepsilon$ with itself: $\varepsilon\varepsilon=1$, which is the {\it fundamental} product in geometric algebra, where, by definition,
\begin{equation}
\varepsilon:={\bf e}_1{\bf e}_2{\bf e}_3{\bf e}_4\not=\pm1. \label{0-def}
\end{equation}
Therefore, consistency requires that any equation that is claimed to follow from $\varepsilon^2=1$ must also use only geometric products on both sides of the equation $||XY||= ||X||\,||Y||$, or on both sides of {\it any} equation for that matter. It would be illogical to use one product rule on the left-hand side of eq.(3) and a different product rule on its right-hand side. 

With this in mind, it is easy to reveal in which steps the critique's and the reviewer's argument involves inconsistency in deriving the alleged contradiction. To that end, the next step in the critique's and the reviewer's argument, namely
\begin{equation}
(\varepsilon-1)(\varepsilon+1)=0, \label{1-res}
\end{equation}
is correct because it only involves geometric product, and consequently the following equation would indeed hold,
\begin{equation}
||(\varepsilon-1)(\varepsilon+1)||=0, \label{2-res}
\end{equation}
provided we only use a single product rule consistently to arrive at it, namely, the fundamental geometric product rule. But if we end up using a scalar product to evaluate the norm in (\ref{2-res}) at any stage, then that would amount to introducing inconsistency by employing two entirely different product rules to evaluate the norm $||(\varepsilon-1)(\varepsilon+1)||$.

Next, the critique and the reviewer correctly infer from (\ref{2-res}) and norm relation $||XY||= ||X||\,||Y||$, or eq.(3), that
\begin{equation}
0=||(\varepsilon-1)(\varepsilon+1)||=||(\varepsilon-1)||\;||(\varepsilon+1)||, \label{3-res}
\end{equation}
and therefore
\begin{equation}
||(\varepsilon-1)||\;||(\varepsilon+1)||= 0. \label{4-res}
\end{equation}
But contrary to the claim made by the reviewer and the claim in the critique [2], the above equation by no means necessitates that either $||(\varepsilon-1)||=0$ or $||(\varepsilon+1)||= 0$ must hold, unless one incorrectly assumes, {\it a priori}, that $\varepsilon$ is a scalar quantity. All it says is that the {\it geometric product} between $||(\varepsilon-1)||$ and $||(\varepsilon+1)||$ must vanish. That is because $\varepsilon$ is not a scalar quantity but a geometric quantity that satisfies $\varepsilon^2=1$ via the geometric product of $\varepsilon$ with itself. Consequently, it by no means follows from (\ref{4-res}) that either $(\varepsilon-1)$ or $(\varepsilon+1)$ must vanish without circularly assuming, {\it a priori}, that $\varepsilon$ is a scalar quantity. And we already know from definition (\ref{0-def}) that $\varepsilon\not=\pm1$. Consequently, neither $(\varepsilon-1)$ nor $(\varepsilon+1)$ can vanish. Moreover, as explained above, no product rule other than geometric product rule must be used to evaluate the norms $||(\varepsilon-1)||$ and $||(\varepsilon+1)||$ in (\ref{4-res}), because what is used to evaluate the left-hand side of (\ref{3-res}), namely, to work out the result (\ref{2-res}), is {\it geometric} product. Now, using $\varepsilon^{\dagger}=\varepsilon$, we can evaluate the norms:
\begin{equation}
||(\varepsilon-1)||=\sqrt{(\varepsilon-1)(\varepsilon-1)^{\dagger}}=\sqrt{(\varepsilon-1)(\varepsilon-1)}=\sqrt{2\,(1-\varepsilon)} \label{30Xx}
\end{equation}
and
\begin{equation}
||(\varepsilon+1)||=\sqrt{(\varepsilon+1)(\varepsilon+1)^{\dagger}}=\sqrt{(\varepsilon+1)(\varepsilon+1)}=\sqrt{2\,(1+\varepsilon)}\,, \label{30Yy}
\end{equation}
which are indeed non-scalar geometric quantities. Consequently, and inevitably, the right-hand side of (\ref{4-res}) vanishes:
\begin{equation}
||(\varepsilon-1)||\;||(\varepsilon+1)||=\left(\sqrt{2\,(1-\varepsilon)}\right)\left(\sqrt{2\,(1+\varepsilon)}\right)=2\sqrt{(1-\varepsilon)(1+\varepsilon)}=2\sqrt{\left(1-\varepsilon^2\right)\,}=2\sqrt{0\,}=0. \label{20XYxy}
\end{equation}
We thus see that, contrary to the claim made by the reviewer and in the critique [2], no contradiction arises unless mathematical manipulations end up involving two different product rules on the two sides of the norm relation (3).

Moreover, given the fact that I have explicitly proved the norm relation $||XY||= ||X||\,||Y||$ in [1] and [3] for any arbitrary multivectors $X$ and $Y$ in ${\cal K}^{\lambda}$, it is not surprising that any attempt to claim contradiction via counterexample would necessarily involve inconsistency of one kind or another, however cleverly hidden in the alleged counterexample.
     
\underbar{\bf Author action}: I have revised Section~2.7 of the manuscript to briefly explain the above points more clearly.}

\bigskip

\bigskip

\hrule 

\bigskip

\underbar{\textcolor{blue}{{\large\bf Reviewer \# 3}}}

A well considered, convincing and concise response.  You have taken a nice approach, referring critics back to original papers.  

I note that the 3-sphere S3 has been identified by various other authors as a key property of various physical theories, and so you do have valid grounds for your  arguments.

{\color{black}{\underbar{\bf Author response}}: I thank the reviewer for these kind remarks and for recommending my manuscript for publication. 

\underbar{\bf Author action}: No action.}

I presume you are looking at extending your ideas to produce an alternate formulation of QM?

{\color{black}{\underbar{\bf Author response}}: Yes, that is correct. In the spirit of Einstein's views, I have argued that a completely local-realistic formulation of quantum mechanics is possible. In my 2018 paper [1] I have taken the first steps in this direction.   

\underbar{\bf Author action}: No action.}

\bigskip

\bigskip

\hrule 

\bigskip

\underbar{\textcolor{blue}{{\large\bf Reviewer \# 4}}}

The author responds to a critique of one of his papers entitled ``Quantum correlations are weaved by the spinors of the Euclidean primitives,'' which was published in the Royal Society Open Science. The paper ``Quantum correlations are weaved by the spinors of the Euclidean primitives'' describes a geometric framework based on a Clifford-algebraic interaction between the quaternionic 3-sphere, or S3, which describes the geometry of physical space, and an octonion-like 7-sphere, or S7, which is the algebraic representation space of the quaternionic 3-sphere. Without resorting to retrocausality or superdeterminism, the framework described in the paper circumvents Bell's theorem by local-realistically recreating all quantum correlations.

According to the author, the critique wrongly asserts that it has several flaws and does not interact with the model described in his paper. The author argues that the criticism is founded on misconceptions and disproves several assertions made in the critique.

The author begins by stating some incorrect claims and points out some of the mistakes in the critique. Some of the claims made by the author are that he has not proposed any locally hidden variable model but a quaternionic 3-sphere model, and neither argued that Bell's proof of his
theorem is mathematically wrong. The author notes that one of the gravest errors in the critique is the assumption that Bell's so-called "theorem" is a mathematically proven theorem, and hence any criticism of it must be incorrect. The author explains that Bell's theorem is not a theorem in the mathematical sense and is based on a variety of implicit and explicit physical assumptions that have been questioned from the past. He further explained that if this theorem is considered a mathematical theorem, actual investigations would not be required to prove its validity, and any loophole would render it invalid.

Further, the author discusses the assumption of the additivity of expectation values and notes that the substitution of the sum of expectation values with the expectation value of the sum, while
accepted in quantum physics, does not hold for hidden variable theories for observables that are not simultaneously measurable, such as those used in Bell-test experiments. This observation has
been noted earlier in the 1930s, Einstein and Grete Hermann in the context of von Neumann's theorem, then Bell, and by several others.

The response delves into the foundational aspects of quantum mechanics concerning Bell's inequality, quantum correlations, geometric algebra, assumption of flat and immutable spacetime, and the S7 model. In my view, this response is worth publishing and will be helpful to the scientific community as it sheds light on several topics and the interplay between them.

{\color{black}{\underbar{\bf Author response}}: I thank the reviewer for the positive review and for recommending my manuscript for publication. 

\underbar{\bf Author action}: No action.}

\bigskip

\bigskip

\hrule 

\bigskip

\underbar{\textcolor{blue}{{\large\bf Reviewer \# 5}}}

I review the ‘Response to ‘Comment on “Quantum correlations are weaved by the spinors of the Euclidean primitives”’’. I neither comment on the qualities of Joy Christian’s article published in RSOS in 2018, nor on the ones of Richard Gill’s comment on it. Concerning Christian’s response, I try to judge its scientific merits, whether it adds new material or arguments to the discussion in the preceding papers and whether it engages with Gill’s comment. It will be enough to consider Christian’s sec.s 2.4 and 2.5., where he discusses two “hitherto underappreciated” assumptions “that
are necessary to derive Bell’s theorem”.

{\color{black}{\underbar{\bf Author response}}: The purpose of my current manuscript is to respond to the Comment paper by Gill on my original 2018 paper. Because Review \# 5 is focused only on Sections~2.4 and 2.5 of the manuscript, it is not surprising that\break some of the comments below by the reviewer lack the benefit of context. My original paper is an indispensable context.    

\underbar{\bf Author action}: In the abstract and Introduction of the revised version of the manuscript I have made the context of the paper more transparent.}

In sec. 2.4, Christian discusses a “hitherto underappreciated” assumption that is “necessary to derive Bell’s theorem”: the “assumption of the additivity of expectation values”. Gill, in his comment, does not say anything about additivity and so Christian’s discussion introduces a new topic. 

{\color{black}{\underbar{\bf Author response}}: This claim by the reviewer is incorrect. What I have discussed in Section~2.4 of the manuscript is a direct response to the claim made in the critique [2] concerning the assumption of the additivity of expectation values. I have quoted this claim verbatim from the critique [2] in my manuscript. It appears as follows: 
\begin{quote}
He also argues [in [17]] that Bell's proof contains a fundamental error in reasoning: the Bell-CHSH inequality involves correlations obtained from different sub-experiments involving
measurements of non-commuting observables, and (he says) therefore cannot be combined. However, in quantum mechanics, even if two observables do not commute, a real linear
combination of those observables is another observable. By the linearity encapsulated in the basic rules of quantum mechanics, \hl{expectation values of linear combinations of non-commuting observables are the same linear combination of the expectation values of each observable separately}. If a local hidden variables model reproduces the statistical predictions of quantum mechanics, then it must reproduce this linearity.
\end{quote}
Evidently, the reviewer seems to have missed the yellow highlighted sentence. It states nothing but the assumption of the additivity of expectation values required in the proofs of both von Neumann's and Bell's theorems. Thus, contrary to the reviewer's claim, I have not introduced ``a new topic'' but responded only to the issue raised in the critique [2].    

\underbar{\bf Author action}: No action.}

Christian claims that one of Gill’s claims (the last sentence of Gill’s sec. 2b and the last sentence of Christian’s quotation of Gill) is “manifestly incorrect”. Christian elaborates:
\begin{quote}
“While \underbar{mathematically correct}, Eq.~(1) is \underbar{physically meaningless within any hidden variable
theory}. […] The problem with Eq.~(1) is that, while the sum of expectation values is \underbar{mathematically the same} as the expectation value of the sum, as in the assumption that allows us to mathematically replace the left-hand side of Eq.~(1) above with its right-hand side, and while this assumption is \underbar{valid in quantum mechanics} […], it is \underbar{not valid} for any hidden variable theory involving dispersion-free states, […]. This makes the replacement of the left-hand side of Eq.~(1) with its right-hand side \underbar{physically invalid}. [… Once the \underbar{fallacious} Eq.~(1)] is removed from Bell’s argument and local realism is implemented correctly, the
bounds on the right-hand side of Eq.~(1) work out to be $\pm2\sqrt{2}$ instead of $\pm2$, thereby mitigating the conclusions of Bells theorem.”  
\end{quote}
Looking at the underlined phrases, we find Christian saying that “Eq.~(1)” is “mathematically correct” and “valid in quantum mechanics” but also “physically meaningless”, even “physically invalid” in certain circumstances and finally “fallacious”. So Christian claims that “Eq.~(1)” is “mathematically correct” but physically false. So, he asserts that some claim is both true and false, which in the present context is unacceptable. Alternatively, we can take Christian as meaning that the equation stands for an argument and that this argument is “mathematically correct” and “valid in quantum mechanics” but also “physically invalid”. It can indeed happen that an argument is valid, given a fixed set of premises, and “invalid”, given another. But for clarity, the different premises must be made explicit. As the discussion stands, it suggests incoherent argumentation and should not be published in a scientific context.

{\color{black}{\underbar{\bf Author response}}: The argument I have presented in Section~2.4 of the manuscript is not one I have invented myself. It cannot be brushed aside as the above comments by the reviewer purports to do. The problem was recognized by Einstein in the context of von~Neumann's theorem against hidden variable theories in the mid 1930s (see Ref.~[1] and [2] below). It was also recognized by Grete Hermann around the same time (see Ref.~[3] below). Thus the problem has been well known in foundations of quantum mechanics for nearly nine decades. One of the most lucid explanations of the problem can be found in Section~3 of Chapter~1 in Bell's book. The chapter has been reprinted from one of his papers (see Ref.~[4] below). The same problem with von~Neumann's theorem is also discussed by several other well known experts in foundations of quantum mechanics (see Refs.~[5] to [10] below).

Briefly, the problem is that, while the sum of expectation values is mathematically the same as the expectation value of the sum, as in the assumption that allows the replacement of the left-hand side of Eq.~(1) in the manuscript by its right-hand side, and while this assumption is valid in quantum mechanics, it is not valid for any hidden variable theory involving dispersion-free states, because the eigenvalue of a sum of operators is not the sum of individual eigenvalues when the constituent operators do not commute, as in the case of Bell-test experiments. This makes Eq.~(1) physically invalid. But without it the absolute upper bound of 2 on the CHSH correlator cannot be derived. Therefore the problem I have pointed out in Section~2.4 is very serious. It invalidates all derivations of the bound~of~2.

[1] J. von Neumann, {\it Mathematical Foundations of Quantum Mechanics} (Princeton University Press, Princeton, 1955).

[2] A. Shimony, {\it Search for a Naturalistic World View}, vol. II (Cambridge University Press, Cambridge, 1993), p. 89.

[3] G. Hermann, Die naturphilosophischen Grundlagen der Quantenmechanik, Abhandlungen der Fries’schen Schule {\bf 6}, 75 (1935).

[4] J. S. Bell, On the problem of hidden variables in quantum mechanics, Rev. Mod. Phys. {\bf 38}, 447 (1966).

[5] J. M. Jauch and C. Piron, Can hidden variables be excluded in quantum mechanics?, Helv. Phys. Acta {\bf 36}, 827 (1963).  

[6] A. Siegel, in {\it Differential Space, Quantum Systems, and Prediction}, eds: N. Wiener, A. Siegel, B. Rankin, and W. T. Martin (MIT Press, Cambridge, MA, 1966).

[7] S. Kochen and E. P. Specker, The problem of hidden variables in quantum mechanics, J. Math. Mech. {\bf 17}, 59 (1967).

[8] M. Jammer, {\it The Philosophy of Quantum Mechanics} (John Wiley and Sons, New York, NY, 1974).

[9] N. D. Mermin, Hidden variables and the two theorems of John Bell, Rev. Mod. Phys. {\bf 65}, 803 (1993).

[10] N. D. Mermin and R. Schack, Homer nodded: von Neumann's surprising oversight, Found. Phys. {\bf 48}, 1007 (2018).

\underbar{\bf Author action}: I have revised Section~2.4 to explain the problem and the above points more clearly.}

I turn to sec. 2.5. It can be summed up by the following quotation:
\begin{quote}
“The second assumption necessary to support Bell’s theorem is that of immutable spacetime. [… If we abandon this assumption and regard spacetime as mutable, then we can] model physical space as a closed and compact quaternionic 3-sphere, or $S^3$, instead of a flat Euclidean space, or ${\mathrm{I\!R}^3}$ […]. But once the physical space is modelled as $S^3$ instead of ${\mathrm{I\!R}^3}$ […], the correlation between the results Aa and Bb […] \underbar{inevitably} turns out to be E(AaBb) = -a$\cdot$b
[…]”.
\end{quote}
(The sphere $S^3$ is a topic briefly mentioned by Gill without any discussion.)

{\color{black}{\underbar{\bf Author response}}: The main defect of the critique [2] is this fact noted by the reviewer. Namely, that ``The sphere $S^3$ is a topic briefly mentioned by Gill without any discussion.'' In other words, the critique [2] is published with disregard for the actual model proposed in my original 2018 paper [1]. Unfortunately, this review comment has the same issue.

\underbar{\bf Author action}: No action.}

Here, the last sentence claims that, given the assumption of S3, ``the correlation between the results Aa and Bb [...] \underbar{inevitably} turns out to be E(AaBb) = -a$\cdot$b''. Now, given a suitable choice of vectors A, A', B, B', this equation “E(AaBb) = -a · b” implies the right side of “Eq. (1)” above with the values “$\pm2\sqrt{2}$ instead of $\pm2$” as bounds. Given such a choice of vectors, both results are equivalent. But this raises the question why we need two arguments -- the argument from S3 and the one from `false Eq. (1)' -- to disable Bell’s theorem. Introducing S3 is claimed to be sufficient (``inevitably'') for this, so the preceding discussion of additivity seems superfluous. On the other hand, the introduction of S3 is contentious and apparently unnecessary for the argument in sec. 2.4, which deals with Eq. (1) and an undisclosed alternative. The argument in sec. 2.4 suggests that just by removing the `false Eq. (1)', we can derive the bounds on the right-hand side of Eq. (1) [...] to be ``$\pm2\sqrt{2}$ instead of $\pm2$''. Accordingly, Christian cannot make plausible that 2.4. and 2.5 are two ``hitherto underappreciated'' assumptions ``that are [both] necessary to derive Bell's theorem''. Instead, one of these assumptions is claimed to be itself sufficient to disable the theorem, for the other this is at least suggested.

{\color{black}{\underbar{\bf Author response}}: The manuscript is a response to a comment paper on my 2018 paper. It is not a comprehensive review paper on my quaternionic 3-sphere model for understanding quantum correlations. That model, as a constructive counterexample to Bell's theorem, is presented in the references [1], [3], [4], [5], [6], [7], [8], [9], [14], [17], and [19] cited in the manuscript. Moreover, the length of my response paper is editorially restricted to no more than the length of the published comment paper [2]. My responses in Sections~2.4 and 2.5 of the manuscript are a direct response to\break two of the issues raised in the comment paper [2]. Among these, Section~2.4 concerns a formal aspect of the proof of Bell's theorem that has been discussed in the comment paper [2]. That formal aspect does not concern the quaternionic 3-sphere model of the quantum correlations, viewed as a counterexample to Bell's theorem. In the manuscript I have not mixed the issues concerning the formal aspect of Bell's theorem and those concerning my specific counterexample. It is unclear to me why the reviewer has tried to mix up these two aspects and tried to read between the lines. I suspect that is because of the reviewer's acknowledged neglect of my original 2018 paper and the comment paper [2]. 

\underbar{\bf Author action}: I have added reference [21] in the paper. It addresses some of the questions raised by the reviewer.}

All in all, the author’s response contains two key sections that are scientifically insufficient and do not address the comment to which it responds. The response seems fruitless and should not be published. I doubt whether a revised version could improve the argument.

{\color{black}{\underbar{\bf Author response}}: It should be evident from my response above that neither of the two claims made by the reviewer hold in the context of my paper. Consequently, the reviewer's conclusion regarding my response paper does not follow. It is important that my response paper is published because it brings out the oversights contained in the critique [2] of my 2018 paper [1].   

\underbar{\bf Author action}: No action.}

\bigskip

\bigskip

\hrule 

\bigskip

\underbar{\textcolor{blue}{{\large\bf Reviewer \# 6}}}

This is not a comprehensive report on the response by Joy Christian, since there are aspects of both the original paper, and the response, that I have not gone through in detail.

However, the aspects that I have gone through in detail, and that I’m familiar with, are sufficient to make it quite clear that the original paper should not have been published in the form it was, and that many of the same grave errors are present in the Christian response under present consideration. No further useful clarifications or statements that could change the conclusions about these errors are made in the response, hence what I will do here is list these same errors again, and point out where in the response they are repeated. They are easily serious enough to say that the response should not be published.

{\color{black}{\underbar{\bf Author response}}: I do not agree with the reviewer's assessment of either my original 2018 paper or of the current manuscript under consideration. In what follows I demonstrate that each of the claim of error made by the reviewer does not hold, just as the claims made about my 2018 paper in [2] and in [13] do not hold. As noted in my manuscript, in [3] and [14] I have already brought out all of the oversights in [13] in considerable detail. Since these oversights are repeated in this review, in what follows I bring out the oversights in [2] and in [13] again.

\underbar{\bf Author action}: No action}

The foremost and easiest to discuss problem, is that JC still believes that in the algebra he calls ${\cal K}^{\lambda}$, then for any multivectors $X$ and $Y$ within it, the following composition law for the norms holds (his equation (3) in the current response and (2.40) in the original paper):
\begin{equation}
||XY||= ||X||\,||Y|| \tag{1.1} \label{L-1.1}
\end{equation}
This seems to be important to JC, and a good part of the first part of the original paper is devoted to ‘establishing’ it, and in the response, it occupies about a page.

{\color{black}{\underbar{\bf Author response}}: The above equation, which is equation (2.40) in the original 2018 paper, is not based on a ``belief." There are no ``beliefs'' in my paper. Every mathematical equation in it is proved rigorously. The above equation is explicitly proved in Section 2.5 of the 2018 paper, with more details of the proof provided in Appendix~A.1 of [3] and in Appendix~B of \url{https://arxiv.org/pdf/1908.06172.pdf}. To date, {\it no one} has identified a mistake in my proof of (2.40). In particular, Gill in [2], Lasenby in [13], and the present reviewer have not identified any mistake in my proof of (2.40) presented in [1] and [3]. All they have managed to do is allege a counterexample, which I have demonstrated in [3] and [14] to be inapplicable. It is based on internal inconsistency. The contradiction is arrived in it by\break inconsistent use of the product rules on the two sides of equation (2.40) [which, as noted, is the same as (1.1) above].

The outline of the proof of (2.40) is straightforward to follow. The algebra ${\cal K}^{\lambda}$ is the eight-dimensional even subalgebra of the sixteen-dimensional Clifford algebra $\mathrm{Cl}_{4,0}$. As such, any general multivectors $X$ and $Y$ in ${\cal K}^{\lambda}$ are of the form
\begin{equation}
X=\, {\bf q}_{r1} + {\bf q}_{d1}\,\varepsilon \label{genele-1}
\end{equation}
and
\begin{equation}
Y=\, {\bf q}_{r2} + {\bf q}_{d2}\,\varepsilon\,, \label{genele-2}
\end{equation}
where ${\bf q}_{r1}$ and ${\bf q}_{d1}$ constituting $X$ are two independent quaternions and $\varepsilon:={\bf e}_1{\bf e}_2{\bf e}_3{\bf e}_4\not=\pm1$ that satisfies $\varepsilon^2=1$, with its reverse $\varepsilon^{\dagger}$ satisfying $\varepsilon^{\dagger}=\varepsilon$. As a result, the geometric product $XX^{\dagger}$ is of the form
\begin{align}
XX^{\dagger}\,&=\left({\bf q}_{r1} + {\bf q}_{d1}\,\varepsilon\right)\left({\bf q}_{r1} + {\bf q}_{d1}\,\varepsilon\right)^{\dagger}
\label{10aa}\\
&=\left({\bf q}_{r1}\,{\bf q}^{\dagger}_{r1}\,+\,{\bf q}_{d1}\,{\bf q}^{\dagger}_{d1}\right)+\left({\bf q}_{r1}\,{\bf q}^{\dagger}_{d1}\,+\,{\bf q}_{d1}\,{\bf q}^{\dagger}_{r1}\right)\varepsilon \label{10bb} \\
&=\,\text{(a scalar)} \,+\, \text{(a scalar)}\,\varepsilon\,, \label{off}
\end{align}
and likewise for $YY^{\dagger}$. Note that this quantity resembles a split complex or hyperbolic number. Consequently, when product rules are applied consistently on the two sides of equation (2.40) using only geometric products, both sides turn out to be identical to the square-root of the following quantity, which also resembles a split complex number: 
\begin{align}
&\left\{\left(\varrho^2_{r1}+\varrho^2_{d1}\right)\left(\varrho^2_{r2}+\varrho^2_{d2}\right) \,+\,\left({\bf q}_{r1}\,{\bf q}^{\dagger}_{d1}+{\bf q}_{d1}\,{\bf q}^{\dagger}_{r1}\right)\left({\bf q}_{r2}\,{\bf q}^{\dagger}_{d2}+{\bf q}_{d2}\,{\bf q}^{\dagger}_{r2}\right)\right\} \notag \\
&\;\;\;\;\;\;\;\;\;\;\;\;\;\;\;\;\;\;\;\;+\left\{\left( \varrho^2_{r1}\,+\,\varrho^2_{d1}\right) \left({\bf q}_{r2}\,{\bf q}^{\dagger}_{d2}+{\bf q}_{d2}\,{\bf q}^{\dagger}_{r2}\right)
+ \left({\bf q}_{r1}\,{\bf q}^{\dagger}_{d1}+{\bf q}_{d1}\,{\bf q}^{\dagger}_{r1}\right)\left( \varrho^2_{r2}\,+\,\varrho^2_{d2}\right)\right\}\,\varepsilon\,, \label{R-10}
\end{align}
where $\varrho_{r1}=\sqrt{{\bf q}_{r1}\,{\bf q}^{\dagger}_{r1}}\,$, {\it etc.}, are scalars, and therefore the quantities appearing in the two curly brackets are also scalars. This is a completely general result. It holds for any multivectors $X$ and $Y$ in the algebra ${\cal K}^{\lambda}$. And therefore the norm relation $||XY|| = ||X||\,||Y||$ holds for any multivectors $X$ and $Y$ in the algebra ${\cal K}^{\lambda}$. I stress again that, to date, no one has found a mistake in the above proof. It takes less than half an hour, with some familiarity in geometric algebra and quaternions, to verify the proof provided in [1] and [3]. The normalization or orthogonality condition (2.54) in [1], namely ${\bf q}_{r}\,{\bf q}^{\dagger}_{d}+{\bf q}_{d}\,{\bf q}^{\dagger}_{r}=0$, then reduces the hyperbolic number (\ref{R-10}) to the scalar quantity
\begin{equation}
\sqrt{\left(\varrho^2_{r1}+\varrho^2_{d1}\right)\left(\varrho^2_{r2}+\varrho^2_{d2}\right)}\,,
\end{equation}
giving the result
\begin{equation}
||XY|| = \sqrt{\left(\varrho^2_{r1}+\varrho^2_{d1}\right)\left(\varrho^2_{r2}+\varrho^2_{d2}\right)\,} = ||X||\,||Y|| \label{R-11not}
\end{equation}
as a special case. Consequently, since the norm relation (2.40) holds for any multivectors $X$ and $Y$ in ${\cal K}^{\lambda}$ and thus even for those with scalar values for their norms, if $X$ and $Y$ happen to be unit multivectors so that $||X||=1$ and $||Y||=1$, then (\ref{R-11}) necessitates that their product $Z = XY$ will also be a unit multivector: $||Z||=||XY||=||X||\,||Y||=1$.

\underbar{\bf Author action}: In the revised Section~2.7 of the manuscript I have explained the above facts more clearly.}

If it were true, then JC would have found an associative version of the 8 dimensional normed division algebra that is normally represented by the non-associative octonions. Indeed this part of the original paper was also written up separately in the paper: {\it ‘Eight-dimensional octonion-like but associative normed division algebra’} (arXiv:1908.06172) which was published in ‘Communications in Algebra’ in 2020. The editors of Communications in Algebra have since retracted the article, having realised that what it describes in the title is specifically ruled out by Hurwitz’s theorem, and the reference is now vol. 49, no. 2, 905–914 (2021), which points to the retracted version.

{\color{black}{\underbar{\bf Author response}}: The reviewer does not seem to have all the facts behind the retraction from {\it Communications in Algebra} of my paper \url{https://doi.org/10.48550/arXiv.1908.06172}. The paper underwent more than six months and two rounds of peer-review process, and remained published online for about two months, until Richard D. Gill, the author of critique [2], complained to the journal with the goal of having it retracted. A retraction of any paper, however, is not a proof of a flaw in the paper, unless such a proof is provided by the journal. Indeed, to date {\it no one}, including any of the reviewers of the current manuscript, has identified a flaw in the {\it actual} proof of the norm relation $||XY||= ||X||\,||Y||$ I have presented in my papers [1] and [3]. The only objection to it is raised by means of a counterexample. However, in what follows I demonstrate that the alleged counterexample in [13] is misplaced.

For verification, I had also sent my retracted paper to one of the well known experts on the subject of normed division algebras; namely, to Prof. Tevian Dray of Oregon State University in the US, who is a coauthor of the well known book on the subject: {\it The Geometry of the Octonions}. I asked him (1) whether he was able to confirm that my proof was correct, and if so, (2) whether he was surprised by it. His answers were the following:
\begin{quote}
(1) As you are likely aware, this algebra is H$\times$C', that is, the tensor product of the quaternions with the split complex numbers, but with an interesting twist. Specifically, just as with the complexified quaternions, you assume that conjugation affects H, but not C -- or, in this case, C'.  For the complexified quaternions, the resulting norm is complex, rather than real, \hl{but the composition law still holds}; here, the norm is split complex, but again, \hl{the composition law remains intact}.

(2) \hl{No, I wasn't particularly surprised} -- once I had figured out what you were doing.  The Hurwitz theorem as usually stated assumes that the underlying coefficient algebra is real; for you, it is the split complex numbers.
\end{quote}
By ``composition law'' Prof. Dray meant the norm relation (\ref{L-1.1}) under discussion, or Eq.~(2.40) of my paper [1]. 

\underbar{\bf Author action}: No action.}

Specific counterexamples to equation (1.1) (equation (3) in JC’s response), have been given by Gill and Lasenby, and are discussed and allegedly refuted in the response.

So how is it that JC thinks his claims are still defensible, and that he has refuted the concrete counterexamples? Clearly his own analysis in the response must contain mathematical errors, so we need to find these.

{\color{black}{\underbar{\bf Author response}}: On the contrary, as I demonstrate below, the mathematical errors are in fact contained in the alleged counterexample and all of its variants. This should not be surprising, because, as I have noted before, equation (2.40) in [1] is rigorously proven in [1] and [3]. Therefore any alleged counterexample must necessarily contain errors or inconsistencies, however hidden they may be. The only remaining task is to bring out these inconsistencies explicitly. 

\underbar{\bf Author action}: No action.}

Shortly after (his) equation (3), he quotes from Gill’s comments the following passage:
\begin{quote}
{\it However, the author's algebra has an element called the ``pseudo-scalar'', I will denote it by $M$, such that $M^2 = 1$. It follows that $0 =M^2 - 1 = (M - 1)(M + 1)$. Taking norms, $0 = ||M - 1||.||M + 1||$. Hence $||M - 1|| = 0$ or $||M + 1|| = 0$. Therefore $M - 1 = 0$ or $M + 1 = 0$, which implies that $M = 1$ or $M = -1$. That is a contradiction.}
\end{quote}
JC then says
\begin{quote}
{\it But it is not difficult to see that this alleged counterexample harbours several elementary errors. The first error in the quoted claim is immediately obvious. It starts out with the equation $M^2=1$ and ends with the equations $M=1$ or $M=-1$. And then it claims that ``That is a contradiction.''}
\end{quote}
This is a completely fallacious point of course. The original object $M$ is meant to be the pseudoscalar of the space, hence finding out that $M$ has to be the scalar 1 or -1, is indeed a contradiction, amounting to saying that the pseudoscalar $=\pm1$, which is impossible in any dimension above one.

{\color{black}{\underbar{\bf Author response}}: Far from being ``fallacious'', what is quoted above by the reviewer from the manuscript is merely my first observation of an elementary logical mistake in the counterexample alleged in the critique [2], and it is a perfectly valid observation. The argument in the critique that $M^2=1\implies M=1$ or $M=-1$ is not a contradiction by itself. To allege a contradiction one must assume that $M$ is a pseudoscalar. But then it by no means follows from  
\begin{equation}
||M - 1||\,||M + 1||=0 \label{R-14}
\end{equation}
that either $||M - 1|| = 0$ or $||M + 1|| = 0$ must hold, as alleged in the critique [2]. On the contrary, all the equation (\ref{R-14}) says is that the {\it geometric} product between $||M - 1||$  and $||M + 1||$ must vanish, precisely because $||M - 1||$  and $||M + 1||$ are no longer scalars. And the geometric product does vanish, as I have demonstrated in the manuscript. I\break have explained this point in greater detail in my response to Reviewer \# 2 [see discussion around Eqs.~(\ref{0-res}) to (\ref{20XY})]. 

\underbar{\bf Author action}: In the revised version of Section~2.7, I have made the above points more clear.}

Next JC considers the assignments $X = \epsilon - 1$, $Y = \epsilon +1$, where $\epsilon$ is his notation for the pseudoscalar. 

{\color{black}{\underbar{\bf Author response}}: Just to be clear, I have not considered such two-dimensional multivectors in my original paper [1]. They do not play any role whatsoever in the 7-sphere framework I have proposed in [1]. They have been mistakenly assumed in [13] and [2] to construct a counterexample. Therefore, even if such {\it ad hoc} two-dimensional objects lead to ``contradiction'' as alleged in [2] and [13], that would have no effect on or consequences for the 7-sphere framework.

\underbar{\bf Author action}: No action.}

He purports to show that the norm relation
(1.1) is satisfied in this case, since each side is zero. This would then disprove Gill’s claims for this case, which would assign non-zero norms to both $X$ and $Y$, but have the norm of their product as zero. 

{\color{black}{\underbar{\bf Author response}}: That is correct.

\underbar{\bf Author action}: No action.}

Now, in equation (2.8) of the original paper, JC correctly defines the norm of an object $\Omega$ as
\begin{equation}
||\Omega|| = \sqrt{\langle\Omega\,\tilde{\Omega}\rangle_s} \tag{1.2} \label{L-1.2}
\end{equation}
where $\tilde{\Omega}$ is the reverse of $\Omega$ (JC writes $\Omega^{\dagger}$ instead, but it is definitely defined as the reverse) and $\langle\dots\rangle_s$ denotes taking the scalar part of the object in the angle brackets.

{\color{black}{\underbar{\bf Author response}}: Here the adjective ``correctly'' is not what I have used  anywhere to describe the {\it ad hoc} definition (\ref{L-1.2}) of a norm of an isolated multivector in geometric algebra. The definition (2.8) of a norm in my paper [1] is from the introductory section in which I summarize the standard geometric algebra going back to Clifford and Grassmann.

What the reviewer does not mention is that (\ref{L-1.2}) above, or (2.8) in [1], is not the definition of the norm that is used in section 2.5 of my paper [1], where the norm relation $||XY||= ||X||\,||Y||$ is proved that evidently involves a geometric product of two multivector X and Y on its left-hand side. The equivalent definition used to prove the norm relation (2.40) is explicitly stated in (2.54) of [1] in which the non-scalar part of the geometric product $XX^{\dagger}$ is set to zero, with explanation of why that definition is used giving the same value for the norm. It is unclear to me why the reviewer insists on using definition (2.8) from the introductory section and ignores definition (2.54) discussed in Section 2.5.

To begin with, the definition (2.8) is {\it ad hoc} because it goes against the very essence of geometric algebra in which the fundamental product is geometric product, not the scalar product being picked out in $\langle\dots\rangle_s$ by disregarding the rich algebra and geometry contained in the non-scalar part prematurely. In addition, definition (2.8) leads to serious inconsistency when products of two or more multivectors are involved in an equation such as $||XY||= ||X||\,||Y||$. Notice that, if we were to use definition (2.8) to evaluate the norms on the two sides of this equation, then on the left-hand side of it we would end up using two different product rules, the geometric product rule between $X$ and $Y$ and the scalar product rule for evaluating the norm $||XY||$. Whereas on its right-hand side we would end up using only the scalar product rule for evaluating $||X||$ and $||Y||$. This inconsistent application of two different product rules on the two sides of the norm relation is the root cause of the contradiction in the alleged counterexample in [13].  

It is also worth noting that no such inconsistency arises in verifying the same norm relation for the complex numbers $c$, quaternions ${\bf q}$, and octonions ${\bf O}$, for which the norms are defined by $||{c}||=\sqrt{{c}\,{c}^{\dagger}}$, $||{\bf q}||=\sqrt{{\bf q}{\bf q}^{\dagger}}$, and $||{\bf O}||=\sqrt{{\bf O}{\bf O}^{\dagger}}$, respectively. In all three cases, only geometric products are used on both sides of the relation $||XY||= ||X||\,||Y||$. In particular, nowhere in the evaluation of these norms scalar products are used in an {\it ad hoc} manner as done in [13].

\underbar{\bf Author action}: No action.}

By taking this scalar part, we obtain something that, provided it is non-negative, we can validly take a square root of, and in fact for all objects in the algebra JC is considering, $\langle\Omega\,\tilde{\Omega}\rangle_s$ will be greater or equal to zero, and the norm is well-defined.

{\color{black}{\underbar{\bf Author response}}: The norm defined by setting the non-scalar part of $X$ equal to zero is equally well-defined. For details see Section 2.5 of [1] and Sections 3.3 and 3.4 of [3]. As noted, both definitions give the same value for norm. 

\underbar{\bf Author action}: No action.}

However, in the current response, JC says that norms are defined as (see line after his equation (3))
\begin{equation}
||X||=\sqrt{X\tilde{X}} \tag{1.3} \label{L-1.3}
\end{equation}
i.e. he has missed out taking the scalar part of the geometric product $X\tilde{X}$ before the square root is taken. This is an incorrect version of the norm, and it is because of using this new (and incorrect) definition that JC gets the results in relation to (1.1) that are claimed. 

{\color{black}{\underbar{\bf Author response}}: By now it should be evident that these claims are misplaced. As noted above, (\ref{L-1.3}) is not a ``new'' definition of the norm I have invented just for my response to the critique [2]. And I have not ``missed out taking the scalar part of the geometric product $X\tilde{X}$ before the square root is taken.'' The definition (\ref{L-1.3}) employing the fundamental geometric product is the correct definition of the norm that does not discard non-scalar parts of geometric quantities prematurely, and is consistently applicable on both sides of the norm relation $||XY||= ||X||\,||Y||$. 

\underbar{\bf Author action}: No action.}

Specifically, in his equation (6), the chain of equalities should be
\begin{align}
||X||= ||(\epsilon - 1)||
&= \sqrt{\langle(\epsilon - 1)\widetilde{(\epsilon - 1)}\rangle_s} = \sqrt{\langle \epsilon^2 - 2\epsilon + 1\rangle_s} \notag \\
&= \sqrt{ \langle 2(1-\epsilon)\rangle_s} =\sqrt{2} \tag{1.4} \label{L-1.4}
\end{align}
and similarly we obtain $||Y||=\sqrt{2}$ also, meaning $||X||\,||Y||= 2$, in contradiction to (1.1) since $||XY||= ||\epsilon^2 - 1||= 0$.

{\color{black}{\underbar{\bf Author response}}: The inconsistency in this argument is contained in the last, unnumbered equation, namely, in
\begin{equation}
||XY||= ||\varepsilon^2 - 1||= 0.
\end{equation}
Notice that, while in (\ref{L-1.4}) the norms $||X||=\sqrt{2}$ and $||Y||=\sqrt{2}$ belonging to the right-hand side of the norm relation $||XY||= ||X||\,||Y||$ are computed using scalar products, the above equation belonging to the left-hand side of the norm relation is computed using both geometric product between $X$ and $Y$ and a scalar product for the norm $||XY||$. This mixing of different product rules on the two sides of norm relation is inevitable for the reviewer, because a new multivector $Z=XY$ belonging to the algebra ${\cal K}^{\lambda}$ can only be obtained by a geometric product between $X$ and $Y$. Thus the alleged contradiction in the reviewer's argument has been achieved by inconsistently applying of the product rules on the two sides of the norm relation (\ref{L-1.1}). The only way to avoid that inconsistency is by using geometric products on both sides of (\ref{L-1.1}), as I have done in the manuscript. But then no contradiction arises and the alleged counterexample in the critiques [2] and [13] fails, as it must in the light of the general result (2.40) proved in [1, 3].

In summary, the only consistent way to derive the norm relation $||XY||= ||X||\,||Y||$ with all scalar values is via the procedure I have described in the paragraph containing equations (\ref{10a}) to (\ref{R-11}) and proved rigorously in [1]~and~[3].

\underbar{\bf Author action}: No action.}

Thus the claims by JC in his response are demonstrably false, and rest on an incorrect definition of the norm. It is worth emphasising that it is incorrect {\it even according to JC}, since the correct definition is given [in] his original paper (equation (2.8) there, as already referred to).

{\color{black}{\underbar{\bf Author response}}: It should be evident by now that the demonstrations I have provided refuting the counterexamples in [2] and [13] are correct. Moreover, as already noted, the definition (2.8) of a norm in my 2018 paper [1] is from the introductory section in which I summarize the standard geometric algebra going back to Clifford and Grassmann. That definition is not used in Section 2.5 of [1] where the norm relation $||XY||= ||X||\,||Y||$ is proved in full generality. 

\underbar{\bf Author action}: No action.}

Note the expressions that JC gives for the norms of $X$ and $Y$ individually, in equations (6) and (7) of the response, namely $\sqrt{2(1-\epsilon)}$ and $\sqrt{2(1+\epsilon)}$, are clearly incorrect, since they involve square roots of scalar and
pseudoscalar combinations, which we could work out if desired, but cannot be norms since they are not non-negative scalars.

{\color{black}{\underbar{\bf Author response}}: The fact that the norms $||X||=\sqrt{2(1-\varepsilon)}$ and $||Y||=\sqrt{2(1+\varepsilon)}$ computed in my manuscript are not scalars does not prove anything. It only demonstrates the trivial fact that the {\it ad hoc} invention of two-dimensional multivectors $X=\varepsilon-1$ and $Y=\varepsilon+1$ leads to non-scalar norms, and therefore those $X$ and $Y$ are not a part of the set of multivectors that are normalizable to scalars and constitute the 7-sphere constructed in my original paper [1].  

To put this differently, the two-dimensional objects $\varepsilon-1$ and $\varepsilon+1$ used to allege a counterexample form merely a side show. They do not play any role whatsoever in the 7-sphere framework proposed in my original 2018 paper [1]. As a\break result, even if we accept the contradiction alleged in [2] and [13], it would not affect the framework proposed in [1]. 

\underbar{\bf Author action}: To be fair to the reviewer's point of view, in the revised version of the manuscript I have added the following footnote on page 7 for the benefit of the readers: ``Some reviewers unjustifiably defended this counterexample during the review process of this paper. In addition to [3], my detailed rebuttal to their defence is available online in the Review History that is published along with this paper. All variants of the alleged counterexample depend on inconsistent application of product rules on the two sides of the norm relation (5).''}

Another area it is worth commenting on is Section 2.6 of JC’s response. Here he is responding to Gill’s criticism that:
\begin{quote}
{\it A curious elementary mathematical error is that he defines two algebras, built from two 8-dimensional real vector spaces ${\cal K}^{+}$ and ${\cal K}^{-}$ by specifying a vector space basis for each algebra and multiplication tables for the 8 basis elements of each algebra. But they are the {\it same} algebra. The linear spans of those two bases are trivially the same. The multiplication operation is the same.}
\end{quote}
JC’s reply in Section 2.6 of the current response is:
\begin{quote}
{\it However, there is no such error in [1] [the original RSOS paper]. I have neither defined two different algebras, nor claimed that ${\cal K}^{+}$ and ${\cal K}^{-}$ span different spaces.}
\end{quote}
However, in equations (2.33) and (2.34) of the original paper, JC writes ${\cal K}^{+}$ as the span of a certain set of elements, and then ${\cal K}^{-}$ as the span of the same set but with some minus signs introduced. He thus seems to believe that ${\cal K}^{+}$ and ${\cal K}^{-}$ are different, since if not, why have a different symbol for them, but of course Gill’s point is correct in that as defined they are identical. This is a trivial matter, but it is telling that JC is unable to accept that what he has written here in the original paper doesn’t make sense mathematically. Of course we are free to define two different orientations for sets of basis elements, but what is being pointed out is that his definitions (2.33) and (2.34) fail to do this, and it is no good his reply saying that ‘there is no such
error in [1]’ --- there is.

{\color{black}{\underbar{\bf Author response}}: What is raised in the above comments by the reviewer is a non-issue. As I have explained in the manuscript, there is no error in [1] in this regard.  The only difficulty here is that [2], [13], and the present reviewer have overlooked a straightforward concept in linear algebra and differential geometry. Namely, regarding an orientable vector space such as ${\cal K}$ (or a manifold such as $S^7$) with unspecified orientation $\lambda$. I have already explained this concept adequately in Section~2.3 of my original paper [1] and in Section~2.6 of the current manuscript. No further comment is necessary. However, it may be helpful to note a comment from Reviewer \# 1 in support of my view on this matter:
\begin{quote}
Reviewer \# 1: ``\textcolor{blue}{Section~2.6 makes a valid point, both Lasenby and Gill have misunderstood that you are referring to differently oriented vector spaces not to different vector spaces regardless of orientation.}''
\end{quote}
As noted in [1] and the manuscript, since $\lambda$ plays the role of a hidden variable as understood within Bell's local-realistic framework, ${\cal K}^{+}$ and ${\cal K}^{-}$ specified in the equations (2.33) and (2.34) of my 2018 paper [1] are physically {\it not} identical. 

\underbar{\bf Author action}: No action.}

Finally, it is maybe worth commenting on some points that JC makes
in Section 2 of his response.

In 2.1(3) he says:
\begin{quote}
{\it The next mistake in the Introduction is more serious. It is claimed that in my paper [1] I connect the 3-sphere, or $S^3$, ``to special relativity, specifically to the solution of Einstein's field equations known as Friedmann-Robertson-Walker spacetime with a constant spatial curvature.'' This is not a misprint or oversight. The same claim appears in several preprint versions of the critique [2] that have been posted on arXiv. The quoted sentence from [2] thus exhibits a lack of understanding of the difference between the special and general theories of relativity and how my proposed quaternionic 3-sphere model fits into the Friedmann-Robertson-Walker solution of Einstein's field equations of general relativity.}
\end{quote}
While I agree that ‘general relativity’ should have been said instead of `special relativity', there is no way in which this can be brought forward as a serious error, and thought worth highlighting in this way. It is a trivial error, and has no consequences for any of Gill’s criticisms.

{\color{black}{\underbar{\bf Author response}}: I do not agree with the reviewer's opinion. To confuse general relativity with special relativity is a serious mistake. And it is not the first time the author of the critique [2] has made such a mistake. It is unfortunate that it is now published in a Royal Society journal. It is even more serious considering that it appears in the context of my work that is fundamentally based on a non-flat geometry of the physical space. Much of the confusion in the critique [2] stems from its failure to understand the difference between the strong correlations within a flat spacetime such as ${\mathrm{I\!R}}\times{\mathrm{I\!R}}^3$ and a curved spacetime such as ${\mathrm{I\!R}}\times S^3$. It is therefore entirely appropriate for me to draw attention to\break this mistake. Moreover, I am not willing to let the readers go away with thinking that the mistake is contained in my original paper. It is my scientific duty to point out that the mistake is in the critique and not in my original paper.

\underbar{\bf Author action}: No action.}

In 2.2(2) and (3) JC says:

\begin{quote}
{\it In the Introduction of [2] [the Gill comment] it is claimed that other papers have also refuted my work. But to date, no one has refuted any part of my work, or undermined it in any way. To be sure, there have been attempts of refutation, but I have elucidated the errors in all such claims, for example in [4-6] and references cited therein. See, especially, Chapters 9 to 12 in [4]}
\end{quote}
\begin{quote}
{\it It is further claimed in the Introduction of [2] that Lasenby in [13] has independently made the same claims about the algebraic core of the 7-sphere framework proposed in [1]. But that is not surprising, because, as acknowledged in [13], it has largely borrowed its claims from the claims made online by the author of [2]. More importantly, the claims made in [13] are equally incorrect, as I have demonstrated point by point in [3] and [14].}
\end{quote}
First of all, there is no statement in the paper by Lasenby [13] that he has `borrowed his claims from the claims made online by the author of [2]'. Instead [13] says: `Note several of the points made here have been made independently by Richard D. Gill and others in the discussion thread attached to the Royal Society paper'.

{\color{black}{\underbar{\bf Author response}}: Fair enough. It is of course possible to make the same oversights independently. 

\underbar{\bf Author action}: I have revised item (3) of Section~2.2 in the manuscript and replaced my statement with the following:
\begin{quote}
It is further claimed in the Introduction of [2] that Lasenby in [13] has independently made the same claims about the algebraic core of the 7-sphere framework proposed in [1]. However, in [13] it is acknowledged that ``... several of the points made [in [13]] have been made independently by [the author of [2]] and others in the discussion thread attached to the Royal Society paper ...'' Thus it is not surprising that the claims in [13] are similar to those in [2]. More importantly, in [3] and [14] I have refuted the claims made in [13].
\end{quote}}

Secondly, since all JC's alleged refutations are along the lines of what is in his current response, i.e. based on demonstrably fallacious mathematics, it is needless to say that all his assertions in the two paragraphs just referred to, concerning lack of refutation of his work, are incorrect.

{\color{black}{\underbar{\bf Author response}}: By now it is hopefully evident that there is no ``fallacious mathematics'', either in my original paper [1] or in the current manuscript. I have demonstrated in my previous comments, in the manuscript, and in [3] and [14] that the claims made in [2], [13], and the present review are not correct. They are based on inconsistent use of product rules on the two sides of (2.40), and misunderstandings of the 7-sphere framework I have proposed in [1].   

\underbar{\bf Author action}: No action.}

\bigskip

\bigskip

\hrule 

\bigskip

\underbar{\textcolor{blue}{{\large\bf Reviewer \# 7}}}

Recommendation: Accept

Comments to the Author(s):

None.

{\color{black}{\underbar{\bf Author response}}: I thank the reviewer for recommending my manuscript for publication.

\underbar{\bf Author action}: No action.}

\bigskip
\bigskip
{\color{black}
\hrule
\bigskip
\hrule}}}

\bigskip

\begin{center}
\underbar{\bf\Large Response to Reviewers --- Round \# 2:}
\end{center}

\underbar{\textcolor{blue}{{\large\bf Reviewer \# 1}}}{\color{blue}{

The response to my critique of Section 2.2 (1) is not adequate. Even according to the 2015 revision of your 2011 preprint, equation 1 states that  $A(a,\lambda^k) = +1$ if $\lambda^k = +1$, $A(a,\lambda^k) = -1$ if $\lambda^k = -1$, $B(b,\lambda^k) = -1$ if $\lambda^k = +1$, $B(b,\lambda^k) = +1$ if $\lambda^k = -1$. From this it immediately follows that $A(a,\lambda^k) = -\,B(b,\lambda^k)$ as Gill pointed out. You have attempted to deny this conclusion in your various papers including reference [6] but the arguments are not cogent and amount to denying the transitivity of the equality relation which is not sensible at all.

{\color{black}{\underbar{\bf Author response}}: The above argument by the reviewer is not correct. It is based on a misreading the 3-sphere model inspired by Gill's misinterpretation of the model. The measurement functions ${\mathscr A}({\bf a},\,{\lambda^k})$ and ${\mathscr B}({\bf b},\,{\lambda^k})$ in the model are defined as results of interactions between spin bivectors ${{\bf L}({\bf s},\,\lambda^k)}$ and detector bivectors ${{\bf D}({\bf a})}$ and ${{\bf D}({\bf b})}$:
\begin{align}
S^3\ni\,{\mathscr A}({\bf a},\,{\lambda^k})\,:=\,\lim_{{\bf s}_1\,\rightarrow\,{\bf a}}\left\{-\,{\bf D}({\bf a})\,{\bf L}({\bf s}_1,\,\lambda^k)\right\}&=\,      
\begin{cases}
+\,1\;\;\;\;\;{\rm if} &\lambda^k\,=\,+\,1 \\
-\,1\;\;\;\;\;{\rm if} &\lambda^k\,=\,-\,1
\end{cases} \notag \\
\text{and}\;\;\;\;S^3\ni\,{\mathscr B}({\bf b},\,{\lambda^k})\,:=\,\lim_{{\bf s}_2\,\rightarrow\,{\bf b}}\left\{+\,{\bf L}({\bf s}_2,\,\lambda^k)\,{\bf D}({\bf b})\right\}&=\,
\begin{cases}
-\,1\;\;\;\;\;{\rm if} &\lambda^k\,=\,+\,1 \\
+\,1\;\;\;\;\;{\rm if} &\lambda^k\,=\,-\,1\,,
\end{cases} \notag
\end{align}
where the orientation ${\lambda}$ of ${S^3}$ is assumed to be a random variable with 50/50 chance of being ${+1}$ or ${-\,1}$ at the moment of the creation of the spin pair that originates at the source situated in the overlap of the backward light-cones of Alice and Bob. As such, the measurement results are limiting scalar points of a quaternionic 3-sphere, taken to model the geometry of the physical space in which the experiments are taking place. The joint result ${\mathscr A}{\mathscr B}({\bf a},\,{\bf b},\,{\lambda^k})$ is therefore a limiting scalar point of the product of the quaternions $\left\{-\,{\bf D}({\bf a})\,{\bf L}({\bf s},\,\lambda^k)\right\}$ and $\left\{+\,{\bf L}({\bf s},\,\lambda^k)\,{\bf D}({\bf b})\right\}$. Since $S^3$ remains closed under multiplication, the product of these quaternions is a third quaternion, and therefore its limiting scalar point ${\mathscr A}{\mathscr B}({\bf a},\,{\bf b},\,{\lambda^k})$ cannot be equal to $-1$ for all settings ${\bf a}$ and ${\bf b}$ as claimed by Gill and the reviewer, unless the conservation of the zero spin angular momentum that originates from the source is violated:
\begin{align}
{\mathscr A}{\mathscr B}({\bf a},\,{\bf b},\,{\lambda^k}) = \lim_{\substack{{\mathbf s}_1\,\rightarrow\,{\mathbf a} \\ {\mathbf s}_2\,\rightarrow\,{\mathbf b}}}\left\{-\,{\bf D}({\bf a})\,{\bf L}({\bf s}_1,\,\lambda^k)\right\}\left\{+\,{\bf L}({\bf s}_2,\,\lambda^k)\,{\bf D}({\bf b})\right\} 
\longrightarrow
\begin{cases}
-1 &\text{if}\;\,{\mathbf s}_1\not={\mathbf s}_2 \\
-\,{\mathbf a}\cdot{\mathbf b} &\text{if}\;\,{\mathbf s}_1={\mathbf s}_2,
\end{cases} \notag
\end{align}
because all bivectors are unit bivectors and ${\bf L}({\bf s}_1,\,\lambda^k)\,{\bf L}({\bf s}_2,\,\lambda^k) = -1$ for 
${\mathbf s}_1={\mathbf s}_2$ so that ${\bf D}({\bf a}){\bf D}({\bf b})\longrightarrow-\,{\mathbf a}\cdot{\mathbf b}$ [7].

I have explained this many times before in detail; for example, in my previously published responses [4, 6, 7]. The 3-sphere model is not a model of after-the-event analysis followed by experimenters ({\it i.e.}, of phenomenology) but an {\it ontological} model of what the correlations will be {\it if} the geometry of the physical space is that of $S^3$ instead of $\mathrm{I\!R}^3$. 

\underbar{\bf Author action}: I have revised the relevant sentences in the manuscript at the end of Section 2.2 (1) to: ``It also does not follow mathematically from any other equations I have written down anywhere without violating the conservation of the initial zero spin angular momentum, as I have explained, for example, in Subsection~IV~E of [6], Subsection~III~E of [7], and Section~VIII of [9]. There are also other oversights in [13], which I have brought out in [4--7]. In particular, [13] makes the same mathematical mistakes I bring out in detail below in Subsection~2.7.''}

The response to my and reviewer 5's critiques of Section 2.4 is not entirely adequate - if two expressions are mathematically equal and one of the expressions represents a correct physical value, then so does the other expression as it is the same value. What we have is that Eq.~1 is not even mathematically valid for Bell type experiments as it is assuming that the hidden variables $\Lambda$ form a single joint probability space for all spin measurements and that one can integrate over this same $\Lambda$ in all cases. That is indeed not the case if we consider eigenvalues of non-commuting operators as statistical random variables, so I am not making a different argument. However the wording of section~2.4 has been changed in the latest version of the paper and appears more adequate now -- I disagree with reviewer~2 that your argument rules out your own model.

{\color{black}{\underbar{\bf Author response}}: Apart from the historical precedence of the argument I have presented in Section~2.4 and the universal agreement among the experts in foundations of quantum mechanics on the invalidity of the assumption of additivity of expectation values for hidden variable theories, the advantage of my argument is that it remains valid even if a single joint probability space for all spin measurements is assumed as insisted on by the adherents of Bell's theorem. One has to adapt double standards -- one standard for von~Neumann's theorem and another for Bell's theorem -- to deny the invalidity of the assumption of additivity of expectation values for hidden variable theories.

I am pleased to know that the present reviewer disagrees with the claim by Reviewer \# 2 that my argument in Section 2.4 rules out my own model. 

\underbar{\bf Author action}: I have further improved the wording of Section~2.4 to make my argument more comprehensible.}

Reviewer 2 raises the question as to how the geometry of space-time is relevant to Bell type experiments. This is something that you have indeed repeatedly tried to explain in your papers and as noted in my own critique whether your model is sensible and rigorous in this regard remains questionable.

{\color{black}{\underbar{\bf Author response}}: In my experience those who question the 3-sphere model of quantum correlations do so because they do not realize that it is {\it not} a model of after-the-event analysis followed by the experimenters but an {\it ontological} model that predicts what the correlations would be {\it if} the geometry of the physical space is that of a quaternionic 3-sphere instead of a flat ${\mathrm{I\!R}^3}$. The present reviewer's claim that the 3-sphere model implies ``$A(a,\lambda^k) = -\,B(b,\lambda^k)$'' for all settings $a$ and $b$ ``as Gill pointed out'' is ultimately because of this category error in understanding the model. 

\underbar{\bf Author action}: In Section~2.1~(3) of the revised manuscript, I have added the following sentence: ``In fact, much of the confusion in [2] stems from its failure to understand the difference between strong correlations within flat spacetime ${\mathrm{I\!R}}\times{\mathrm{I\!R}}^3$ and curved spacetime ${\mathrm{I\!R}}\times S^3$.''}

The response to my critique of the calculation involving epsilon in Section 2.7, and similar concerns by reviewers 2 and 6 are far from adequate. In the original paper on which Gill has commented the norm is defined as $|\Omega| := \Omega\cdot\Omega^{\dagger} \equiv \sqrt{(s)}$ -- the square root of the scalar part of $\Omega\cdot\Omega^{\dagger}$, as reviewer 6 also points out. 

{\color{black}{\underbar{\bf Author response}}: This is not correct. In the original paper [1] the orthogonality condition ${\bf q}_{r}\,{\bf q}^{\dagger}_{d}+{\bf q}_{d}\,{\bf q}^{\dagger}_{r}=0$ is used as a normalization condition, defined in Eq.~(2.54) of Section~2.5, to consistently obtain scalar values for the norms on both sides of the norm relation (2.59). The scalar values thus obtained happen to be same as those obtained using the traditional definition $||\Omega||= \sqrt{\Omega\cdot\Omega^{\dagger}}$ stated in Eq.~(2.8) of Section~2.5. Reviewer~\#~2 and \#~6 are also mistaken regarding this point, as I explained in detail in the previous round of reviews. Moreover, Reviewer~\#~3 has now verified my proof of the norm relation in detail, whose reports I reproduce below verbatim.  

\underbar{\bf Author action}: In Section 2.8 of the revised manuscript I have summarized the proof again for convenience.}

This is scalar valued and positive definite and the standard rules for scalar multiplication and positive definite norms apply and the square root being used is the standard one of arithmetic. Thus placing an expression within a square root so understood implies that the expression is a scalar. 

{\color{black}{\underbar{\bf Author response}}: All of the above properties remain true for the norms obtained using the orthogonality condition ${\bf q}_{r}\,{\bf q}^{\dagger}_{d}+{\bf q}_{d}\,{\bf q}^{\dagger}_{r}=0$ as a normalization condition. I have shown this in detail in Refs.~[3] and [25] cited in the manuscript.

\underbar{\bf Author action}: In Section~2.8 of the manuscript I have added a proof of the positive definiteness of the norm.}

But in your attempt to refute Gill's counterexample you have changed the definition of the norm to a non-scalar geometric magnitude as well as changed to working with a geometric square root, which is not what appears in the original paper.

{\color{black}{\underbar{\bf Author response}}: This is not correct. I have not changed the definition of the norm I have used in the original paper [1], as I explained above. The definition I have used is stated in Eq.~(2.54) of Section~2.5 in the paper [1].

More importantly, the supposed counterexample by Gill and Lasenby fails as I have demonstrated in the manuscript as well as in the previous round of reviews. Reviewer \# 3 has also verified in detail that the counterexample fails. 

\underbar{\bf Author action}: No action.}

Additionally the meaning of this square root that is being used needs further elucidation, only a reference to an unpublished paper appears.

{\color{black}{\underbar{\bf Author response}}: In addition to the elucidations in Refs.~[3] and [25] cited in the manuscript, Reviewer~\#~3 has also now elucidated the meaning of the square root I have used for the hyperbolic or split complex numbers. 

\underbar{\bf Author action}: In Section~2.8 of the revised manuscript, I have added the following footnote: ``During the review\break process of this paper, one of the reviewers independently verified this proof of the norm relations (5)  and (27) with detailed calculations and comments, which are available online in the Review History published with this paper.''}

Section 2.8 has been improved.

{\color{black}{\underbar{\bf Author response}}: I thank the reviewer for this observation.

\underbar{\bf Author action}: No action.}

\bigskip

\bigskip

\hrule 

\bigskip 

\underbar{\textcolor{blue}{{\large\bf Reviewer \# 2}}}

Dear Editor,

I recommend that this paper will not be accepted for publication. Here is a list of some reasons that support this opinion.

1. ``Bell's theorem is not a  theorem in mathematical sense"

I cannot even understand what a sentence like this means: in which other sense can a theorem be a theorem?

{\color{black}{\underbar{\bf Author response}}: Reviewer has quoted the heading of Section 2.3 of the manuscript, while omitting the quotation marks on the word ``theorem.'' In the first two paragraphs of Section 2.3 I have very clearly explained why Bell's so-called ``theorem'' is not a theorem in mathematical sense. Unlike it, a mathematical theorem does not depend on physical and metaphysical assumptions such as locality, realism, ``free will'' of experimenters, {\it etc.}, and does not require physical experiments for its validity. Any loophole (or ``gap'') would render a mathematical theorem invalid.    

\underbar{\bf Author action}: No action.}

Bell's theorem can be precisely formulated as a theorem, even if John Bell didn't do it by himself, and its conclusions can be trivially proved. Of course, like any other theorem, Bell's theorem does rely on certain assumptions, and it could certainly be the case that these assumptions are not fulfilled in the actual experiments that test the Bell inequalities. That would make the Bell theorem irrelevant for understanding or interpreting the experiments, but it does not make the theorem wrong. The theorem would be wrong only if not all the assumptions needed to prove it were properly stated.

{\color{black}{\underbar{\bf Author response}}: I have not claimed that the inequalities usually attributed to Bell (which were in fact derived by George Boole some one hundred and eleven years before Bell's work) are wrong. The reviewer seems to have succumb to the same confusion between Bell's inequalities and Bell's theorem as the author of the critique [2] has. 

\underbar{\bf Author action}: No action.}

Let me give a very simple example that could clarify the situation. Consider the following theorem:

Theorem: Under the assumptions of Newton's laws, the distance covered by a body that falls freely with constant acceleration g is given by the expression
\begin{equation}
\mathrm{d} = 0.5 * \mathrm{g} * \mathrm{t}^2 + \mathrm{v} * \mathrm{t}, \notag
\end{equation}
where $\mathrm{t}$ is the time elapsed and $\mathrm{v}$ is the body's initial velocity.

This theorem is precisely formulated and can be trivially proven under the assumptions that appear in the statement of the theorem. Nonetheless, if we could perform an experiment in which we would let a body fall freely from an altitude of 3,000 Km, we would find that the results of the experiment do not fit the predictions of the above theorem. Of course, the reason why the theorem fails to describe the said experiment is not that ``the theorem is not a theorem in the mathematical sense", but the fact that the assumption of constant acceleration under which the theorem was derived does not hold for a body that falls freely from very high altitude.

The author fails to clearly make this simple observation and instead makes an extremely confusing argumentation about Bell's theorem not being a theorem even though it can be trivially proven. The reader, obviously, gets absolutely confused and learns nothing from the author's paragraph.

{\color{black}{\underbar{\bf Author response}}: The example presented by the reviewer does not depend on metaphysical assumptions such as ``realism'' and ``free will'', and therefore it does not correspond to the problems with Bell's so-called theorem. I have explained the problem with Bell's supposed theorem very clearly in the Sections 2.3, 2.4, and 2.5 of the manuscript. 

\underbar{\bf Author action}: No action.}

2. ``Assumption of the additivity of the expectation values''

A successful model for a physical system, e.g. Bell's experiment, is a precisely defined mathematical object that reproduces the characteristic features of the collected experimental data about the system that we want to describe or understand.

In a Bell experiment, four different physical observables are measured. Let me call them E11, E12, E21, E22. The measured values for these magnitudes are, of course, rational numbers. These four numbers are then added up (in the usual way in which rational numbers are added up) and a fifth number S=abs(E11+E12+E21-E22) is obtained.

A successful model for (any settings of) this experiment must reproduce all these five numbers, and not only the last one. The author, nonetheless, argues that his model's prediction for S does not equal the usual addition of its predictions for E11, E12, E21  E22. If his claim is correct, then his model is automatically ruled out because it does not reproduce the experimental data.

{\color{black}{\underbar{\bf Author response}}: The last claim made by the reviewer is not correct. This is also observed by Reviewer~\#~1 who has noted that ``\textcolor{blue}{I disagree with reviewer 2 that your argument rules out your own model.}'' The problem with Bell argument is that what the reviewer calls ``a fifth number S=abs(E11+E12+E21-E22)'' is not a correct eigenvalue of the corresponding quantum mechanical operator, and therefore it does not belong to the right-hand side of Eq.~(1) in any hidden variable theory based on dispersion-free states. It is thus the same mistake that invalidated von~Neumann's theorem against hidden variable theories. I have explained this very clearly with necessary details in Section~2.4 of the manuscript. When the sum E11+E12+E21-E22 is replaced with the correct eigenvalue of the corresponding quantum mechanical operator, the bounds on the CHSH correlator exceeds from $\pm2$ to $\pm2\sqrt{2}$, thereby permitting models like the one I have proposed to reproduce strong correlations instead of ruling them out.

\underbar{\bf Author action}: I have improved some wording in Section~2.4 to make my argument clearer. To explain the above points more clearly, I have also added the example of Stern-Gerlach magnet Bell has used in Chapter~1 of his book.}

3. ``Assumption of a flat and immutable spacetime"

The author claims that his model succeeds to reproduce the statistical features of the experimental data collected in actual Bell's experiments by modeling the spatial geometry of the Universe as a closed sphere. He further argues that collected cosmological data even supports this geometry. 

The question then is: how is it possible that the geometry of the whole Universe is relevant in order to describe a `tabletop' experiment? Violation of the Bell inequality has been observed in experiments performed between two stations separated at most a few hundred kilometers. The visible patch of our Universe is more than twenty orders of magnitude larger than this distance!!

Since a sphere is different from a homogeneous and isotropic flat manifold only in their local curvature and their topology, the author's model for the Bell experiment must be sensible to either the curvature of the sphere, its topology, or both. According to the cosmological evidence cited by the author, the curvature of the Universe is at most only very slightly different from zero. Hence, the author's model must be sensible to the topology of the sphere (in fact, this is already obvious from the author's example with a Moebius strip, which is different from a usual closed strip only in their topologies).

That means, that the author's model for the tabletop Bell's experiment do not comply with the usual causality constraints, since the correlations generated in the short time interval elapsed between the emission and detection of the pair of entangled particles  are sensible to the topology of the whole Universe.

{\color{black}{\underbar{\bf Author response}}: 

I have addressed the issue raised by the reviewer in Answer 4 of Appendix B of Ref.~[10] cited in the manuscript. The 3-sphere model presented in [1] does not require Bell-test experiments to be performed over cosmological distances. They could be performed on a table top and the 3-sphere model would then still provide a viable explanation of the observed results. I have proposed just such a table top experiment in Refs~[8] and [22] cited in the manuscript. The curvature of the Universe is not of direct relevance for the 3-sphere model or the proposed experiment. The geometry of the Universe is indeed irrelevant for the Bell-test experiments performed on Earth. The 3-sphere model concerns the {\it intrinsic} geometry of the physical space, since it is a part of a solution of the field equations of general relativity. In the toy model of the M\"obius world I have discussed in Ref.~[10], the twist in the strip responsible for the strong correlations is an extrinsic twist in the geometry of the strip, while the twists in the Hopf bundle of $S^3$ are intrinsic to $S^3$. Consequently, unlike in the M\"obius world where relative handedness of spins depend on the revolutionary distance between them, in the real world the relative handedness of quaternions that constitute the 3-sphere reflects their intrinsic spinorial characteristics, independently of any distance between them. Therefore the strong correlations predicted by the 3-sphere model and observed in the Bell-test experiments performed in terrestrial\break laboratories can be taken as evidence that the intrinsic geometry of the physical space is that of $S^3$ and not of ${{\mathrm{I\!R}}^3}$. 

\underbar{\bf Author action}: No action.}

4. ``Algebra ${\cal K}^{\lambda}$ used in [1] is not incompatible with Hurwitz's theorem"

The author argues that he has proven this point somewhere else, but no clear explanation can be found neither in the document that he cites, so I will again raise the question:

{\color{black}{\underbar{\bf Author response}}: This claim by the reviewer is not credible. A very clear and explicit proof of the norm relation in question is already provided in Section 2.5 of the original paper [1], and again with exhaustive detail in [3] and [25]. Moreover, now Reviewer \# 3 has independently verified my proof explicitly, which I include and discuss below. 

\underbar{\bf Author action}: In Section 2.8 of the revised manuscript I have summarized the proof again for convenience.}

Let me assume, as the author claims, that in the said algebra the following identity
\begin{equation}
||\mathrm{X} \mathrm{Y} || = || \mathrm{X} ||\;|| \mathrm{Y} ||, \notag
\end{equation}
holds for any pair of objects X,Y that belong to the said algebra, where by definition $||X||$ and $||Y||$ are non-negative real numbers. As I said, I am willing to accept that this identity indeed holds for the said algebra.

The author further claims that in his algebra ${\cal K}^{\lambda}$ there exist an element epsilon different from I (the identity of the algebra) and different from -I, that fulfills the identity
\begin{equation}
\mathrm{epsilon}^2 = \mathrm{I}. \notag
\end{equation}
Now, using the fact that epsilon is an element of the algebra we can write the last identity as
\begin{equation}
\text{(epsilon + I) (epsilon - I) = 0} \notag
\end{equation}
and taking X=epsilon + I and Y=epsilon - I and using the above identity, which as I said I did not dispute, we find
\begin{equation}
0 = || 0 || = || \text{(epsilon + I) (epsilon-I)} || = || \text{epsilon + I}|| \, ||\text{epsilon - I}||, \notag
\end{equation}
which implies that either epsilon+I = 0 or epsilon-I = 0, since for a normed algebra $||X||$ = 0 if and only if X=0.

That is, either epsilon=I or epsilon= -I, in contradiction with the author's statement in eq.(2.51) that there exists in the algebra ${\cal K}^{\lambda}$ and element epsilon different from I and -I for which holds the identity epsilon$^2$=I.

This derivation shows that the only way to escape this contradiction is that either eq.(2.40) does not hold, or the statement $||X||$ = 0 iif X = 0. Since the author claims that eq.(2.40) holds (and I did not dispute his claim), the only option left is that [there] exists X.neq.0 for which $||X||$ = 0, but in such a case ${\cal K}^{\lambda}$ is not a normed algebra.

{\color{black}{\underbar{\bf Author response}}: In the last round of reviews I have already pointed out the mathematical oversights in the above argument in considerable detail. Unfortunately, the reviewer has repeated the same oversights again without recognizing them. Since I have also discussed the issue raised by the reviewer in the Sections~2.7 and 2.8 of the manuscript, and since Reviewer~\#~3 has now verified my proof of the norm relation $|| \mathrm{X} \mathrm{Y} || = || \mathrm{X} ||\;|| \mathrm{Y} ||$ in detail, I will only point out the main oversight in the reviewer's argument. The oversight is evident in the following equation:
\begin{equation}
0 = || 0 || = || \text{(epsilon + I) (epsilon-I)} || = || \text{epsilon + I}|| \, ||\text{epsilon - I}||. \notag
\end{equation}
Notice that a {\it geometric} product between $\text{X = (epsilon + I)}$ and $\text{Y = (epsilon - I)}$ is used by the reviewer to obtain XY = 0 on the left-hand side of this equation, whereas scalar products are tacitly used to guess scalar-valued norms on its right-hand side (without defining what norm $||\cdot||$ is). Indeed, it is inevitable to use a geometric product on the left-hand side because otherwise the resulting multivector Z = XY would not be a member of the algebra ${\cal K}^{\lambda}$. Moreover, if the scalar products are {\it not} used to evaluate norms on the right-hand side but geometric products are used instead to match the use of geometric product on the left-hand side, then the norms would not be scalar-valued. They would be equal to $||\text{epsilon + I}|| = \text{I + epsilon}$ and $||\text{epsilon - I}|| = \text{I - epsilon}$, giving zero right-hand side:  
\begin{equation}
||\text{epsilon + I}|| \,||\text{epsilon - I} || = \text{(I + epsilon) (I - epsilon)} = \text{I - epsilon}^2 = \text{I - I = 0}. \notag
\end{equation}
Therefore the conclusion drawn by the reviewer that either  (I + epsilon) = 0 or  (I - epsilon) = 0 does not follow. Consequently, the alleged contradiction drawn by the reviewer that either epsilon = I or epsilon = -I does not follow.

One can always draw a contradiction from any mathematical equation by employing two different rules on the two sides of the equation (in this case product rules). The oversight in the reviewer's argument is thus quite elementary.

\underbar{\bf Author action}: To be fair to the reviewer's point of view, in a footnote in Section~2.8 I have added the following footnote: ``Some reviewers unjustifiably defended this counterexample during the review process of this paper. In addition to [3], my detailed rebuttal to their defence is available online in the Review History published along with this paper.''}

I think that I have enumerated a long enough list of serious criticisms that justifies my recommendation of rejection.

{\color{black}{\underbar{\bf Author response}}: I have demonstrated above that none of the criticisms by the reviewer is correct. They are based\break on technical oversights. Therefore the reviewer's recommendation for the rejection of my manuscript is not justified. 

\underbar{\bf Author action}: No action.}

\bigskip

\bigskip

\hrule 

\bigskip

\underbar{\textcolor{blue}{{\large\bf Reviewer \# 3}}}

I have confirmed your norm for the dual quaternions, and it is interesting how it works, and that the unit dual, epsilon, squares to +1 and commutes with quaternions.  The norm of a dual quaternion always giving a dual number a + epsilon b, a,b Real.

{\color{black}{\underbar{\bf Author response}}: I am grateful to the reviewer for verifying the norm relation $\|X Y\|=\|X||\,\| Y \|$ in full detail.

Below I reproduce the reviewer's calculations verbatim, with some comments. It is surprising that while the validity of the norm relation has been challenged by Gill, Lasenby, and others despite its proof I have presented in [1], [3], and [25], the present reviewer is the only person I am aware of who has bothered to verify the proof in detail to appreciate how it actually works. Others have merely alleged a counterexample, or that the norm relation contradicts Hurwitz's\break theorem, without digging into details or pointing out an error in the proof I have presented in
[1], [3], and [25]. 

\underbar{\bf Author action}: In Section 2.8 of the revised manuscript I have reproduced a succinct version of the proof.}

1. Not sure therefore why you are requiring the two quaternions (inside a dual quaternion) to be orthogonal?  The subscript `s' on your norm actually confused me, seems unnecessary.

{\color{black}{\underbar{\bf Author response}}: The orthogonality condition ${\bf q}_{r}\,{\bf q}^{\dagger}_{d}+{\bf q}_{d}\,{\bf q}^{\dagger}_{r}=0$ is the \underbar{normalization condition} I have used in Eq.~(2.54) of the Section 2.5 of the original paper [1] to consistently obtain scalar values for the norms on both sides of the norm relation (2.59). This then allows us to construct a 7-sphere with a scalar radius, as in Eq.~(2.60) of [1].

By contrast the standard definition (2.8) of the norm (stated in the preamble of the introductory Section 2 of [1] that summarizes the basic concepts of Geometric Algebra) picks out the scalar part of the geometric product $\Omega\Omega^{\dagger}$, which is indicated by the subscript `s'. This definition is {\it not} used in calculating the norms in [1] because it is not consistently applicable in an equation like $\|X Y\|=\|X||\,\| Y \|$
that involves geometric product on at least one side.

\underbar{\bf Author action}: No action.}

2. Can your S3 space be viewed as describing the wave function of the two particles?

{\color{black}{\underbar{\bf Author response}}: In my view wave functions and entanglement are merely placeholders for the observed quantum correlations. They are not fundamental to Nature. The $S^3$ space facilitates a local-realistic model for the observed correlations between two or many particles without needing a wave function. Thus it does the job that is traditionally accomplished by quantum mechanical wave functions via the usual Hilbert space formulation of quantum mechanics. 

\underbar{\bf Author action}: No action.}

3. Your S3 space  could also be viewed as arising from the space of unit quaternions used to rotate vectors in 3D space?

{\color{black}{\underbar{\bf Author response}}: Yes, $S^3$ is a set of all unit quaternions. It is a quaternionic 3-sphere that can be defined as
\begin{equation}
S^3:=\left\{\,{\bf q}(\theta,\,{\mathbf r})=\varrho_r\left[\cos\left(\frac{\theta}{2}\right)+{\mathbf J}({\mathbf r})\,\sin\left(\frac{\theta}{2}\right)\right]
\Bigg|\;\left|\left|\,{\bf q}\left(\theta,\,{\mathbf r}\right)\,\right|\right|=\varrho_r\right\}\!, \notag
\end{equation}
where ${{\mathbf J}({\mathbf r})=I_3{\mathbf r}}$ is a unit bivector (or a pure quaternion) rotating about an axis vector ${{\bf r}\in{\mathrm{I\!R}}^3}$ with rotation angle ranging from ${0\leq\theta < 4\pi}$, $\varrho_r$ is the radius of the corresponding 3-sphere, and $I_3={\mathbf e}_x{\mathbf e}_y{\mathbf e}_z$ is the standard trivector.

\underbar{\bf Author action}: No action.}

I decided to check the criticisms of the other reviewers regarding JC's norm that he defined for the dual quaternion, pdf document attached.

My calculations show that the norm is correct as defined by JC, coupled with his condition of orthogonal quaternions, which then creates a norm within the Real numbers. Hence, the counter-example proposed with $\mathrm{e}^2-1=(\mathrm{e}-1)(\mathrm{e}+1)=0$, does not invalidate this norm, as both $\mathrm{e}-1$ and $\mathrm{e}+1$ are not orthogonal quaternions. Nevertheless, for this counter-example, they actually do satisfy the norm anyway, provided the square root is taken properly, see attached.  I also explicitly calculate the square root of $\mathrm{e}-1$ and $\mathrm{e}+1$, to help show this, which JC does not do.

{\color{black}{\underbar{\bf Author response}}: I am grateful to the reviewer for verifying my proof of the norm relation and recognizing that the alleged counterexample does not invalidate it. I agree with the reviewer's calculations reproduced below. It is easy to see that they mostly parallel the proofs I have presented in [1], [3], and [25], apart for the useful addition of explicit square roots of $1 - \varepsilon$ and $1 + \varepsilon$.

I do calculate explicit square roots for the general elements of ${\cal K}^{\lambda}$ in Appendix~A.1 of Ref.~[3] using the exponential method for taking square roots of hyperbolic numbers. See equations (A.22) to (A.25) in Appendix~A.1 of [3]. 

\underbar{\bf Author action}: I have adapted the reviewer's explicit expressions for the square roots of $1 - \varepsilon$ and $1 + \varepsilon$ in Section~2.8 of the revised manuscript.}

Your have created an impressive theory, and I hope the ultimate test of an experiment can be performed soon, as you suggest.

{\color{black}{\underbar{\bf Author response}}: I thank the reviewer for these kind words.

\underbar{\bf Author action}: No action.}

\parindent 0.7cm

\begin{center}
\Large{\bf Calculation by Reviewer \# 3 --- Dual quaternions and their norm:}
\end{center}
\section{Introduction}
It is shown here that JC's generalized norm is satisfied, both in the specific case, using the expansion $\varepsilon^{2}-1=(\varepsilon-1)(\varepsilon+1)=0$ as well as in the general case, provided that the square root operation uses the principal square root.

Extra clarity is added to this issue through explicitly calculating each individual square root.

\section{Analysis}

\subsection{The special case with $\varepsilon^{2}-1=(\varepsilon-1)(\varepsilon+1)=0$}

We have the algebraic expansion $\varepsilon^{2}-1=(\varepsilon-1)(\varepsilon+1)=0$, and selecting $Z=\varepsilon^{2}-1$, and $X=\varepsilon-1, Y=\varepsilon+1$ we wish to show that the norm relation
\begin{equation}
\|Z\|=\|X Y\|=\|X||\| Y \|  
\end{equation}
is satisfied where $\|Z\|=\sqrt{Z Z^{\dagger}}$. We also have defined $\varepsilon=e_{x} e_{y} e_{z} e_{\infty}$, where $\varepsilon^{\dagger}=\varepsilon$ and $\varepsilon^{2}=1$

Now
\begin{align}
&\|X\|=\sqrt{(\varepsilon-1)(\varepsilon-1)^{\dagger}}=\sqrt{(\varepsilon-1)(\varepsilon-1)}=\sqrt{2(1-\varepsilon)}, \\
&\|Y\|=\sqrt{(\varepsilon+1)(\varepsilon+1)^{\dagger}}=\sqrt{(\varepsilon+1)(\varepsilon+1)}=\sqrt{2(1+\varepsilon)}. \notag
\end{align}
We can now explicitly calculate the square root of each individual term. That is
\begin{align}
&\sqrt{1-\varepsilon}=\left(\frac{1}{\sqrt{2}}-\frac{\varepsilon}{\sqrt{2}}\right), \\
&\sqrt{1+\varepsilon}=\left(\frac{1}{\sqrt{2}}+\frac{\varepsilon}{\sqrt{2}}\right). \notag
\end{align}
Its obviously important for consistency to take the principal, positive square root in both cases. These square roots can be confirmed through squaring, that is
\begin{equation}
\left(\frac{1}{\sqrt{2}} \pm \frac{\varepsilon}{\sqrt{2}}\right)^{2}=1 \pm \varepsilon,
\end{equation}
as required.

Hence
\begin{align}
&\|X\|=\sqrt{2} \sqrt{(1-\varepsilon)}=\sqrt{2}\left(\frac{1}{\sqrt{2}}-\frac{\varepsilon}{\sqrt{2}}\right)=1-\varepsilon \\
&\|Y\|=\sqrt{2} \sqrt{(1+\varepsilon)}=\sqrt{2}\left(\frac{1}{\sqrt{2}}+\frac{\varepsilon}{\sqrt{2}}\right)=1+\varepsilon. \notag
\end{align}
{\color{black}{\underbar{\bf Author comment}}: I have adapted the above expressions for the square roots in Section~2.8 of the manuscript.}\break
\\
Therefore
\begin{equation}
\|X\|\|Y\|=(1-\varepsilon)(1+\varepsilon)=1-\varepsilon^{2}=0,
\end{equation}
using $\varepsilon^{2}=1$. We also have $\|Z\|=\left\|\varepsilon^{2}-1\right\|=\|0\|=0$.

Hence, we have shown explicitly that $\varepsilon^{2}-1=(\varepsilon-1)(\varepsilon+1)=0$ satisfies the norm relation.

\subsection{Confirming the norm in the general case}

We firstly confirm that $\varepsilon$ commutes with all quaternions, such as
\begin{align}
&q_{r}=a+v_{1} e_{y} e_{z}+v_{2} e_{z} e_{x}+v_{3} e_{x} e_{y}=a+I_{3} \boldsymbol{v}, \\
&q_{d}=b+w_{1} e_{y} e_{z}+w_{2} e_{z} e_{x}+w_{3} e_{x} e_{y}=b+I_{3} \boldsymbol{w},
\end{align}
where $I_{3}=e_{x} e_{y} e_{z}$. For clarity we have used regular Cartesian vectors $\boldsymbol{v}=$ $v_{1} e_{x}+v_{2} e_{y}+v_{3} e_{z}$ and $\boldsymbol{w}=w_{1} e_{x}+w_{2} e_{y}+w_{3} e_{z}$. We can confirm that
\begin{equation}
I_{3} \boldsymbol{v}=e_{x} e_{y} e_{z}\left(v_{1} e_{x}+v_{2} e_{y}+v_{3} e_{z}\right)=v_{1} e_{y} e_{z}+v_{2} e_{z} e_{x}+v_{3} e_{x} e_{y},   
\end{equation}
as required. Then we can write a dual quaternion
\begin{equation}
Q=q_{r}+\varepsilon q_{d}
\end{equation}
and the norm squared of $Q$ is
\begin{equation}
\|Q\|^{2}=Q Q^{\dagger}=\left(q_{r}+\varepsilon q_{d}\right)\left(q_{r}^{\dagger}+\varepsilon q_{d}^{\dagger}\right)=q_{r} q_{r}^{\dagger}+q_{d} q_{d}^{\dagger}+\varepsilon\left(q_{r} q_{d}^{\dagger}+q_{d} q_{r}^{\dagger}\right).
\end{equation}
Now
\begin{equation}
q_{r} q_{r}^{\dagger}+q_{d} q_{d}^{\dagger}=a^{2}+\boldsymbol{v}^{2}+b^{2}+\boldsymbol{w}^{2}=a^{2}+v_{1}^{2}+v_{2}^{2}+v_{3}^{2}+b^{2}+w_{1}^{2}+w_{2}^{2}+w_{3}^{2},
\end{equation}
a non-negative scalar, giving the Pythagorean distance in eight dimensions. Also
\begin{equation}
q_{r} q_{d}^{\dagger}+q_{d} q_{r}^{\dagger}=\left(a+I_{3} \boldsymbol{v}\right)\left(b-I_{3} \boldsymbol{w}\right)+\left(b+I_{3} \boldsymbol{w}\right)\left(a-I_{3} \boldsymbol{v}\right)=2(a b+\boldsymbol{v} \cdot \boldsymbol{w}),
\end{equation}
giving a pure scalar result, as the dot product of two vectors is a scalar. Hence, the norm squared of the dual quaternion is a commuting dual number (also called a hyperbolic or split complex number) $c+\varepsilon d$, where $c, d \in \Re$.\break
\\
{\color{black}{\underbar{\bf Author comment}}: I have adapted some of this language in my proof presented in Section~2.8 of the manuscript.}

Now for two dual quaternions $X, Y$, we have the norm squared
\begin{equation}
\|X Y\|^{2}=(X Y)(X Y)^{\dagger}=X Y Y^{\dagger} X^{\dagger}=X\|Y\|^{2} X^{\dagger}=X X^{\dagger}\|Y\|^{2}=\|X\|^{2}\|Y\|^{2}
\end{equation}
a completely general result, relying on the norm being a commuting dual number. The square root of a dual number is
\begin{equation}
\sqrt{c+\varepsilon d}=\frac{b}{\sqrt{2} \sqrt{c \pm \sqrt{c^{2}-d^{2}}}}+\frac{\varepsilon}{\sqrt{2}} \sqrt{c \pm \sqrt{c^{2}-d^{2}}}   
\end{equation}

However, we can see from the norm, that the real part is always greater or equal to the pseudoscalar part, and hence $c>=d$. Hence, the roots of these dual numbers will also be dual (and not complex), although they will be two-valued, in general.

We thus need care when distributing the square root over two dual numbers $\sqrt{\|X\|^{2}\|Y\|^{2}}$, through selecting the principal root in each case. The double valued nature of the square root implies the equation will be satisfied at two different values.

\subsection{The orthogonality condition}

If we specify an orthogonality condition of $q_{r} q_{d}^{\dagger}+q_{d} q_{r}^{\dagger}=0$ then the norm will always be a non-negative real number and so we can immediately write
\begin{equation}
\|X Y\|=\|X\|\|Y\|,   
\end{equation}
as we will only be taking roots over non-negative Real numbers.

If we consider again the special test case of $\varepsilon^{2}-1=(\varepsilon-1)(\varepsilon+1)=0$, we can see immediately that with multivectors $X=\varepsilon-1$ and $Y=\varepsilon+1$ that they do not satisfy the orthogonality condition and so are not in the restricted class of multivectors with a non-negative real valued norm. Nevertheless, as shown, they still satisfy the norm relation.

\section{Discussion}

One of the points of contention with $\mathrm{JC}$ 's 'norm' is that it is not conventional to call it a norm, which should be a non-negative scalar quantity. However, accepting JC's definition of a generalized norm over the dual numbers, it does appear that the norm relation $\|X Y\|=\|X\|\|Y\|$ is valid in general, provided that the principal root is selected (similar to complex numbers). Specifically, the roots of dual numbers are two-valued, and so they need to be matched up on both sides of the equation.

We show explicitly that the alleged counter example using $\varepsilon^{2}-1=(\varepsilon-$ 1) $(\varepsilon+1)=0$ does not disprove this norm relation.\break
\\
{\color{black}{\underbar{\bf Author comment}}: Moreover, using the orthogonality condition ${\bf q}_{r}\,{\bf q}^{\dagger}_{d}+{\bf q}_{d}\,{\bf q}^{\dagger}_{r}=0$ as a normalization condition, the resulting norms reduced to scalar-valued norms, as explained in more detail in Section~2.8 of the manuscript.\break
\\
\underbar{\bf Author action}: In Section~2.8 of the revised manuscript, I have added the following footnote: ``During the review\break process of this paper, one of the reviewers independently verified this proof of the norm relations (5)  and (27) with detailed calculations and comments, which are available online in the Review History published with this paper.''}

\begin{center}
\Large{\bf Another calculation by Reviewer \# 3 --- Dual quaternions:}
\end{center}
{\color{black}{\underbar{\bf Author comment}}: The following calculation by Reviewer \# 3 appears to be an earlier version of the one above. It is less detailed than the previous version, but I am including it here for the sake of completeness of the records.}
\setcounter{equation}{0}
\section*{1\;\;\;\,Introduction}
We have $\varepsilon=e_{x} e_{y} e_{z} e_{\infty}$, where $\varepsilon^{\dagger}=\varepsilon$ and $\varepsilon^{2}=1$ and $\varepsilon$ commutes with all quaternions, such as
\begin{align}
&q_{r}=a+v_{1} e_{y} e_{z}+v_{2} e_{z} e_{x}+v_{3} e_{x} e_{y}=a+I_{3} \boldsymbol{v}, \\
&q_{d}=b+w_{1} e_{y} e_{z}+w_{2} e_{z} e_{x}+w_{3} e_{x} e_{y}=b+I_{3} \boldsymbol{w}, \notag
\end{align}
where $I_{3}=e_{x} e_{y} e_{z}$. For clarity we can use regular Cartesian vectors $\boldsymbol{v}=v_{1} e_{x}+$ $v_{2} e_{y}+v_{3} e_{z}$ and $\boldsymbol{w}=w_{1} e_{x}+w_{2} e_{y}+w_{3} e_{z}$. We can then confirm that
\begin{equation}
I_{3} \boldsymbol{v}=e_{x} e_{y} e_{z}\left(v_{1} e_{x}+v_{2} e_{y}+v_{3} e_{z}\right)=v_{1} e_{y} e_{z}+v_{2} e_{z} e_{x}+v_{3} e_{x} e_{y},    
\end{equation}
as required. Then we can write a dual quaternion
\begin{equation}
Q=q_{r}+\varepsilon q_{d}   
\end{equation}
and the norm squared of $Q$ is
\begin{equation}
\|Q\|^{2}=Q Q^{\dagger}=\left(q_{r}+\varepsilon q_{d}\right)\left(q_{r}^{\dagger}+\varepsilon q_{d}^{\dagger}\right)=q_{r} q_{r}^{\dagger}+q_{d} q_{d}^{\dagger}+\varepsilon\left(q_{r} q_{d}^{\dagger}+q_{d} q_{r}^{\dagger}\right).  
\end{equation}
Now
\begin{equation}
q_{r} q_{d}^{\dagger}+q_{d} q_{r}^{\dagger}=\left(a+I_{3} \boldsymbol{v}\right)\left(b-I_{3} \boldsymbol{w}\right)+\left(b+I_{3} \boldsymbol{w}\right)\left(a-I_{3} \boldsymbol{v}\right)=2(a b+\boldsymbol{v} \cdot \boldsymbol{w}),
\end{equation}
giving a pure scalar result, as the dot product of two vectors is a scalar. Hence, the norm squared of the dual quaternion is a commuting dual number $c+\varepsilon d$, where $c, d \in \Re$.

Now for two dual quaternions $X, Y$, we have the norm squared
\begin{equation}
\|X Y\|^{2}=(X Y)(X Y)^{\dagger}=X Y Y^{\dagger} X^{\dagger}=X\|Y\|^{2} X^{\dagger}=X X^{\dagger}\|Y\|^{2}=\|X\|^{2}\|Y\|^{2}   
\end{equation}
a completely general result, relying on the norm being a commuting dual number.

Taking square roots, in order to find the norm could be problematic over the dual numbers, as the dual numbers may not contain the square root (if they require a square root of minus one) and care will be needed to distribute the square root over two dual numbers when finding $\sqrt{\|X\|^{2}\|Y\|^{2}}$.

However, if we specify an orthogonality condition, of $q_{r} q_{d}^{\dagger}+q_{d} q_{r}^{\dagger}=0$ then the norm will always be a non-negative real number and so we can write
\begin{equation}
\|X Y\|=\|X\|\|Y\|. 
\end{equation}

If we consider the special test case of $\varepsilon^{2}-1=(\varepsilon-1)(\varepsilon+1)=0$, we can see immediately that with multivectors $X=(\varepsilon-1)$ and $Y=(\varepsilon+1)$ they do not satisfy the orthogonality condition, and so are not valid dual quaternions to use in the norm. However, we can proceed regardless to see if the norm is satisfied in this case. We find the norm
\begin{equation}
\|X\|=\sqrt{(\varepsilon-1)(\varepsilon-1)}=\sqrt{2(1-\varepsilon)}.
\end{equation}
Now, in this case the square root does exist within the dual numbers, with $\sqrt{1-\varepsilon}=\frac{1}{\sqrt{2}}-\frac{\varepsilon}{\sqrt{2}}$ and so $\|X\|=1-\varepsilon$, and we can also find $\|Y\|=1+\varepsilon$. Therefore $\|X\|\|Y\|=(1-\varepsilon)(1+\varepsilon)=0$, as required, and so satisfies the norm. However, the individual norms $\|X\|,\|Y\|$, are not scalars because they do not satisfy the orthogonality constraint.

\bigskip

\bigskip

\hrule 

\bigskip

\parindent 0pt

\underbar{\textcolor{blue}{{\large\bf Reviewer \# 4}}}

The author has satisfactorily responded to the reviewers' comments and suggestions and appropriately modified the manuscript. I recommend the comment for publication.

{\color{black}{\underbar{\bf Author response}}: I thank the reviewer for recommending my manuscript for publication. 

\underbar{\bf Author action}: No action.}

\bigskip

\bigskip

\hrule 

\bigskip

\underbar{\textcolor{blue}{{\large\bf Reviewer \# 5}}}

With respect to sec. 2.4, I criticized material from the core of the section. I objected to a quotation from Christian as implying that a certain claim is both true and false. The quotation runs:

\begin{quote}
“While \underbar{mathematically correct}, Eq.~(1) is \underbar{physically meaningless within any hidden variable theory}. […] The problem with Eq. (1) is that, while the sum of expectation values is \underbar{mathematically the same} as the expectation value of the sum, as in the assumption that allows us to mathematically replace the left-hand side of Eq. ( 1) above with its right-hand side, and while this assumption is \underbar{valid in quantum mechanics} […], it is \underbar{not valid} for any hidden variable theory involving dispersion-free states, […]. This makes the replacement of the left-hand side of Eq. (1) with its right-hand side \underbar{physically invalid}. [… Once the \underbar{fallacious} Eq.~(1)] is removed from Bell’s argument and local realism is implemented correctly, the bounds on the right-hand side of Eq. (1) work out to be $\pm2\sqrt{2}$ instead of $\pm2$, thereby mitigating the conclusions of Bells theorem.”
\end{quote}

Christian has now cancelled some of the material (“mathematically correct”, “fallacious”) that led to trouble. 

{\color{black}{\underbar{\bf Author response}}: I have not ``cancelled'' any material. I had never used the word ``fallacious'' in the manuscript, for example, so there is no question of ``cancelling'' it. I only improved the wording and presentation of Section~2.4 to make my argument clearer.  

\underbar{\bf Author action}: I have further improved the wording and presentation of Section~2.4 in the revised manuscript.}

The revised quotation runs:

\begin{quote}
“The problem with Eq. (1) is that, while the sum of expectation values is mathematically the same as the expectation value of the sum, as in the assumption that allows us to mathematically replace the left-hand side of Eq. (1) above with its right-hand side, and while this assumption is valid in quantum mechanics as the critique notes, it is not valid for any hidden variable theory involving dispersion-free states, because the eigenvalue of a sum of operators is not the sum of individual eigenvalues when the constituent operators are noncommuting, as in the Bell-test experiments.”
\end{quote}

However, the text still contains the following sentence:

\begin{quote}
“[…] while the sum of expectation values is mathematically the same as the expectation value of the sum […], and while this assumption is valid in [… theory A …], it is not valid for [… theory B.]”
\end{quote}

To understand this sentence, we require clarification. Strictly speaking, a claim (an assumption or proved identity) cannot be valid or invalid but only true or false. Is the present claim (“the sum of expectation values is mathematically the same as the expectation value of the sum”) an assumption or a proved identity? Christian clearly means the latter (“is \underbar{mathematically} the same”) despite his reference to “assumption”. Hence, his claim is that a proved identity ‘is valid in theory A, not valid in theory B’ or, more exactly, ‘is true in theory A, false in theory B’. This is absurd. Assume alternatively that the present claim is an assumption that we just have made, aware of the fact that we do not know whether it is true or false (= a mere assumption). Then, Christian says that the present claim ‘is
a mere assumption that is true in theory A, false in theory B’. This is also absurd.

Hence, Christian cannot dispel the original concern that his presentation implies that a certain claim
is both true and false.

{\color{black}{\underbar{\bf Author response}}: As I explained in the previous round of reviews, the argument I have presented in Section~2.4 of the manuscript is not the one I have invented myself. It cannot be brushed aside as ``absurd'', as the above comments by the reviewer purports to do. The problem was recognized by Einstein in the context of von~Neumann's theorem against hidden variable theories in the mid 1930s. It was also recognized by Grete Hermann around the same time. Thus the problem has been well known in foundations of quantum mechanics for nearly nine decades. One of the most lucid explanations of the problem can be found in Section~3 of Chapter~1 in Bell's book. As I noted with a list of references in the previous round of reviews, the same problem with von~Neumann's theorem is also discussed by several other well known experts in foundations of quantum mechanics. Apart from the historical precedence of the argument I have presented in Section~2.4, there is a near universal agreement among the experts in foundations of quantum mechanics on the invalidity of the assumption of additivity of expectation values for hidden variable theories [18]. One has to adapt double standards -- one standard for von~Neumann's theorem and another for Bell's theorem -- to deny the invalidity of the assumption of additivity of expectation values for hidden variable theories. 

\underbar{\bf Author action}: I have revised Section~2.4 to explain the problem and the above points more clearly with the help of the example of Stern-Gerlach magnet Bell gives in Chapter~1 of his book [17].}

With respect to sec. 2.5, I have criticized that it is unclear from what theory a violation of the Bell-CHSH inequality is derivable. Originally, Christian wrote:

\begin{quote}
“The second assumption \underbar{necessary} to support Bell’s theorem is that of immutable spacetime. [… If we abandon this assumption and regard spacetime as mutable, then we can] model
physical space as a closed and compact quaternionic 3-sphere, or S3 , instead of a flat Euclidean space, or IR3 […]. But once the physical space is modelled as S 3 instead of IR3 […], the correlation between the results Aa and Bb […] \underbar{inevitably} turns out to be E(AaBb) = -a·b […]”.
\end{quote}

To repeat my earlier criticism: the first sentence seems to claim that two things are “[\underbar{jointly?}] \underbar{necessary} to support Bell’s theorem”, the failure of eq.~1 for “operators [that] are non-commuting” and a mutable spacetime. But the last sentence (“[…] once the physical space is modelled as S 3 …], the correlation […] \underbar{inevitably} turns out to be […the quantum-mechanical one]”) suggests that introducing S 3 alone is sufficient to derive the same correlation.

Christian does not address this problem. He responds to the criticism as follows:

\begin{quote}
“[…] Section 2.4 concerns a formal aspect of the proof of Bell’s theorem that has been discussed in the comment paper [2]. That formal aspect does not concern the quaternionic 3-sphere model of the quantum correlations, viewed as a counterexample to Bell’s theorem. In the manuscript I have not mixed the issues concerning the formal aspect of Bell’s theorem and those concerning my specific counterexample. It is unclear to me why the reviewer has
tried to mix up these two aspects and tried to read between the lines. I suspect that is because of the reviewer’s acknowledged neglect of my original 2018 paper and the comment paper [2].” 
\end{quote}

Christian leaves it unclear whether A and B are (jointly?) necessary for C or whether B alone is sufficient for C. Asking which argument exactly Christian wants to present is not the same as mixing up arguments that should be kept separate. Moreover, in the present context it is of great interest to know from what assumptions precisely a violation of the Bell-CHSH inequality follows.

{\color{black}{\underbar{\bf Author response}}: As I noted in my response to the above comments in the previous round of reviews, the current manuscript is a response to a comment paper [2] on my 2018 paper [1]. It is not a comprehensive review paper on my quaternionic 3-sphere model for understanding quantum correlations. That model, as a constructive counterexample to Bell's theorem, is presented in the references [1], [3], [5], [6], [7], [8], [9], [10], [15], [18], and [22] cited in the manuscript. My discussion in Sections~2.4 and 2.5 of the manuscript are a direct response to two of the issues raised in the comment paper [2]. Among these issues, Section~2.4 concerns a formal aspect of the proof of Bell's theorem that has been discussed in the comment paper [2]. That formal aspect does not concern the quaternionic 3-sphere model\break of the quantum correlations, viewed as a counterexample to Bell's theorem. In the manuscript I have not mixed the issues concerning the formal aspect of Bell's theorem and those concerning my specific counterexample in [1].

Had the reviewer read my original paper [1] instead of opting not to read it before reviewing the manuscript, he/she would have noticed Section~3.3 of the paper entitled ``Derivation of Tsirel'son's bounds on the correlation strength'' in which I explicitly derive violations of Bell inequalities within the 7-sphere framework presented therein, thereby precisely reproducing the quantum mechanical predictions. Unfortunately, the reviewer has instead opted to argue abstractly in terms of `theory A' and `theory B' without substance and without specifying what these theories are.

\underbar{\bf Author action}: No action.}

Neither response to the two criticisms is convincing. Moreover, they are representative for Christian’s style of argumentation. In a response to Gill’s criticisms, these criticisms should be properly addressed, not dismissed in favor of more text concerning other topics and ideas.

I still do not recommend the publication of this response.

{\color{black}{\underbar{\bf Author response}}: Both issues raised by the reviewer are somewhat tangential to the main goal of the manuscript, which is to address the specific issues raised in the critique [2]. The reviewer has focused only on Sections~2.4 and 2.5 of the current manuscript and acknowledged (in the previous round) that he/she has not read or intend to read either the original paper [1] or its critique [2]. It is therefore not surprising that the reviewer's comments lack any substantive context of the details of the constructive model presented in the original paper [1] or its critique in [2].  

The reviewer has instead commented on my style of argumentation. However, in my view the substantive scientific issues I have elucidated in Sections~2.4 and 2.5 are independent of my style of argumentation or other idiosyncrasies. I therefore contend that the reviewer's recommendation for the rejection of my manuscript is not justified. 

\underbar{\bf Author action}: No action.}

\bigskip

\bigskip

\hrule 

\bigskip

\underbar{\textcolor{blue}{{\large\bf Reviewer \# 6}}}

My comments relate to the author's responses to my own first round comments as a referee (number 6 in the list).

$\quad\quad$ There has been a fractional step forward here, in that the author now recognises that what he is saying does not work for the standard definition of norm.

{\color{black}{\underbar{\bf Author response}}: The above statement is not accurate. It gives an incorrect impression of what I actually said. What I explained, with demonstration, in the previous round of reviews is that the standard definition of norm in Geometric Algebra that picks out the scalar part of the geometric product in an {\it ad hoc} manner (as in Eq.~(2.8) of [1]) is not applicable to an equation such as $\|X Y\|=\|X\|\,\|Y\|$ that involves geometric product at least on one of its sides, because it leads to mixing of two different product rules of the two sides of the equation. That amounts to introducing inconsistency. That is what the reviewer's alleged counterexample did. It inadvertently introduced inconsistency in the equation to derive the alleged contradiction. One can always draw a contradiction from any mathematical equation by employing two different rules on the two sides of the equation (in our case product rules).

The reviewer's statement gives the incorrect impression that I have recognized something new ``now'', thanks to the reviewer's comments in the previous round, or that I have changed my mind about something in the light of the critique [2]. That is not at all the case. I have not invented new definitions just for my response to the critique [2]. My arguments were already very clearly stated in the original paper in 2018. Nothing new has transpired since.        

\underbar{\bf Author action}: No action.}

In particular, he says the norm he quotes in equation (2.8) in the original RSOS paper (2018) (which is the standard one), is not the one he is using by the time he gets to Section $2.5$ of that paper.

{\color{black}{\underbar{\bf Author response}}: That is correct, but it is by no means a new statement. It should not come as a surprise to anyone who has read the original {\it RSOS} paper [1], even if they are unaware of my repeatedly pointing out to this fact in the discussion thread of the paper and in my replies [3] and [15] to the critique [14]. Let me reproduce the relevant paragraph form Section~2.5 of [1] with yellow highlights for convenience in case the reviewer has missed it:
\begin{quote}
Now the normalization of ${{\mathbb Q}_z}$ in fact necessitates that every ${{\bf q}_r}$ be orthogonal to its dual ${{\bf q}_d}$:
\begin{equation}
\hll{||{\mathbb Q}_z|| = \sqrt{2}\,\varrho\;\;\Longrightarrow\;\;{\bf q}_r\,{\bf q}^{\dagger}_d+{\bf q}_d\,{\bf q}^{\dagger}_r = 0}\,, \tag{2.54}
\end{equation}
or equivalently, ${\langle\,{\bf q}_r\,{\bf q}^{\dagger}_d\,\rangle_s = 0}$ ({\it i.e.}, ${{\bf q}_r\,{\bf q}^{\dagger}_d}$ is a pure quaternion). We can see this by working out the product of ${{\mathbb Q}_{z}}$ with ${{\mathbb Q}^{\dagger}_{z}}$ while using ${\varepsilon^2=+1}$, which gives
\begin{equation}
{\mathbb Q}_{z}\,{\mathbb Q}^{\dagger}_{z}\,=\;\left({\bf q}_{r}\,{\bf q}^{\dagger}_{r}\,+\,{\bf q}_{d}\,{\bf q}^{\dagger}_{d}\right)\,+\,\left({\bf q}_{r}\,{\bf q}^{\dagger}_{d}\,+\,{\bf q}_{d}\,{\bf q}^{\dagger}_{r}\right)\,\varepsilon\,. \tag{2.55}
\end{equation}
Now, using the definition of ${\bf q}$ in (2.16), it is not difficult to see that 
${{\bf q}_{r}\,{\bf q}^{\dagger}_{r}={\bf q}_{d}\,{\bf q}^{\dagger}_{d}=\varrho^2}$, reducing the above product to
\begin{equation}
{\mathbb Q}_{z}\,{\mathbb Q}^{\dagger}_{z}\,=\;2\,\varrho^2\,+\,\left({\bf q}_{r}\,{\bf q}^{\dagger}_{d}\,+\,{\bf q}_{d}\,{\bf q}^{\dagger}_{r}\right)\,\varepsilon\,. \tag{2.56}
\end{equation}
It is thus clear that \hl{for ${{\mathbb Q}_{z}\,{\mathbb Q}^{\dagger}_{z}}$ to be a scalar ${{\bf q}_{r}\,{\bf q}^{\dagger}_{d}+{\bf q}_{d}\,{\bf q}^{\dagger}_{r}}$ must vanish, or equivalently ${{\bf q}_r}$ must be orthogonal to ${{\bf q}_d}$.} 

$\quad\;\;$\hl{But there is more to the normalization condition  ${{\bf q}_{r}\,{\bf q}^{\dagger}_{d}+{\bf q}_{d}\,{\bf q}^{\dagger}_{r}=0}$ then meets the eye. It also leads to the crucial norm relation (2.40)}, which is at the very heart of the only possible four normed division algebras associated with the four parallelizable spheres ${S^0}$, ${S^1}$, ${S^3}$, and ${S^7}$ (cf. appendix~A). 
\end{quote}
In the rest of the Section~2.5 I then sketch the proof of the norm relation (2.40), leaving details for exercise. 

\underbar{\bf Author action}: No action.}

This seems contradicted by explicit statements in Section $2.4$ of the paper, which is directly connected to Section 2.5, and contains very explicitly the key result in terms of norms, which the author spends a lot of time seeking to prove, namely equation (2.40). In Section (2.4), it introduces the expansion of general quantities in the even subalgebra in terms of coefficients multiplied by basis elements (equations (2.36) and (2.37)). Then it says that using (2.8) then after normalisation the square of their norms is given by the expressions in (2.38), which read
\begin{equation}
\|X\|^{2}=\sum_{\mu=0}^{7} X_{\mu}^{2}=1 \tag{1.1}
\end{equation}
with a similar expression for $Y$, and then this result is repeated in (2.41) for their product $Z$. 

{\color{black}{\underbar{\bf Author response}}: There is no contradiction in what I have written in Section~2.4 and Section~2.5 of the paper. The norm relation (2.40) holds for all elements of the algebra ${\cal K}^{\lambda}$, {\it without exception}. I have proved (2.40) in [1] and in Refs.~[3], and [25] cited in the manuscript. And
now Reviewer~\#~3 has also independently verified my proof of (2.40). As evident from Eq.~(2.54), the scalar values for the norms obtained by using (2.8) and (2.54) are identical:
\begin{equation}
\sqrt{\,\langle\,{\mathbb Q}_z\,{\mathbb Q}_z^{\dagger}\rangle_s}\;\equiv\sqrt{\left({\mathbb Q}_{z}{\mathbb Q}^{\dagger}_{z}\right)\!\Big|_{\,{\bf q}_{r}\,{\bf q}^{\dagger}_{d}+{\bf q}_{d}\,{\bf q}^{\dagger}_{r}\,=\,0}\,}\;. \notag
\end{equation}
Therefore all the statements made in Sections~2.4 and 2.5 of the paper are correct and compatible with Eq.~(2.54).

\underbar{\bf Author action}: No action.}

There are only two ways these expressions can apply to general non-zero elements. The first is that the definition (2.8) is being used (as in fact stated by the author at this point in the paper) - this involves taking the scalar part of e.g. $X \tilde{X}$ before taking the square root. The second is to work with quantities $X$ of a type which when we form $X \tilde{X}$ then this product does not contain non-scalar parts. In the $4 \mathrm{~d}$ Euclidean Clifford algebra, the extra part which could be present in this product is a grade- 4 pseudoscalar, so a quantity of this form would have to satisfy
\begin{equation}
\langle X \tilde{X}\rangle_{4}=0 \tag{1.2}
\end{equation}
in order for (2.38) to follow, and similarly for $Y$. 

{\color{black}{\underbar{\bf Author response}}: The reviewer's equation (1.2) is the normalization condition I have used in my original paper [1].\break This is evident from Eqs.~(2.54) to (2.56) in the yellow highlighted paragraph I have quoted above from the paper. It is not difficult to appreciate that both definitions (2.8) and (2.54) in [1] give the same scalar value for the norm.

\underbar{\bf Author action}: No action.}

Such an object is called a scaled rotor in Geometric Algebra, and the assertion made in the key equation (2.40) that
\begin{equation}
\|X Y\|=\|X\|\,\|Y\| \tag{1.3} 
\end{equation}
does indeed follow for this type of object, and corresponds to the fact that the composition of rotors is another rotor.

$\quad\quad$ However, the author claims that (2.40) is true for any even multivectors $X$ and $Y$, 

{\color{black}{\underbar{\bf Author response}}: Yes, the norm relation (2.40) holds for any multivectors $X$ and $Y$ within the algebra ${\cal K}^{\lambda}$, without exception. I have proved the norm relation (2.40) in Refs.~[1], [3], and [25] cited in the manuscript, and now Reviewer~\#~3 (whose calculations I have reproduced above verbatim) has independently verified my proof.  

\underbar{\bf Author action}: In Section 2.8 of the revised manuscript I have summarized the proof again for convenience.}

and I think we both now agree that for the norm which is apparently, and also stated to be used at this point, then this statement is false (as demonstrated by the counterexamples) - it only works (in a general sense) for the scaled rotors.

{\color{black}{\underbar{\bf Author response}}: No, I do not agree with the above claims. The norm relation (2.40) holds for any multivectors $X$ and $Y$ within the algebra ${\cal K}^{\lambda}$, without exception. No counterexample to (2.40) can or does exist. In the last round of reviews, in [3], and in the manuscript, I have comprehensively refuted the alleged counterexample to (2.40). 

\underbar{\bf Author action}: No action.}

$\quad\quad$ At points in the current reply by the author, there is a recognition that the product rule for norms only works for the special case where $\langle X \tilde{X}\rangle_{4}=0$, and the equivalent for $Y$, and is not general, but this is coupled with an insistence that this restriction does not limit the type of object considered. It does, to scaled rotors.

{\color{black}{\underbar{\bf Author response}}: I do not agree with the above claim. The confusion in it stems from a failure to appreciate that $\langle X \tilde{X}\rangle_{4}=0$ is a normalization condition (cf. Eq.~(2.54) of [1] quoted above). Normalization conditions do not impose any restrictions on the algebra itself. This is not difficult to understand. But since the confusion has persisted over four years, let me give an elementary example from geometry. Consider a vector space $\mathrm{I\!R}^3$ spanned by vectors of the form ${\bf v}=x\,{\bf e}_x+y\,{\bf e}_y+z\,{\bf e}_z$ from the origin. Then a 2-sphere of a fixed radius $\rho$ embedded in $\mathrm{I\!R}^3$ can be constructed as a surface $S^2$ by normalizing the vectors in a familiar manner:
\begin{equation}
\rho = \sqrt{{\bf v}\cdot{\bf v}}=\sqrt{x^2+y^2+z^2}. \notag
\end{equation}
Now this normalization condition is a constraint equation that restricts the length of the vectors to $\rho$. But it by no means puts any restriction on the space $\mathrm{I\!R}^3$. Similarly, the normalization condition $\langle X \tilde{X}\rangle_{4}=0$ I have used to construct $S^7$ in my paper may seem unorthodox, but it by no means puts any restriction on the algebra ${\cal K}^{\lambda}$ itself. 

\underbar{\bf Author action}: No action.}

$\quad\quad$ The remedy currently proposed by the author, in order to rescue the statement that (2.40) applies to all objects in the algebra, is to enlarge the concept of `norm' to be the square root of $X \tilde{X}$ even when this product contains non-scalar parts.

{\color{black}{\underbar{\bf Author response}}: The fact that the norm relation (2.40) holds for all elements of the algebra ${\cal K}^{\lambda}$ does not require either rescue or remedy. It is a proven fact. No counterexample to (2.40) can or does exist. In the previous round of reviews the present reviewer attempted to allege a counterexample to (2.40). However, I demonstrated that the counterexample fails, as it must, because (2.40) has been rigorously proven by me and now also by Reviewer~\#~3. On the other hand, the contradiction in the alleged counterexample was achieved by the present reviewer and the critique [2] by employing two different product rules on the two sides of the equation $\|X Y\|=\|X\|\,\|Y\|$, and also by mixing up two different product rules on its left-hand side. Contradiction can be drawn from any mathematical equation by employing different rules on its two sides. Thus the alleged counterexample was both logically and mathematically flawed. Since no counterexample to the norm relation (2.40) can or does exist, and since explicit proof of it exists, no remedy or rescue is needed for its validity. It holds for all elements $X$ \& $Y$ of the algebra ${\cal K}^{\lambda}$.

In the alleged counterexample that failed, the reviewer considered two-dimensional objects $X=\varepsilon-1$ and $Y=\varepsilon+1$ within ${\cal K}^{\lambda}$ by setting the coordinates of the remaining six dimensions of ${\cal K}^{\lambda}$ to zero. That amounted to the real and dual quaternions not being orthogonal to each other --- {\it i.e.}, that amounted to setting ${\bf q}_r\,{\bf q}^{\dagger}_d+{\bf q}_d\,{\bf q}^{\dagger}_r \not= 0$, and thereby violating the normalization condition (2.54) I have used in my papers to reduce the values of the norms to scalar quantities. This violation of orthogonality or normalization condition in the reviewer's counterexample is also noticed by Reviewer~\#~3 whose calculations I reproduced above. It is then not surprising that the two-dimensional objects considered by the reviewer cannot be normalized to scalar quantities. The reviewer's strategy therefore was to simply pick the scalar parts of the geometric products $XX^{\dagger}$ and $YY^{\dagger}$ and substitute their square roots as norms into $\|X Y\|=\|X\|\,\|Y\|$ to derive contradiction. But that strategy inevitably depends on applying two different product rules on the two sides of the equation $\|X Y\|=\|X\|\,\|Y\|$ because $XY$ on its left-hand side is inevitably a geometric product and not a scalar product. Thus contradiction can be achieved only by introducing inconsistency. 

It is also important to note that word ``currently'' in the above sentence by the reviewer again gives the wrong impression that I have invented a new strategy ``now'' amounting to ``remedy'' or ``rescue'' of (2.40), in the light of the critiques [2] and [14], or during this review process of the current manuscript. But it should be evident from the yellow highlighted paragraph I have quoted above from my original 2018 paper that that impression is not correct.  

\underbar{\bf Author action}: No action.}

This does not have a justification with what is done for complex numbers, quaternions and octonions, despite what the author says. For the first two, the quantity $X \tilde{X}$, is automatically a scalar with no restriction being necessary. For octonions, the conjugate octonion is constructed so that its octonionic product with the original octonion, returns a positive scalar. It is for this reason that restrictions are not necessary in these cases.

{\color{black}{\underbar{\bf Author response}}: The above comments by the reviewer are also mistaken. To begin with, contrary to what the reviewer claims, any normalization process is inevitably a restriction on the coefficients of the multivector such as $X$, regardless of whether that multivector is a complex number, quaternion, octonion, or an element of ${\cal K}^{\lambda}$. In one of my comments above I gave a simple example of ordinary 3-vectors normalized to construct $S^2$ embedded in $\mathrm{I\!R}^3$.

Any multivector $X$ in general, regardless of its algebra, would not be a scalar quantity. The usual strategy to infer its ``length'' is to consider an entirely different multivector $X^{\dagger}$, which is the ``conjugate'' or ``reverse'' of $X$. It is then noted that their geometric product $XX^{\dagger}$ is a scalar quantity, say $\varrho^2$, at least in the case of complex numbers, quaternions, and octonions, so that we can say the ``length'' of $X$ is
\begin{equation}
\|X\| = \sqrt{XX^{\dagger}}=\varrho. \notag
\end{equation}
But this is a constraint equation. Contrary to what is claimed by the reviewer, it puts a restriction on the coefficients of $X$, just as the restriction on the coefficients of the 3-vectors in the case of $S^2$ embedded in $\mathrm{I\!R}^3$ I considered above. 

Thus what I have used in my original paper to infer the length of $X$ is no different in principle from this strategy. For arbitrary multivector ${\mathbb Q}_z=\, {\bf q}_{r} + {\bf q}_{d}\,\varepsilon$ in ${\cal K}^{\lambda}$, in [1] I have used the following definition for a scalar-valued norm:
\begin{equation}
||{\mathbb Q}_z||:=\sqrt{\left({\mathbb Q}_{z}{\mathbb Q}^{\dagger}_{z}\right)\!\Big|_{\,{\bf q}_{r}\,{\bf q}^{\dagger}_{d}+{\bf q}_{d}\,{\bf q}^{\dagger}_{r}\,=\,0}\,}\;=\varrho, \notag
\end{equation}
which gives the same value for the norm as that obtained using the traditional {\it ad hoc} definition in Geometric Algebra (cf. Eq.~(2.8) in [1]). This is explained also with more detail in [25]. As explained in Eqs.~(2.54) to (2.56) in [1], the quadratic form in the above definition with split-complex image is
\begin{equation}
{\mathbb Q}_{z}{\mathbb Q}^{\dagger}_{z}=\left({\bf q}_{r}\,{\bf q}^{\dagger}_{r}+{\bf q}_{d}\,{\bf q}^{\dagger}_{d}\right)+\left({\bf q}_{r}\,{\bf q}^{\dagger}_{d}+{\bf q}_{d}\,{\bf q}^{\dagger}_{r}\right)\varepsilon=\left(\varrho_r^2+\varrho_d^2\right)+\left({\bf q}_{r}\,{\bf q}^{\dagger}_{d}+{\bf q}_{d}\,{\bf q}^{\dagger}_{r}\right)\varepsilon \notag
\end{equation}
and $\varepsilon\cdot\left({\mathbb Q}_{z}{\mathbb Q}^{\dagger}_{z}\right)={\bf q}_{r}\,{\bf q}^{\dagger}_{d}+{\bf q}_{d}\,{\bf q}^{\dagger}_{r}$ is the scalar coefficient of $\varepsilon$. It is easy to see that this definition gives a scalar value for the norm as the Pythagorean distance from the origin in eight dimensions,
\begin{equation}
||{\mathbb Q}_z||=\sqrt{\varrho_r^2+\varrho_d^2\,}=\sqrt{g^2+u_x^2+u_y^2+u_z^2+h^2+v_x^2+v_y^2+v_z^2\,}, \notag
\end{equation}
with the notation ${\bf q}_{r}=g+I_3{\bf u}$ and ${\bf q}_{d}=h+I_3{\bf v}$, and therefore the norms so defined are also positive definite:
\begin{equation}
||{\mathbb Q}_{z}||=0\iff{\mathbb Q}_{z}=0. \notag
\end{equation}

Finally, note that while both definitions of normalization discussed above --- the traditional one and the one I have used in my paper --- are constraint equations and therefore put restrictions on the coefficients of the multivectors $X$ to deduce their lengths, they do not put any restrictions on the corresponding algebras itself. Thus the claim by the reviewer that the definition of norm I have used in my paper puts a restriction on the algebra ${\cal K}^{\lambda}$ is not correct.    

\underbar{\bf Author action}: In Section~2.8 of the manuscript I have added a proof of the positive definiteness of the norm.}

$\quad\quad$ Returning to the $4 \mathrm{~d}$ Euclidean Clifford algebra, if norms really were to be taken as being the square roots of scalar+pseudoscalar quantities, a further point is that we would need to worry about `branch cuts' in the definition of the square roots, in an analogous way to what we need to worry about in taking the square root of a complex number. This doesn't seem to be worried about by the author, maybe because of his belief that the restriction of the type contained in his equation (2.54) (in the RSOS paper), which is the same as my restriction $\langle X \tilde{X}\rangle_{4}=0$ above, somehow doesn't in fact restrict the quantities being considered, and we can apply it in general, and thereby rule out of consideration the quantities $1 \pm \epsilon$ used in the counterexample.

{\color{black}{\underbar{\bf Author response}}: That is correct. As I noted above, normalization conditions do not put restrictions on algebra. 

\underbar{\bf Author action}: No action.}

$\quad\quad$ So my summary so far, is that the author is being disingenuous in saying that he was using the definition $\|X\|=\sqrt{X \tilde{X}}$ in Section $2.5$ of the original paper. This is because Section $2.4$ (in particular equation (2.38)) clearly shows he was using (2.8) at that time, and also because of the fact that he felt it necessary to restrict objects by imposing the condition (2.54) (whilst simultaneously claiming this was not a restriction).

{\color{black}{\underbar{\bf Author response}}: All mathematical statements made is Sections~2.4 and 2.5 of the original paper [1] are correct, unambiguous, and mutually consistent. In particular, Eq.~(2.38) explicitly considers {\it unit} multivectors. In other words, multivectors with unit {\it scalar-valued} lengths. It is quite clear from the discussion in the highlighted paragraph of Section~2.5 reproduced above that the value of the length deduced using the normalization condition (2.54) in Section~2.5 would be identical to that deduced using the traditional normalization condition (2.8) used in Section~2.4:
\begin{equation}
\sqrt{\,\langle\,{\mathbb Q}_z\,{\mathbb Q}_z^{\dagger}\rangle_s}\;\equiv\sqrt{\left({\mathbb Q}_{z}{\mathbb Q}^{\dagger}_{z}\right)\!\Big|_{\,{\bf q}_{r}\,{\bf q}^{\dagger}_{d}+{\bf q}_{d}\,{\bf q}^{\dagger}_{r}\,=\,0}\,}\;. \notag
\end{equation}
Once again, Eq.~(2.54) is a normalization condition. It by no means puts any restriction on the algebra ${\cal K}^{\lambda}$ itself. 

\underbar{\bf Author action}: No action.}

$\quad\quad$ Leaving this all aside, we can provisionally accept that the author now wants to work with the definition $\|X\|=\sqrt{X \tilde{X}}$. The relation $\|X Y\|=$ $\|X\|\|Y\|$ is true for this definition, since we can write
\begin{equation}
\|X Y\|^{2}=(X Y) \widetilde{(X Y)}=X Y \tilde{Y} \tilde{X}=(X \tilde{X})(Y \tilde{Y}) \tag{1.4}
\end{equation}
where we can move the $\tilde{X}$ through the $Y \tilde{Y}$ since the $4 \mathrm{~d}$ pseudoscalar commutes with all even elements. 

{\color{black}{\underbar{\bf Author response}}: I once again object to the adjective ``now'' in the first sentence of the above comment, unless by ``now'' the reviewer means the year 2018. As I noted before, I am not inventing new strategies of defence ``now.''

\underbar{\bf Author action}: No action.}

However, this definition will not allow the author to say anything about octonions (since this isn't how octonion norms work). 

{\color{black}{\underbar{\bf Author response}}: I do not intend to say anything about octonions. My paper concerns an associative algebra ${\cal K}^{\lambda}$,\break which is the even subalgebra of the Clifford algebra $\mathrm{Cl}_{4,0}$, not the Cayley algebra with octonionic product rules. 

\underbar{\bf Author action}: No action.}

Moreover, we do not need this to be the definition in order to prove that the product of scaled rotors is another scaled rotor, since this works using the original (2.8) definition.

{\color{black}{\underbar{\bf Author response}}: I disagree with the above comment. To begin with, the traditional definition (2.8) is {\it ad hoc} because it goes against the very essence of geometric algebra in which the fundamental product is geometric product, not the scalar product being picked out in $\langle\dots\rangle_s$ by disregarding the rich algebra and geometry contained in the non-scalar part prematurely. In addition, definition (2.8) leads to serious inconsistency when products of two or more multivectors are involved in an equation such as $||XY||= ||X||\,||Y||$. If we use definition (2.8) to evaluate the norms on the two sides of this equation, then on the left-hand side of it we would end up using two different product rules, the geometric product rule between $X$ and $Y$ and the scalar product rule for evaluating the norm $||XY||$. Whereas on its right-hand side we would end up using only the scalar product rule for evaluating $||X||$ and $||Y||$. This inconsistent application of two different product rules on the two sides of the norm relation is the root cause of the contradiction in the alleged counterexample in [2] and [14]. Thus definition (2.8) has limited applicability. 

\underbar{\bf Author action}: No action.}

Finally, we need to be clear that this definition of norm does not agree with the usual mathematical one, since in mathematics, a norm takes us from a space of elements of some kind, to the non-negative real numbers, and the new `norm' now favoured by the author, fails to do this.

{\color{black}{\underbar{\bf Author response}}: I disagree with this claim. It is evident that the normalization condition (2.54) I have used in the paper [1] also takes us from a space ${\cal K}^{\lambda}$ of eight-dimensional multivectors to the non-negative real numbers.''

\underbar{\bf Author action}: No action.}

$\quad\quad$ A possible way forward, which would allow the author to continue using the correct norm (2.8), is if the author is willing to admit that the entire mathematical development through to the end of Section $2.5$ of the RSOS paper amounts to the assertion that the product of scaled rotors is another scaled rotor. We would then have nothing more to argue about on this point.

{\color{black}{\underbar{\bf Author response}}: As I explained above, definition (2.8) is not universally applicable. By now it should be evident that there are very good reasons why I have instead introduced the correct normalization condition (2.54) in [1].

\underbar{\bf Author action}: No action.}

$\quad\quad$ This would remove anything of interest connected with octonions of course, since the preservation of the (true) norm would then only apply to a selected subset of elements, and thus remove the claim of having found an associative version of the octonion algebra, which would have to work for all elements, but it may be sufficient mathematical equipment to be able to carry on with what the author needs for his purposes in the second part of the paper. This has yet to be determined.

{\color{black}{\underbar{\bf Author response}}: Nowhere in [1] have I made a claim that I have found an associative version of octonion algebra. My paper does not concern octonions or Cayley algebra. It concerns the eight-dimensional associative algebra ${\cal K}^{\lambda}$ (which is the even subalgebra of the Clifford algebra $\mathrm{Cl}_{4,0}$) and the corresponding 7-sphere embedded within ${\cal K}^{\lambda}$. 

\underbar{\bf Author action}: No action.}

$\quad\quad$ (Note there are several more parts of the author's reply, but these basically amount to statements, without evidence, that he is right, so it is not really necessary to reply to these.

{\color{black}{\underbar{\bf Author response}}: This claim by the reviewer is not correct. It can be easily verified that I have not made any statements in my replies in the previous round without providing evidence and reasons for my disagreements.  

\underbar{\bf Author action}: No action.}

An example would be what I pointed out last time concerning the span of two sets of elements in fact being the same even though they are given different symbols. The sets of elements in question are the same except for signs, and the span of a set of elements consists of all possible linear sums of them, with arbitrary coefficients. This means that two sets that differ only in the signs of the elements must have the same span, but the author continues to deny this. I appreciate fully that he is trying to talk about orientations, but what he has written are spans, and hence are certainly the same.)

{\color{black}{\underbar{\bf Author response}}: As I pointed out in the previous round of reviews, what is raised in the above comments is a non-issue. As I have explained in the manuscript, there is no error in [1] in this regard.  Nowhere have I claimed that ${\cal K}^{+}$ and ${\cal K}^{-}$ span different vector spaces. They differ only in orientation $\lambda$, and this orientation plays a role of a Bell-type hidden variable representing the initial state of the physical system in the 7-sphere model. I have already explained this adequately in Section~2.3 of my original paper [1] and in Section~2.6 of the current manuscript. In the last round, Reviewer \# 1 also made the following observation in support of my view on this matter:
\begin{quote}
Reviewer \# 1: ``\textcolor{blue}{Section~2.6 makes a valid point, both Lasenby and Gill have misunderstood that you are referring to differently oriented vector spaces not to different vector spaces regardless of orientation.}''
\end{quote}
As noted in [1] and the manuscript, since $\lambda$ plays the role of a hidden variable as understood within Bell's local-realistic framework, ${\cal K}^{+}$ and ${\cal K}^{-}$ specified in the equations (2.33) and (2.34) of my 2018 paper [1] are physically {\it not} identical. That is why ${\cal K}^{+}$ and ${\cal K}^{-}$ are specified with two different symbols. This is not difficult to understand. 

\underbar{\bf Author action}: For clarity, at the end of Section~2.6 in the revised manuscript I have added the following sentence: ``Therefore ${{\cal K}^+}$ and ${{\cal K}^-}$ are physically not identical within the 7-sphere framework proposed in [1].'' }

\bigskip

\bigskip

\hrule 

\bigskip

\underbar{\textcolor{blue}{{\large\bf Reviewer \# 7}}}

I believe that the usual and well accepted judgements issued with regard to a mathematical inequality---whose origins can be traced to Boole in 19th century as the author has reminded his readers---on mostly metaphysical questions relating to the nature of physical reality, locality etc., can and should be subjected to questioning and further investigation---particularly in view of the understanding of the nature of space itself that is provided by the power mathematical tools offered by geometric algebra and Einstein's theory of gravity.

It appears to me that a quest for a better understanding of the nature of space is directly connected to addressing the well accepted conclusions, and wisdom, from Bell's theorem regarding physical reality and locality that are presented as the most logical, direct, and doubtless outcomes of the theorem.

I see Christian's work from the past many years as such a quest that has raised hope for more satisfactory answers to difficult questions at the heart of foundations of quantum mechanics.

I believe that while respecting the mathematical core of Bell's theorem, often presented as a system of inequalities, the usual conclusions, and wisdom, regarding the nature of physical reality, locality, etc. can and needs to be subjected to questioning and further investigation. If an improved description of space, as offered by geometric algebra or the powerful mathematical tools of general relativity, can help us in revising these usual conclusions then why should there be any objection or resistance to that and why such an effort should be viewed as a scientific heresy as seems to be the case.

Reviewer \#1 recommends removing the discussion of a mathematically valid and a physically valid equation.

{\color{black}{\underbar{\bf Author response}}: I have dealt with that recommendation by Reviewer \# 1 in my response to it above.

\underbar{\bf Author action}: No action.}

Reviewer \#5 considers the author's discussion on this in Section 2.4 to be ``incoherent argumentation" that should not be published in a scientific context. Referring to a number of earlier mentions of this problem in literature on the foundations of quantum mechanics, the author has kept this discussion in his manuscript, and has further clarified it in his revision, and to which I agree.

Reviewer \#2 has objected to the author's use of geometry of space-time as offered by general relativity (GR) relevant for explaining Bell's experiments? I do not see why such an investigation can or should be criticized in particular if it can offer a hope in answering what many find as the troubling questions, usual implications and accepted wisdom, regarding Bell's theorem.

Reviewer \#6 has made a list of technical remarks/objections to which the author has given what I see as satisfactory responses.

I believe that he has presented convincing and careful replies, with attention to details, to all comments and objections from referees and this manuscript should be accepted in its present form.

{\color{black}{\underbar{\bf Author response}}: I thank the reviewer for positive review and for recommending my manuscript for publication.

\underbar{\bf Author action}: No action.}

\bigskip
\bigskip
{\color{black}
\hrule
\bigskip
\hrule}}}

\end{document}